\documentclass[twoside, 11pt]{article}

\usepackage{jmlr2e}

\usepackage{amsmath}
\usepackage{graphicx,psfrag,epsf}
\usepackage{enumerate}
\usepackage{natbib}
\usepackage{setspace}
\usepackage{microtype}

\usepackage[dvipsnames]{xcolor}
\usepackage{amsfonts, amsmath,amssymb,dsfont}
\usepackage{mathtools}
\usepackage[noend]{algpseudocode}
\usepackage{breakcites}
\usepackage{slashbox}
\usepackage{amsfonts, amsmath,amssymb,dsfont}
\usepackage{array}
\usepackage{cases} 
\DeclareMathAlphabet{\mathpzc}{OT1}{pzc}{m}{it}
\usepackage{multirow, multicol, tabularx, booktabs}
\usepackage{algorithm}
\usepackage[noend]{algpseudocode}
\usepackage{etoolbox}
\usepackage{accents}
\usepackage{array}
\usepackage{cases} 
\usepackage{url} 
\usepackage{caption}
\usepackage{subcaption}
\usepackage{threeparttable}

\usepackage{enumerate}
\newcommand{\p}{{\rm I}\kern-0.18em{\rm P}}
\newcommand{\1}{{\rm 1}\kern-0.24em{\rm I}}
\newcommand{\E}{{\rm I}\kern-0.18em{\rm E}}

\makeatletter
\def\BState{\State\hskip-\ALG@thistlm}
\makeatother

\DeclarePairedDelimiter\floor{\lfloor}{\rfloor}
\DeclarePairedDelimiter\ceil{\lceil}{\rceil}
\newcommand{\tran}{^{\mkern-1.5mu\mathsf{T}}}
\newcommand{\R}{{\rm I}\kern-0.18em{\rm R}}
\newcommand{\N}{{\rm I}\kern-0.18em{\rm N}}

\DeclareMathOperator*{\argmin}{arg\,min}
\usepackage[english]{babel}

\newtheorem{condition}{Condition}

 \RequirePackage{pgfopts}
\newcommand\given[1][]{\:#1\vert\:}

\newcommand{\bd}[1]{\boldsymbol{#1}}
\newcommand{\cN}[1]{\mathcal{N}}
\newcolumntype{P}[1]{>{\centering\arraybackslash}p{#1}}
\newcommand{\uderbar}[1]{\underset{\raise0.3em\hbox{$\smash{\scriptscriptstyle-}$}}{#1}}

\newcommand{\jl}[1]{\textcolor{black}{#1}}

\newcommand{\jjl}[1]{\textcolor{black}{#1}}

\newcommand{\rv}[1]{\textcolor{black}{#1}}

\newcommand{\rvv}[1]{\textcolor{black}{#1}}

\usepackage{lastpage}
\jmlrheading{22}{2021}{1-\pageref{LastPage}}{6/20; Revised
5/21}{5/21}{20-673}{Jingyi Jessica Li, Yiling Elaine Chen, and Xin Tong}
\ShortHeadings{Prediction-based marginal feature ranking}{Li, Chen, and Tong}
\firstpageno{1}

\begin{document}




\title{A flexible model-free prediction-based framework for feature ranking}
\author{\name Jingyi Jessica Li  \email jli@stat.ucla.edu \\
       \addr Department of Statistics\\
University of California, Los Angeles 
\AND 
\name  Yiling Elaine Chen  \email yiling0210@ucla.edu  \\
        \addr  Department of Statistics\\
University of California, Los Angeles 
\AND 
  \name Xin Tong \email xint@marshall.usc.edu \\
       \addr Department of Data Sciences and Operations \\
       Marshall Business School\\
       University of Southern California
}

\editor{Isabelle Guyon}

\maketitle

\begin{abstract}

Despite the availability of numerous statistical and machine learning tools for joint feature modeling, many scientists investigate features marginally, i.e., one feature at a time. This is partly due to training and convention but also roots in scientists' strong interests in simple visualization and interpretability. As such, marginal feature ranking for some predictive tasks, e.g., prediction of cancer driver genes, is widely practiced in the process of scientific discoveries. In this work, we focus on marginal ranking for binary classification, one of the most common predictive tasks.  We argue that the most widely used marginal ranking criteria, including the Pearson correlation, the two-sample $t$ test, and two-sample Wilcoxon rank-sum test, do not fully take feature distributions and prediction objectives into account. To address this gap in practice, we propose two ranking criteria corresponding to two prediction objectives: the classical criterion (CC) and the Neyman-Pearson criterion (NPC), both of which use model-free nonparametric implementations to accommodate diverse feature distributions. Theoretically, we show that under regularity conditions, both criteria achieve sample-level ranking that is consistent with their population-level counterpart with high probability. Moreover, NPC is robust to sampling bias when the two class proportions in a sample deviate from those in the population. This property endows NPC good potential in biomedical research where sampling biases are ubiquitous. We demonstrate the use and relative advantages of CC and NPC in simulation and real data studies. Our model-free objective-based ranking idea is extendable to ranking feature subsets and generalizable to other prediction tasks and learning objectives. 
\end{abstract}


\section{Introduction}

From scientific research to industrial applications, practitioners often face the challenge to rank features for a prediction task.  Among the ranking tasks performed by scientists and practitioners, a large proportion belongs to marginal ranking; that is, ranking features based on the relation between the response variable and one feature at a time, ignoring other available features. For example, to predict cancer driver genes, biomedical researchers need to first extract predictive features from patients' data. Then they decide whether each extracted feature is informative by examining its marginal distributions in tumor and normal tissues, usually by boxplots and histograms \citep{davoli2013cumulative,lyu2020dorge}. Another example is genome-wide association studies, where single nucleotide polymorphisms are ranked by their marginal associations with a phenotype \citep{buniello2019nhgri}. 

From a prediction perspective, marginal feature ranking may seem suboptimal, as multiple features usually have dependence and would thus be jointly predictive of the response variable beyond a simple additive manner. \rv{Hence, interpretation of all features' importance in a multivariate predictive model is an active area of research; popular criteria include the SHAP value \citep{lundberg2017unified} and the feature importance measured by Gini index in the random forest (RF) algorithm.} However, \rv{such ``joint'' feature ranking is too computationally intensive when the candidate feature number is huge, as it requires all candidate features to be input into one multivariate model; also, it does not reflect a feature's marginal predictive power when other highly-correlated candidate features are in the model.} Moreover, the popularity of marginal feature ranking roots in not only researchers' education backgrounds and discipline conventions but also their strong desire for simple interpretation and visualization in the trial-and-error scientific discovery process. As such, marginal feature ranking has been an indispensable data-analysis step in the scientific community, and it will likely stay popular.    

In practice, statistical tests (e.g., two-sample $t$ test and two-sample Wilcoxon rank-sum test) and association measures (e.g., Pearson correlation) are often used to rank features marginally \citep{davoli2013cumulative, lyu2020dorge}. However, these tests and association measures do not reflect the objective of a prediction task. For example, if the classification error is of concern, it is unclear how the significance of these tests or the values of these measures are connected to the classification error. This misalignment of ranking criterion and prediction objective is undesirable: the resulting feature rank list does not reflect the marginal importance of each feature for the prediction objective. Hence, scientists and practitioners call for a marginal ranking approach that meets the prediction objective.

In this work, we focus on marginal ranking for binary prediction, which can be formulated as binary classification in machine learning.   Binary classification has multiple prediction objectives, which we refer to as paradigms here. \jjl{These paradigms include} (1) the \textit{classical} paradigm that minimizes the classification error, i.e., a weighted sum of the type I and type II errors where the weights are the class probabilities \citep{Hastie.Tibshirani.ea.2009, james2013introduction}, (2) the \textit{cost-sensitive learning} paradigm that replaces the \jl{two error weights by pre-determined costs} \citep{Elkan01, ZadLanAbe03}, (3) the \textit{Neyman-Pearson (NP)} paradigm that minimizes the type II error subject to a type I error upper bound \citep{cannon2002learning, scott2005neyman, tong2013plug, tong2016neyman}, and (4) the \textit{global} paradigm that focuses on the overall prediction accuracy under all possible thresholds: the area under the receiver-operating-characteristic curve (AUROC) or the area under the precision-recall  curve (AUPRC). Here we consider marginal ranking of features under the classical and NP paradigms, and we define the corresponding ranking criteria as the classical criterion (CC) and the Neyman-Pearson criterion (NPC). To implement CC and NPC, we take a model-free approach by using nonparametric estimates of class-conditional feature densities. This approach makes CC and NPC more adaptive to diverse feature distributions than existing criteria for marginal feature ranking. The idea behind CC and NPC is easily generalizable to the cost-sensitive learning paradigm and the global paradigm. 

It is worth \jl{highlighting that NPC} is robust to sampling bias; that is, even when the class \jl{proportions in a sample \jjl{deviate} from those in the population, NPC still achieves feature} ranking consistency between sample and population with high probability. This \jl{nice property makes NPC particularly useful for disease diagnosis, where a long-standing obstacle is that the proportions of diseased patients and healthy people in medical records do not reflect the proportions in the population.}   

%
%

The rest of the paper is organized as follows. In Section \ref{sec:background}, we define the population-level CC and NPC as the oracle criteria under the classical and NP paradigms, respectively. In Section \ref{sec:methods}, we define the sample-level CC and NPC, and we develop model-free algorithms to implement them. In Section \ref{sec:theoretical properties}, we derive theoretical results regarding the ranking consistency of the sample-level CC and NPC in relation to their population counterparts. In Section \ref{sec:simulation}, we use simulation studies to demonstrate the performance of sample-level CC and NPC in ranking low-dimensional and high-dimensional features. \rv{We also implement variants of sample-level CC and NPC based on the support vector \rvv{machine} (SVM) algorithm and show that they are less robust than our proposed sample-level CC and NPC to sampling bias.} In Section \ref{simu:realdata}, we apply sample-level CC and NPC to marginal feature ranking in two real datasets. Using the first dataset regarding breast cancer diagnosis, we show that both criteria can identify informative features, many of which have been previously reported; we also provide a Supplementary Excel File for literature evidence. Using the second dataset for prostate cancer diagnosis from urine samples, we demonstrate that NPC is robust to sampling bias. \rv{In both simulation and real-data studies, we compare sample-level CC and NPC with joint feature ranking criteria---the SHAP value and the feature importance measures in the RF algorithm---and commonly-used marginal ranking criteria that may give feature ranking misaligned with the prediction objective, including the Pearson correlation, the distance correlation \citep{szekely2009brownian},\footnote{In binary classification, the response variable is encoded as $0$ and $1$ and treated as a numerical variable in the calculation of of the Pearson and distance correlations.} the two-sample $t$ test, and the two-sample Wilcoxon rank-sum test.} We conclude with a discussion in  Section \ref{sec:conclusions}. Additional materials and proofs of lemmas, propositions, and theorems are relegated to the Appendix.

The code for reproducing the numerical results is available at \url{http://doi.org/10.5281/zenodo.4680067}. The R package \texttt{frc} is available at \url{https://github.com/JSB-UCLA/frc}.

\section{Population-level ranking criteria}\label{sec:background}
%

In this section, we introduce two objective-based marginal feature  ranking criteria, \jjl{on the population level,} under the classical paradigm and the Neyman-Pearson (NP) paradigm. As argued previously, when \jjl{one has} a learning/prediction objective, the feature ranking criterion should be in line with that. Concretely, the $j$-th ranked feature should be the one that achieves the $j$-th best performance based on that objective.   
This objective-based feature ranking perspective is extendable to ranking feature subsets (e.g., feature pairs). Although we focus on marginal feature ranking in this work, to cope with this future extension, our notations in the methodology and theory development are compatible with ranking feature subsets.

\subsection{Notations and classification paradigms}

We first introduce essential mathematical notations to facilitate our discussion. Let $\left(\bd X,Y\right)$ be a pair of random observations where $\bd X \in \mathcal{X} \subseteq {\R}^d$ is a vector of features and $Y\in \left\{ 0,1 \right\}$ indicates the class label of $\bd X$. A \textit{classifier} $\phi:\mathcal{X}\rightarrow  \left\{ 0,1 \right\}$ maps from the feature space to the label space. A \textit{loss function} assigns a cost to each misclassified instance $\phi(\bd X) \neq Y$, and the \textit{risk} is defined as the expectation of this loss function with respect to the joint distribution of $\left( \bd X,Y\right)$. We adopt in this work a commonly used loss function, the $0$-$1$ loss: $\1\left(\phi(\bd X)\neq Y \right)$, where $\1(\cdot)$ denotes the indicator function. Let $\p$ and $\E$ denote the generic probability distribution and expectation, whose meaning depends on specific contexts. With the choice of the $0$-$1$ loss function, the risk is the classification error: $R(\phi) = \E \left[ \1\left( \phi(\bd X)\neq Y\right) \right] = \p \left( \phi(\bd X)\neq Y\right)$, which is aligned with most practitioners' interest in classifier evaluation. \rv{Note that in this work, the $0$-$1$ loss is only used as an evaluation criterion in our development of marginal ranking criteria, not as a loss function for training a classifier from data.}

In this paper, we call the learning objective of minimizing $R(\cdot)$ the \textit{classical paradigm}.  Under \jjl{this} paradigm, one aims to mimic the \textit{classical oracle classifier} $\varphi^{*}$ that minimizes the population-level classification error, 
$$
\varphi^{*}=\argmin \limits_{\phi: \R^d\rightarrow \{0, 1\}} R\left( \phi\right)\,.
$$ 
It is well known in literature that the classical oracle $\varphi^*(\cdot) = \1 (\eta (\cdot) > 1/2)$,  where $\eta(\bd x) = \E (Y|\bd X=\bd x)$ is the regression function \citep{koltchinskii2011introduction}. Equivalently, $\varphi^*(\cdot) = \1(p_1(\cdot)/p_0(\cdot)>\pi_0/\pi_1)$, where $\pi_0 =\p(Y=0)$, $\pi_1 =\p(Y=1)$, $p_0$ is the probability density function of $\bd X|(Y=0)$, and $p_1$ is the probability density function of $\bd X|(Y=1)$.  Note that the risk can be decomposed as follows:
\begin{align*}
 	R(\phi) &= \p(Y=0)\cdot\p\left( \phi(\bd X) \neq Y \given Y=0\right) + \p(Y=1)\cdot \p\left( \phi(\bd X) \neq Y \given Y=1\right)\\
 	 &= \pi_0 R_0\left(\phi\right)+ \pi_1 R_1\left(\phi\right)\,,
 \end{align*} where $R_j\left(\phi\right) = \p\left( \phi(\bd X) \neq Y \given Y=j\right)$, for $j= 0 \text{ or } 1$. The notations $R_0(\cdot)$ and $R_1(\cdot)$ denote the population-level type I and II errors respectively. Note that minimizing $R(\cdot)$ implicitly \jjl{imposes} a weighting of $R_0$ and $R_1$ by $\pi_0$ and $\pi_1$. This is not always desirable.  For example, when practitioners know the explicit costs for making type I and II errors: $c_0$ and $c_1$, they may want to optimize the criterion $c_0R_0(\cdot) + c_1 R_1(\cdot)$, which is often referred to as \textit{the cost-sensitive learning paradigm}.   
 
\jjl{In parallel to the classical paradigm, we consider the \textit{NP paradigm}, which} aims to mimic the \textit{level-$\alpha$ NP oracle classifier} \jjl{that minimizes the type II error while constraining the type I error under $\alpha$, a user-specified type I error upper bound,} 
 \begin{align}\label{eq:NP_oracle}
 \varphi^{*}_{\alpha} = \argmin \limits_{\varphi: R_0(\varphi) \leq \alpha} R_1(\varphi)\,.
\end{align} 
Usually, \jjl{$\alpha$ is a small value (e.g., $5\%$ or $10\%$), reflecting a conservative attitude towards the type I error.}  As the development of classifiers under the NP paradigm is relatively new, \jjl{here we review the NP oracle classifier} $\varphi^*_{\alpha}(\cdot)$. \jjl{Motivated by the classic NP Lemma (Appendix \ref{sec::np lamma}) and a correspondence between classification and statistical hypothesis testing,} $\varphi^*_{\alpha}$ in \eqref{eq:NP_oracle} can be constructed by thresholding $p_{1}(\cdot)/p_{0}(\cdot)$ at a proper level  $C^*_{\alpha}$ \citep{tong2013plug}:
 \begin{equation}\label{equ: neyman_pearson}
 	\varphi_{\alpha}^*(\bd x) = \1\left(p_1(\bd x)/p_0(\bd x) > C_\alpha^*\right)\,, \end{equation}
\rv{where $C^*_{\alpha}$ is such that \rvv{$\p(p_1(\bd X)/p_0(\bd X)>C^*_{\alpha}|Y=0) = \alpha$}, and the estimation of $C^*_{\alpha}$ is introduced in Section \ref{sec: construction of NP}.}

In addition to the above three paradigms, a common practice is to evaluate a classification algorithm by its AUROC or AUPRC, which we refer to as the \textit{global paradigm}. In contrast to the above three paradigms that lead to a single classifier, which has its corresponding type I and II errors, the global paradigm evaluates a classification algorithm by aggregating its all possible classifiers with type I errors ranging from zero to one. For example, the oracle AUROC is the area under the curve
$ \left\{ \left(R_0(\varphi_\alpha^*),\, 1-R_1(\varphi_\alpha^*)\right): \alpha \in [0,1]
	\right\}$.

\subsection{Classical and Neyman-Pearson criteria on the population level}\label{sec:NPC_population}
Different learning/prediction objectives in classification induce distinct feature ranking criteria. We first define the population-level CC and NPC. Then we show that these two criteria lead to different rankings of features in general, and that NPC may rank features differently at different $\alpha$ values. We denote by $\varphi^*_{A}$ and $\varphi^*_{\alpha A}$, respectively, the classical oracle classifier and the level-$\alpha$ NP oracle classifier that only use the features indexed by $A \subseteq \{1,\ldots, d \}$.  This paper focuses on the case when $|A| = 1$.   
Concretely, under the classical paradigm, the classical oracle \jjl{classifier on index set $A$, $\varphi^*_{A}$,} achieves 
\begin{equation*}
	R \left(\varphi^*_{A}\right) = \min_{\varphi_A} R \left(\varphi_{A}\right) = \min_{\varphi_A} \p (\varphi_{A}(\bd X)\neq Y)\,,
\end{equation*} 
where $\varphi_A: \mathcal X \subseteq \R^d \rightarrow \{0, 1\}$ is any mapping that first projects $\bd X\in \R^d$ to its $|A|$-dimensional sub-vector $\bd X_A$, which comprises of the coordinates of $\bd X$ corresponding to the index set $A$, and then maps from $\bd X_A\in \R^{|A|}$ to $\{0, 1\}$. Analogous to $\varphi^*(\cdot)$, we know   
\begin{align}\label{eqn:classical oracle}
\varphi^*_{A}(\bd x) = \1(\eta_A(\bd x_A) > 1/2) = \1(p_{1A}(\bd x_A)/p_{0A}(\bd x_A) > \pi_0 / \pi_1)\,, 
\end{align}
where $\eta_A(\bd x_A)  = \E (Y|\bd X_A=\bd x_A)$ is the regression function using only features in the index set $A$, and $p_{1A}$ and $p_{0A}$ denote the class-conditional probability density functions of the features $\bd X_A$.  Suppose that candidate feature subsets denoted by $A_1, \ldots, A_J$ are provided, which may be enumerated by a computational algorithm or curated by domain experts. \jjl{We define the \textit{population-level classical criterion} (p-CC) of $A_i$ as its \textit{optimal} risk $R\left(\varphi^*_{A_i}\right)$; i.e., $A_1, \ldots, A_J$ will be ranked based on $\left\{R \left(\varphi^*_{A_1}\right), \ldots, R \left(\varphi^*_{A_J}\right) \right\}$, with the smallest being ranked the top}. The prefix ``p" in p-CC indicates ``population-level."
 Note that \jjl{$R(\varphi^*_{A_i})$ represents} $A_i$'s best achievable performance measure under the classical paradigm and \jjl{does} not depend on any specific models \jjl{assumed for} the distribution of $(\bd X, Y)$.

Under the NP paradigm, the NP oracle \jjl{classifier} defined on the index set $A$, $\varphi^*_{\alpha A}$,  achieves 
\begin{equation}\label{ideaL_sormulation_np}
	R_1 \left(\varphi^*_{\alpha A}\right) = \min_{\substack{\varphi_{A}  \\ R_0 \left(\varphi_{\alpha A}\right)\leq\alpha}} R_1 \left(\varphi_{\alpha A}\right) = \min_{\substack{\varphi_{A}  \\ \p(\varphi_{A} (\bd X) \neq Y | Y=0)\leq\alpha}} \p(\varphi_{A} (\bd X) \neq Y | Y=1)\,.
\end{equation} 
By the NP Lemma,
\begin{equation}\label{eqn: np oracle}
\varphi^*_{\alpha A}(\bd x) = \1 \left(p_{1A}(\bd x_A)/p_{0A}(\bd x_A) > C^*_{\alpha A}\right)\,,
\end{equation}
\rv{where $C^*_{\alpha A}$ is a constant such that $\p(p_{1A}(\boldsymbol{X}_A)/p_{0A}(\boldsymbol{X}_A)>C^*_{\alpha A}|Y=0) = \alpha$. }

For a given level $\alpha$, \jjl{we define the \textit{population-level Neyman-Pearson criterion} (p-NPC) of $A_i$ as its \textit{optimal} type II error $R_1 \left(\varphi^*_{\alpha A_i}\right)$; i.e., $A_1, \ldots, A_J$ will be ranked based on $\left\{R_1 \left(\varphi^*_{\alpha A_1}\right), \ldots, R_1 \left(\varphi^*_{\alpha A_J}\right) \right\}$, with the smallest being ranked the top}.

\begin{figure}[h!]
    \centering
    \makebox{\includegraphics[width = 0.75\textwidth]{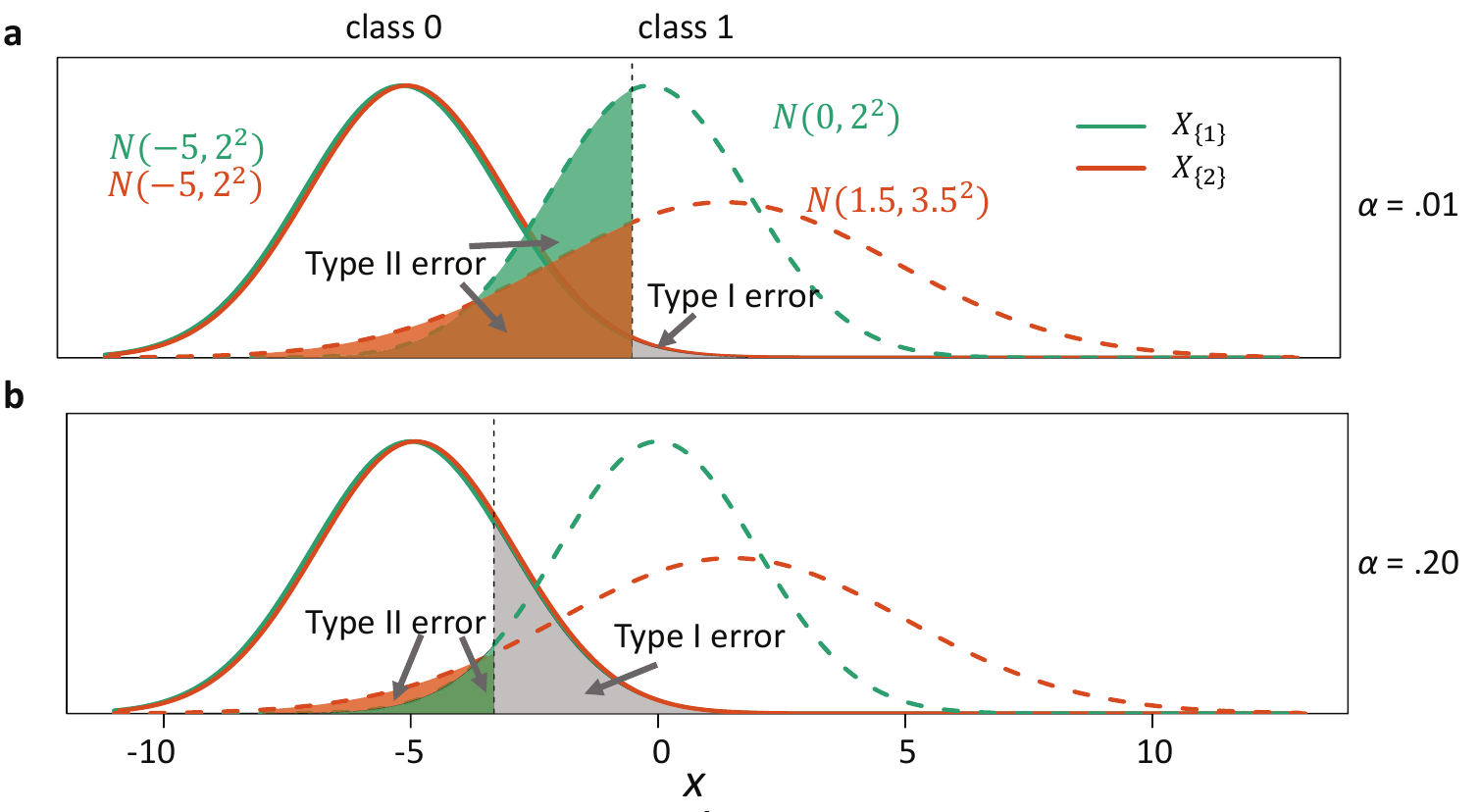}}
    \caption{\small{A toy example in which feature ranking under p-NPC changes as $\alpha$ varies. \textbf{Panel a}: $\alpha=.01$. The NP oracle classifier based on feature $1$ (or feature $2$) has the type II error $.431$ (or $.299$). \textbf{Panel b}: $\alpha=.20$. The NP oracle classifier based on feature $1$ (or feature $2$) has the type II error $.049$ (or $.084$).}}\label{fig:toy example 1}
\end{figure}

For a graphical illustration of \rv{marginal feature ranking by p-CC and p-NPC, please see Figure~\ref{fig:marginal_ranking_illustration} in Appendix~\ref{sec:add_tabs_figs}. It is worth \rvv{noting} that p-CC and p-NPC do not always give the same feature ranking. For a toy example,} we compare two features $\bd X_{\{1\}}, \bd X_{\{2\}} \in \R$,\footnote{Usually, we denote the two scalar-valued features by $X_1$ and $X_2$, but here we use $\bd X_{\{1\}}$ and $\bd X_{\{2\}}$ to be consistent with the notation $\bd X_{A}$.} whose class-conditional distributions are \jjl{the following Gaussians}: 
\begin{align}\label{eq:toy_example}
	\bd X_{\{1\}} \given (Y=0) &\sim \mathcal{N}(-5, 2^2)\,, & \bd X_{\{1\}}\given (Y=1) &\sim \mathcal{N}(0, 2^2)\,,\\
	\bd X_{\{2\}} \given (Y=0) &\sim \mathcal{N}(-5, 2^2)\,, & \bd X_{\{2\}} \given (Y=1) &\sim \mathcal{N}(1.5, 3.5^2)\,, \notag
\end{align}
and the class priors are equal, i.e., $\pi_0 = \pi_1 =  .5$. 
It can be calculated that $R \left(\varphi^*_{{\{1\}}}\right) = .106$ and $R \left(\varphi^*_{{\{2\}}}\right)= .113$. Therefore,  $R \left(\varphi^*_{{\{1\}}}\right) < R \left(\varphi^*_{{\{2\}}}\right)$, and \jjl{p-CC ranks feature $1$ higher than feature $2$}. \jjl{The comparison is more subtle for p-NPC}. If we set $\alpha =.01$, $R_1 \left(\varphi^*_{\alpha \{1\}}\right) = .431$ is \textit{larger} than $R_1 \left(\varphi^*_{\alpha \{2\}}\right) = .299$. However, if we set $\alpha = .20$, $R_1 \left(\varphi^*_{\alpha \{1\}}\right) = .049$ is \textit{smaller} than $R_1 \left(\varphi^*_{\alpha \{2\}}\right)= .084$. Figure \ref{fig:toy example 1} illustrates the NP oracle classifiers for \jjl{these $\alpha$'s.}

This example suggests a general phenomenon that feature ranking \jjl{depends} on the user-chosen criteria. For some \jjl{$\alpha$} values (e.g., $\alpha =.20$ in the example), p-NPC and p-CC  agree on the ranking, while for others (e.g., $\alpha = .01$ in the example), they disagree.\footnote{Under special cases, however, we can derive conditions under which p-NPC gives an $\alpha$-invariant feature ranking \jjl{that always agrees with the ranking by} p-CC. In Lemma \ref{lem: toy example1} of the Appendix, interested readers can find such a condition under Gaussian distributions.} This observation calls for the development of sample-level CC and NPC.

\section{Sample-level ranking criteria} \label{sec:methods}

In \jjl{the following text, we refer to sample-level CC and NPC as} ``s-CC" and ``s-NPC" respectively.  In the same model-free spirit of the p-CC and p-NPC definitions, we use model-free nonparametric techniques to construct s-CC and s-NPC.  Admittedly, such construction would be impractical when the feature subsets to be ranked have large cardinalities.  However, since we are mainly interested in marginal feature ranking, with intended extension to small subsets such as feature pairs, model-free nonparametric techniques are appropriate.


In the methodology and theory sections, we assume the following sampling scheme. Suppose we have a training dataset $\mathcal{S} = \mathcal{S}^0 \cup \mathcal{S}^1 $, where  $\mathcal{S}^0= \left\{\bd {X}_{1}^{0}, \dots,  \bd {X}_{m}^{0} \right\}$ are \jjl{independent and identically distributed (i.i.d.)} class $0$ observations, $\mathcal{S}^1= \left\{\bd {X}_{1}^{1}, \dots,  \bd {X}_{n}^{1} \right\}$ are i.i.d. class $1$ observations, and $\mathcal{S}^0$ is independent of $\mathcal{S}^1$. The sample sizes $m$ and $n$ are considered as \jjl{fixed positive integers}. \jjl{The construction of both s-CC and s-NPC involves} splitting the class $0$ and class $1$ observations. To increase stability, \jjl{we perform multiple random splits. In detail,} we randomly divide $\mathcal{S}^0$ for $B$ times into two halves $\mathcal{S}_{\rm ts}^{0(b)} = \left\{ \bd X_{1}^{0(b)}, \dots,  \bd X_{m_1}^{0(b)} \right\}$ and ${\mathcal{S}}_{\rm lo}^{0(b)} = \left\{ \bd {X}_{m_1+ 1}^{0(b)}, \dots,  \bd {X}_{m_1+m_2}^{0(b)} \right\}$, where $m_1 + m_2 = m$,   the subscripts ``ts" and ``lo" stand for \textit{train-scoring} and \textit{left-out} respectively, and the superscript $b\in\{1,\ldots, B\}$ indicates the $b$-th random split. \jjl{We also randomly split} $\mathcal{S}^1$ \jjl{for $B$} times into $\mathcal{S}_{\rm ts}^{1(b)}  = \left\{ \bd X_1^{1(b)}, \dots,  \bd X_{n_1}^{1(b)} \right\}$ and $\mathcal{S}_{\rm lo}^{1(b)} = \left\{\bd {X}_{n_1 + 1}^{1(b)}, \dots,  \bd {X}_{n_1+n_2}^{1(b)} \right\}\,$, where $n_1+n_2=n$ and $b\in\{1, \ldots, B\}$.  \jjl{In this work, we make equal-sized splits: $m_1 = \lfloor m/2 \rfloor$ and $n_1 = \lfloor n/2 \rfloor$. We leave the possibility of data-adaptive splits to future work.}

As in the definitions of p-CC and p-NPC, we use notations to allow \jjl{for the extension to ranking} feature subsets.  For $A\subseteq\{1, \ldots, d\}$ with $|A| = l$, recall the classical oracle classifier restricted to $A$, $\varphi^*_A(\bd x)$, defined in \eqref{eqn:classical oracle} and the NP oracle classifier restricted to $A$, $\varphi^*_{\alpha A}(\bd x)$, defined in \eqref{eqn: np oracle}. Although these two oracles have different thresholds, $\pi_0 / \pi_1$ vs. $C^*_{\alpha A}$, the class-conditional density ratio $p_{1A}(\cdot)/ p_{0A}(\cdot)$ \jjl{is involved in} both oracles.  The densities $p_{0A}$ and $p_{1A}$ can be estimated respectively from  $\mathcal{S}^{0(b)}_{\rm ts}$ and $\mathcal{S}^{1(b)}_{\rm ts}$ by kernel density estimators
\begin{align}\label{eqn:kernel density estimates}
\hat{p}_{0A}^{(b)}(\bd x_A)=\frac{1}{m_1h_{m_1}^l}\sum_{i=1}^{m_1} K\left(\frac{\bd X^{0(b)}_{iA}-\bd x_A}{h_{m_1}}\right) \quad \text{ and } \quad \hat{p}_{1A}^{(b)}(\bd x_A)=\frac{1}{n_1h_{n_1}^l}\sum_{i=1}^{n_1} K\left(\frac{\bd X_{iA}^{1(b)}-\bd x_A}{h_{n_1}}\right)\,,
\end{align}
where $h_{m_1}$ and $h_{n_1}$ denote the bandwidths, and $K(\cdot)$ is a  kernel in $\R^l$.

\subsection{Sample-level classical ranking criterion}

To define s-CC, we first construct plug-in classifiers $\hat\phi_A^{(b)}(\bd x) = \1\left( \hat{p}_{1A}^{(b)}(\bd x_A)/ \hat{p}_{0A}^{(b)}(\bd x_A) > m_1/n_1\right)$ on $\mathcal S_{\rm ts}^{0(b)} \cup \mathcal S_{\rm ts}^{1(b)}$ for $b\in\{1, \ldots, B\}$. In each classifier, the threshold $m_1/n_1$ mimics $\pi_0 / \pi_1$. If classes $0$ and $1$ are sampled with probabilities $\pi_0$ and $\pi_1$, respectively, then each classifier $\hat\phi_A^{(b)}(\bd x)$ is \jjl{a good plug-in estimate} of $\varphi^*_A(\bd x)$ defined in \eqref{eqn:classical oracle}. However, in the presence of sampling bias, $m_1/n_1$ cannot mimic $\pi_0/\pi_1$, and thus $\hat\phi_A^{(b)}(\bd x)$ is not \jjl{a good estimate} of $\varphi^*_A(\bd x)$. Armed with the classifiers  $\hat\phi_A^{(1)}(\cdot), \ldots, \hat\phi_A^{(B)}(\cdot)$, we define the \textit{sample-level CC} (s-CC) \jjl{of} $A$ as
\begin{align}\label{CC}
	\mathrm{CC}_A &:= \frac{1}{B} \sum_{b=1}^B \mathrm{CC}_A^{(b)}\,,\\\notag
	\text{with } \mathrm{CC}_A^{(b)} &:= \frac{1}{m_2+n_2}\left\{ \sum_{i=m_1+1}^{m_1+m_2} \hat{\phi}_A^{(b)}\left(\bd X_{i}^{0(b)}\right)  + \sum_{i'=n_1+1}^{n_1+n_2} \left[ 1-\hat{\phi}^{(b)}_{A}\left(\bd X_{i'}^{1(b)}\right) \right] \right\}\,.
\end{align}
In other words, $\text{CC}_A$ is the average of the risks of $\hat\phi_A^{(b)}(\cdot)$ on the left-out observations $\mathcal S_{\rm lo}^{0(b)} \cup \mathcal S_{\rm lo}^{1(b)}$ for $b\in\{1, \ldots, B\}$.  

\subsection{Sample-level Neyman-Pearson ranking criterion}\label{sec: construction of NP}


To define s-NPC, we use the same kernel density estimates to \jjl{plug in} $p_{1A}(\cdot)/ p_{0A}(\cdot)$, as in s-CC.  To \jjl{estimate} the oracle threshold $C^*_{\alpha A}$, we use the NP umbrella algorithm \citep{tong2016neyman}. \jjl{Unlike s-CC, in which we use both $\mathcal S_{\rm lo}^{0(b)}$ and $\mathcal S_{\rm lo}^{1(b)}$ to evaluate the constructed classifier, for s-NPC we use $\mathcal S_{\rm lo}^{0(b)}$ to estimate the threshold and only $\mathcal S_{\rm lo}^{1(b)}$ to evaluate the classifier}.

The NP umbrella algorithm finds proper thresholds for all \textit{scoring-type classification methods} (e.g., nonparametric density ratio plug-in, logistic regression, and RF) so that the resulting classifiers achieve a high probability control on the type I error under the pre-specified level $\alpha$. \jjl{A scoring-type classification method outputs a scoring function that maps the feature space $\mathcal X$ to $\R$, and a classifier is constructed by combining the scoring function with a threshold.} To construct an NP classifier given a scoring-type classification method, the NP umbrella algorithm first trains a scoring function $\hat{s}^{(b)}_A(\cdot)$ on $\mathcal{S}^{0(b)}_{\rm ts} \cup \mathcal{S}^{1(b)}_{\rm ts}\,$. In this work, we specifically use $\hat{s}^{(b)}_A(\cdot) = \hat{p}_{1A}^{(b)}(\cdot)/ \hat{p}_{0A}^{(b)}(\cdot)$, in which the numerator and the denominator are defined in \eqref{eqn:kernel density estimates}.  Second, the algorithm applies $\hat{s}^{(b)}_A(\cdot)$ to $\mathcal{S}^{0(b)}_{\rm lo}$ to obtain  scores $\left\{T_i^{(b)} = \hat{s}^{(b)}_A\left(\bd X^{0(b)}_{m_1+i}\right),  i=1,\dots, m_2\right\}$, which are \jjl{then} sorted in an increasing order and denoted by  $\left\{T_{(i)}^{(b)}, i=1,\dots, m_2\right\}$. Third, for a user-specified type I error upper bound $\alpha \in (0,1)$ and a violation rate $\delta_1 \in(0,1)$\jjl{, which refers to the probability that the type I error of the trained classifier exceeds} $\alpha$, the algorithm chooses the order  
\begin{align*}
	k^* = \min\limits_{k=1,\dots, m_2} \left\{k:\sum_{j=k}^{m_2} \binom{m_2}{j} (1-\alpha)^j \alpha^{m_2-j}\leq \delta_1\right\}\,.
\end{align*} 
When $m_2 \geq \frac{\log \delta_1}{\log(1-\alpha)}\,,$ a finite $k^*$ exists,\footnote{If one were to assume a parametric model, one can get rid of the minimum sample size requirement on $m_2$ \citep{Tong.Xia.Wang.Feng.2020}.  However, we adopt the non-parametric NP umbrella algorithm \citep{tong2016neyman}  to achieve the desirable mode-free property of our feature ranking framework.} and the umbrella algorithm chooses the threshold of the estimated scoring function as 
$
	\widehat{C}_{\alpha A}^{(b)} = T_{(k^*)}^{(b)}. 
$
Hence, the resulting NP classifier is
\begin{align}\label{eq:NP_classifier}
	\hat{\phi}_{\alpha A}^{(b)}(\cdot) = \1 \left(\hat{s}^{(b)}_A (\cdot) >  \widehat{C}_{\alpha A}^{(b)} \right)\,.
\end{align}

Proposition 1 in \cite{tong2016neyman} states that there is no more than $\delta_1$ probability for the type I error of $\hat{\phi}_{\alpha A}^{(b)}(\cdot)$ to exceed $\alpha$:  
\begin{equation}
\p \left(R_0 (\hat{\phi}_{\alpha A}^{(b)}) > \alpha\right)  \leq\sum_{j=k^*}^{m_2} \binom{m_2}{j} (1-\alpha)^j \alpha^{m_2-j}\leq \delta_1\,, \label{ineq:npc}
\end{equation} 
for every $b = 1,\ldots, B$.  We evaluate the type II error of the  $B$ NP classifiers $\hat{\phi}^{(1)}_{\alpha A}, \ldots, \hat{\phi}^{(B)}_{\alpha A}$ on the left-out class $1$  sets $\mathcal S_{\rm lo}^{1(1)},\ldots,\mathcal S_{\rm lo}^{1(B)}$ respectively.    Our \textit{sample-level NPC} (s-NPC) of $A$ at level $\alpha$, denoted by $\mathrm{NPC}_{\alpha A}$, computes the average of these type II errors: 
\begin{align}\label{Npscore}
 	  \mathrm{NPC}_{\alpha A} &:=\frac{1}{B} \sum_{b=1}^{B} \mathrm{NPC}_{\alpha A}^{(b)}\,,\\\notag
 	  \text{with } \mathrm{NPC}_{\alpha A}^{(b)} &:= \frac{1}{n_2} \sum_{i= n_1 + 1}^{n_1+n_2} \left[ 1-\hat{\phi}^{(b)}_{\alpha A}\left(\bd X_i^{1(b)}\right) \right] = \frac{1}{n_2} \sum_{i=n_1 +1}^{n_1+n_2}\1\left( \hat{s}^{(b)}_{A}\left(\bd{X}_{iA}^{1(b)}\right) \le \widehat{C}^{(b)}_{\alpha A}\right)\,,
 \end{align} 
 where $\hat{s}^{(b)}_{A}(\cdot) = \hat{p}_{1A}^{(b)}(\cdot)/ \hat{p}_{0A}^{(b)}(\cdot)$ is the kernel density ratios constructed on $\mathcal{S}_{\rm ts}^{0(b)} \cup\mathcal{S}_{\rm ts}^{1(b)}$ using only the features indexed by $A$, and $\widehat{C}^{(b)}_{\alpha A} = T_{(k^*)}^{(b)}$ is found by the NP umbrella algorithm. 
 
\vspace{.1in}

\rvv{The time complexity is $O\left((m_1+n_1)\cdot (m_2+n_2)\right)$ for calculating s-CC, and an additional complexity of $O\left(m_2 \log m_2\right)$ is needed for calculating s-NPC; both time complexities can be reduced to $O(m+n)$ if approximate kernel density estimation is used and $m_2$ is bounded. We discuss the calculation details of the time complexities in Appendix~\ref{sec::time_complexity}, and we illustrate the calculation of s-CC and s-NPC in Figure~\ref{fig:illustration}.}

\rvv{For the implementation of s-NPC and s-CC, we use the \texttt{kde()} function with default arguments in the R package \texttt{ks}. By default, the function uses the Gaussian kernel and the bandwidth selected by the univariate plug-in selector of \citep{wand1994multivariate}.} 

\subsection{Revisiting the toy example at the sample level} 

With the above definitions of s-CC and s-NPC, we demonstrate how they rank the two features in the toy example (Figure \ref{fig:toy example 1}) and that their ranks are consistent with their population-level counterparts p-CC and p-NPC, respectively, with high probability. 

We simulate $1000$ samples, each of size $N=2000$ (two classes combined), from the two-feature distribution (\ref{eq:toy_example}) in the toy example (Figure \ref{fig:toy example 1}). 
With $B=11$, we apply s-CC (\ref{CC}) and s-NPC with $\delta_1 = .05$ (\ref{Npscore}) to each sample to rank the two features, and we calculate the frequency of each feature being ranked the top among the $1000$ ranking results. 
Table \ref{tab:toy_example} shows that s-NPC ($\alpha = .01$) ranks feature $2$ the top with high probability, while s-CC and s-NPC ($\alpha = .20$) prefer feature $1$ with high probability. This is consistent with our population-level result in Section \ref{sec:NPC_population}: p-NPC ($\alpha=.01$) prefers feature $2$, while p-CC and p-NPC ($\alpha=.20$) find feature $1$ better. 

\begin{table}[htbp]
\caption{\label{tab:toy_example}The frequency of each feature being ranked the top by each criterion among $1,000$ samples in the toy example (Figure \ref{fig:toy example 1}).}
\centering
\begin{tabular}{lrr}
\hline
Criterion & Feature $1$ & Feature $2$\\
\hline
s-CC & $78.0\%$ & $22.0\%$ \\
s-NPC ($\alpha = .01$) & $1.6\%$ & $98.4\%$ \\
s-NPC ($\alpha = .20$) & $99.0\%$ & $1.0\%$\\
\hline	
\end{tabular}
\end{table}


\section{Theoretical properties}\label{sec:theoretical properties}

In this section, we investigate the ranking properties of s-CC and s-NPC.  Concretely, we will address this question: for \jjl{$J$} candidate feature index sets $A_1, \ldots, A_J$ of size $l$, is it guaranteed that s-CC and s-NPC have ranking agreements with p-CC and p-NPC respectively with high probability? In our theory development, we consider $J$ as a fixed number; for simplicity, we assume the number of random splits to be $B = 1$ in s-CC and s-NPC, thus removing the super index $(b)$ in all notations in this section and the Appendix proofs. 

In addition to investigating ranking consistency, we discover a property unique to s-NPC: the robustness against sampling bias. Concretely,  as long as the sample sizes $m$ and $n$ are large enough, s-NPC gives ranking consistent with p-NPC even when the class size ratio $m/n$ in the sample is far from the ratio $\pi_0 / \pi_1$ in the population. In contrast, s-CC is not robust against sampling bias, except when the population class proportion ratio $\pi_0 / \pi_1$ is known and we replace the thresholds in the plug-in classifiers $\hat\phi_A^{(1)}(\cdot), \ldots, \hat\phi_A^{(B)}(\cdot)$, which are used for s-CC, by this \jjl{ratio}.

\subsection{Definitions and key assumptions}
We assume that the candidate index sets $A_1,\ldots,A_J$ have a moderate size $l$ $(\ll d)$. 
 Following \cite{Audibert05fastlearning}, for any multi-index $\bd t=\left(t_1, \ldots, t_l \right)\tran \in \N^l$ and $\bd x= \left( x_1, \ldots, x_l\right)\tran \in \R^l$, we define $|\bd t| = \sum_{i=1}^{l}t_i$, $\bd t! = t_1!\cdots t_l!$, $\bd x^{\bd t}=x_1^{t_1} \cdots x_l^{t_l}$, $\left\| \bd x\right\| = \left( x_1^2 + \ldots + x_l^2 \right)^{1/2}$, and the differential operator $D^{\bd t} = \frac{\partial^{t_1 + \cdots + t_l}}{\partial {x_1^{t_1}} \cdots \partial {x_l^{t_l}}}$.   For all the theoretical discussions, we assume the domain of $p_{0A}$ and $p_{1A}$, \jjl{i.e.,} the  class-conditional densities of $\bd X_A|(Y=0)$ and $\bd X_A|(Y=1)$,  is $[-1,1]^l$, where $l = |A|$. We denote the distributions of $\bd X_A|(Y=0)$ and $\bd X_A|(Y=1)$ by $P_{0A}$ and $P_{1A}$ respectively.   
 
\begin{definition}[H\"{o}lder function class]\label{def:holder_function_class}
	Let $\beta>0$. Denote by $\floor*{\beta}$ the largest integer strictly less than $\beta$. For a $\floor*{\beta}$-times continuously differentiable function $g: \R^l \rightarrow \R$, we denote by $g_{\bd x}$ its Taylor polynomial of degree $\floor*{\beta}$ at a value $\bd x \in \R^l$:
$$g_{\bd x}^{(\beta)}(\cdot) = \sum_{{\left| {\bd t}\right|}\leq \floor*{\beta}} \frac{\left(\cdot - {\bd x}\right)^{\bd t}}{{\bd t}!} D^{\bd t}g\left({\bd x}\right).$$ 
For $L >0 $, the $\left( \beta, L, \left[-1, 1\right]^l\right)$-H\"{o}lder function class, denoted by $\Sigma\left( \beta, L, \left[-1, 1\right]^l\right)$, is the set of $\floor*{\beta}$-times continuously differentiable functions $g: \R^l \rightarrow \R$ that satisfy the following inequality:
$$\left| g\left( {\bd x}\right) -g_{\bd x}^{(\beta)}\left( {\bd x}^{\prime}\right) \right| \leq L\left\| {\bd x}- {\bd x}^{\prime} \right\|^{\beta}\,, \quad \text{ for all } {\bd x}, {\bd x}^{\prime} \in \left[-1, 1\right]^l\,.$$
\end{definition}

\begin{definition}[H\"{o}lder density class]\label{def:holder_density_class}
	The $\left( \beta, L, \left[-1, 1\right]^l\right)$-H\"{o}lder density class is defined as $$\mathcal{P}_{\Sigma} \left( \beta, L, \left[-1, 1\right]^l\right)= \left\{ p: p \geq 0, \int p=1, p \in \Sigma\left( \beta, L, \left[-1, 1\right]^l\right)\right\}\,.$$ 
\end{definition}

The following $\beta$-valid kernels are multi-dimensional analogs of univariate higher order kernels.
\begin{definition}[$\beta$-valid kernel]\label{definition1}
Let $K(\cdot)$ be a real-valued kernel function on $\R^l$ with the support $[-1,1]^l$\,. For a fixed $\beta>0$\,, the function $K(\cdot)$ is a $\beta$-valid kernel if it satisfies (1) $\int |K|^q <\infty$ for any $q\geq 1$, (2) $\int \|\bd u \|^\beta|K(\bd u)|d\bd u <\infty$, and (3) in the case $\floor* \beta \geq 1$\,, $\int \bd u^{\bd t} K(\bd u)d\bd u = 0 $ for any $\bd t =(t_1, \dots, t_l) \in \mathbb N^l$ such that $1\le |\bd t| \le\floor* \beta$\,.
\end{definition}

One example of $\beta$-valid kernels is the product kernel whose ingredients are kernels of order $\beta$ in $1$ dimension:
$$
K (\bd x) = K_1(x_1)K_1(x_2)\cdots K_1(x_l)\1(\bd x\in[-1,1]^l)\,,
$$
where $K_1(\cdot)$ is a 1-dimensional $\beta$-valid kernel and is constructed based on Legendre polynomials. Such kernels have been considered in \cite{RigVer09}.  When a $\beta$-valid kernel is constructed out of Legendre polynomials, it is also Lipschitz and bounded. For simplicity, we assume that all the $\beta$-valid kernels considered in the theory discussion are constructed from Legendre polynomials.

\begin{definition}[Margin assumption]\label{def: margin_assumpion}
	A function $f(\cdot)$ satisfies the margin assumption of the order $\bar{\gamma}$ at the level $C$, with respect to the probability distribution $P$ of a random vector $\bd X$, if there exist positive constants $\bar{C}$ and $\bar{\gamma}$, such that for all $\delta \geq 0$,
$$P \left(\left| f\left(\bd X\right) - C \right| \leq \delta\right) \leq \bar C \delta^{\bar{\gamma}}\,.$$
\end{definition}

The above condition for density functions was first introduced in \citet{polonik1995measuring}, and its counterpart in the classical binary classification was called the margin condition \citep{MamTsy99}, which is a low noise condition.  
Recall that the set $\{\bd x: \eta(\bd x)=1/2\}$ is the decision boundary of the classical oracle classifier, and the margin condition in the classical paradigm is a special case of Definition \ref{def: margin_assumpion} by taking $f(\cdot) = \eta(\cdot)$ and $C=1/2$. Unlike the classical paradigm where the optimal threshold $1/2$ on regression function $\eta(\cdot)$ is known, the optimal threshold under the NP paradigm is unknown and needs to be estimated, thus suggesting the necessity of having sufficient data around the decision boundary to detect it. This concern motivated \cite{tong2013plug} to formulate a detection condition that works as an opposite force to the margin assumption, and \cite{zhao2016neyman} improved the condition and proved the condition's necessity in bounding the excess type II error of an NP classifier. To establish ranking consistency properties of s-NPC, a bound on the excess type II error is an intermediate result, so we also need this \jjl{detection condition} for our current work.  


\begin{definition}[Detection condition \citep{zhao2016neyman}]\label{def:detection_assumption}
	A function $f(\cdot)$ satisfies the detection condition of the order $\underaccent{\bar}{\gamma}$ at the level $(C, \delta^*)$ with respect to the probability distribution $P$ of a random vector $\bd X$, if there exists a positive constant $\underaccent{\bar}C$, such that for all $\delta\in\left(0, \delta^*\right) $,
$$P\left( C \leq f\left(\bd X\right) \leq  C + \delta \right) \geq  \underaccent{\bar}C \delta^{\underaccent\bar{\gamma}} \,.$$
\end{definition}

\subsection{A uniform deviation result of the scoring function}

For $A\subseteq\{1, \ldots, d\}$ and $|A| = l$, recall that we estimate $p_{0A}(\cdot)$ and $p_{1A}(\cdot)$ respectively from  $\mathcal{S}^0_{\rm ts}$ and $\mathcal{S}^1_{\rm ts}$ by kernel density estimators $\hat{p}_{0A}(\cdot)$ and $\hat{p}_{1A}(\cdot)$ defined \eqref{eqn:kernel density estimates},
where $K(\cdot)$ is a $\beta$-valid kernel in $\R^l$. We are interested in deriving a high probability bound for $\left\| \hat p_{1A}(\bd x_A)/\hat p_{0A}(\bd x_A) - p_{1A}(\bd x_A)/p_{0A}(\bd x_A)\right\|_{\infty}$. 


\begin{condition}\label{condition: 1}
Suppose that for all $A\subset\{1 \ldots, d\}$ satisfying $|A|=l$,
\begin{itemize}
\item[(i)] there exist positive constants $\mu_{\min}$ and $\mu_{\max}$ such that $\mu_{\max}\geq p_{0A}(\cdot) \geq \mu_{\min}$ and $\mu_{\max}\geq p_{1A}(\cdot) \geq \mu_{\min}$;
\item[(ii)] there is a positive constant $L$ such that $p_{0A}(\cdot),\, p_{1A}(\cdot)\in\mathcal{P}_{\Sigma}(\beta, L, [-1, 1]^{l})$.   
\end{itemize}

\end{condition}

%

 \begin{proposition}\label{lem:bound_s_shat_for_plugin}
Assume Condition \ref{condition: 1} and let the kernel $K(\cdot)$ be $\beta$-valid and $L^\prime$-Lipschitz. Let $A \subseteq\{1, \ldots, d\}$ and $|A| = l$. Let $\hat p_{0A}(\cdot)$ and $\hat p_{1A}(\cdot)$ \jjl{be} kernel density estimates defined in \eqref{eqn:kernel density estimates}. Take the bandwidths $h_{m_1}=\left(\frac{\log m_1}{m_1}\right)^{\frac{1}{2\beta+l}}$ and $h_{n_1}=\left(\frac{\log n_1}{n_1}\right)^{\frac{1}{2\beta+l}}$.   For any $\delta_3 \in (0,1)$, if sample \jjl{sizes} $m_1 = |\mathcal{S}_{\rm ts}^0|$ and $n_1 = |\mathcal{S}_{\rm ts}^1|$ satisfy \[
 	\sqrt{\frac{\log\left(2m_1/\delta_3\right)}{m_1h_{m_1}^{l}}} < 1\wedge \frac{\mu_{\min}}{2 C_0}  \,, \quad \sqrt{\frac{\log\left(2n_1/\delta_3\right)}{n_1h_{n_1}^{l}}}< 1, \quad n_1 \wedge m_1 \geq 2/\delta_3\,,\quad 
 	\] 
\rv{where $\wedge$ denotes the minimum}, $C_{0}=\sqrt{48c_{1}} + 32c_{2}+2Lc_{3}+L'+L+C\sum_{1\leq|\bd q|\leq\lfloor\beta\rfloor}\frac{1}{\bd q!}$, in which $c_{1}=\mu_{\max}\|K\|^2$, $c_{2}=\|K\|_{\infty}+\mu_{\max}+\int|K||\bd t|^{\beta}d\bd t$, $c_{3}=\int |K||\bd t|^{\beta}d\bd t$ and $C$ is such that\\ $C \geq \sup_{1\leq|\bd q|\leq\lfloor \beta\rfloor}\sup_{\bd x_A\in[-1, 1]^l}|D^{\bd q}p_{0A}(\bd x_A)|$. Then there exists a positive constant $\widetilde{C}$ that does not depend on $A$, such that we have with probability at least $1-\delta_3$, \[
 	\left\| \hat p_{1A}(\bd x_A)/\hat p_{0A}(\bd x_A) - p_{1A}(\bd x_A)/p_{0A}(\bd x_A)\right\|_{\infty} \leq \widetilde{C}\left[\left( \frac{\log m_1}{m_1}\right)^{\beta/(2\beta+l)} + \left( \frac{\log n_1}{n_1}\right)^{\beta/(2\beta+l)} \right]\,.
 	\]

 \end{proposition}


\subsection{Ranking property of s-CC}\label{sec:theoretic_plug-in-CC}

To study the ranking agreement between s-CC and p-CC, an essential step is to develop a concentration result between $\text{CC}_A$ and $R(\varphi^*_A)$, where $\varphi^*_{A}$ was defined in \eqref{eqn:classical oracle}, based on Proposition \ref{lem:bound_s_shat_for_plugin}.

\begin{theorem}\label{prop: CC1}
Let $\delta_3, \delta_4, \delta_5\in (0, 1)$. In addition to the assumptions of Propositions \ref{lem:bound_s_shat_for_plugin}, assume that the density ratio $s_A(\cdot) = p_{1A}(\cdot)/p_{0A}(\cdot)$ satisfies the margin assumption of order $\bar\gamma$ at level $\pi_0 / \pi_1$ (with constant $\bar C$) with respect to both $P_{0A}$ and $P_{1A}$ (the distributions of $\bd X_A | (Y=0)$ and $\bd X_A | (Y=1)$), that $m_2 \geq (\log\frac{2}{\delta_5})^2$ and $n_2 \geq (\log\frac{2}{\delta_4})^2$,  and that $m / n = m_1 / n_1 = \pi_0 / \pi_1$,  
then we have with probability at least $1-\delta_3-\delta_4-\delta_5$, 
$$
\left| \mathrm{CC}_{A} - R \left( {\varphi}^*_{A} \right)\right|\leq \widetilde C_1 \left[\left( \frac{\log m_1}{m_1}\right)^{\frac{\beta\bar\gamma}{2\beta+l}} + \left( \frac{\log n_1}{n_1}\right)^{\frac{\beta\bar\gamma}{2\beta+l}} + m_2^{-\frac{1}{4}} + n_2^{-\frac{1}{4}} \right]\,,
$$	 
for some positive constant $\widetilde C_1$ that does not depend on $A$. 
\end{theorem}

\rv{Under smoothness, regularity, and sample size conditions, \rvv{Theorem} \ref{prop: CC1} shows the concentration of $\mathrm{CC}_{A}$ around $R \left( {\varphi}^*_{A}\right)$ with probability at least $1-(\delta_3+\delta_4+\delta_5)$. The user-specified violation rate $\delta_3$ accounts for the uncertainty in training the scoring function $\hat s_A(\cdot)$ on a finite sample; $\delta_4$ represents the uncertainty of using left-out class $1$ observations $\mathcal{S}^1_{\rm lo}$ to estimate $R_1(\hat\phi_{A})$; $\delta_5$ represents the uncertainty of using left-out class $0$ observations $\mathcal{S}^0_{\rm lo}$ to estimate $R_0(\hat\phi_{A})$ (Recall that $R(\cdot) = \pi_0 R_0(\cdot) + \pi_1 R_1(\cdot)$).  Like the constant $C_0$ in Proposition \ref{lem:bound_s_shat_for_plugin}, the generic constant $\widetilde C_1$ in \rvv{Theorem} \ref{prop: CC1} can be provided more explicitly, but it would be too cumbersome to do so.  More discussion about \rvv{Theorem} \ref{prop: CC1} can be found after its counterpart for NPC, i.e.,  Theorem \ref{thm:1}. }

\rvv{Theorem} \ref{prop: CC1} leads to the ranking consistency of s-CC.

\begin{theorem}\label{thm:selection_consistency_cc}
Let $\delta_3$, $\delta_4$, $\delta_5\in (0,1)\,,$ $A_1, \ldots, A_J \subseteq\left\{1,\ldots, d \right\}$, $|A_1| = |A_2|=\cdots = |A_J| = l$, and $R(\varphi^*_{A_1}) < \cdots < R(\varphi^*_{A_J})$, ordered by p-CC. Both $J$ and $l$ are constants that do not diverge with the sample sizes. In addition to the assumptions in \rvv{Theorem} \ref{prop: CC1}, assume that  the \jjl{p-CC's} of $A_1, \ldots, A_J$ are separated by some margin $g>0$; in other words,  
$$
	 \min \limits_{i \in \{1,\dots, J-1\}}\left\{ R\left( {\varphi}^*_{A_{i+1}}\right) - R\left( {\varphi}^*_{A_i}\right) \right\} > g\,. 
$$ 
In addition, assume that $m_1, m_2, n_1, n_2$ satisfy
\begin{equation}\label{eqn:sample size requirement CC}
\widetilde C_1 \left[\left( \frac{\log m_1}{m_1}\right)^{\frac{\beta\bar\gamma}{2\beta+l}} + \left( \frac{\log n_1}{n_1}\right)^{\frac{\beta\bar\gamma}{2\beta+l}} + m_2^{-\frac{1}{4}} + n_2^{-\frac{1}{4}} \right]  < \frac{g}{2}\,, 
\end{equation}
where $\widetilde C_1$ is the generic constant in \rvv{Theorem} \ref{prop: CC1}.  
Then with probability at least $1 - J(\delta_3+\delta_4+\delta_5)$, $\mathrm{CC}_{A_i} < \mathrm{CC}_{A_{i+1}}$ for all $i = 1, \ldots, J-1$. That is, \jjl{s-CC} ranks $A_1, \ldots, A_J$ the same as \jjl{p-CC} does. 
\end{theorem}

\begin{remark}
\rv{If the sample size ratio $m/n$ ($= m_1/n_1$) is far from $\pi_0/\pi_1$, we cannot expect a concentration result on $\left| \mathrm{CC}_{A} - R \left( {\varphi}^*_{A} \right)\right|$, such as \rvv{Theorem} \ref{prop: CC1}, to hold. The rationale is, if we replace the trained scoring function $\hat p_{1A}(\cdot) / \hat p_{0A}(\cdot)$ in s-CC by the optimal $ p_{1A}(\cdot) / p_{0A}(\cdot)$ (think of $m$ and $n$ extremely large), then s-CC is based on the classifier $\1(p_{1A}(\cdot) / p_{0A}(\cdot) > m_1/n_1)$. In contrast, p-CC is based on the oracle  $\1(p_{1A}(\cdot) / p_{0A}(\cdot) > \pi_0/\pi_1)$. When $m_1/n_1$ is far from $\pi_0/\pi_1$, clearly the classification errors of these two classifiers would not be close, so  we would not have  $\left| \mathrm{CC}_{A} - R \left( {\varphi}^*_{A} \right)\right|$ small. As \rvv{Theorem} \ref{prop: CC1} is a cornerstone to ranking consistency between s-CC and p-CC, we conclude that the classical criterion is not robust \jjl{to} sampling bias.  }	
\end{remark}

\subsection{Ranking property of s-NPC}\label{sec:theoretic_plug-in}

To establish ranking agreement between s-NPC and p-NPC, an essential step is to develop a concentration result of $\mathrm{NPC}_{\alpha A}$ around $R_1(\varphi^*_{\alpha A})$,  where $\varphi^*_{\alpha A}$ is defined in \eqref{ideaL_sormulation_np}.  Recall that  $\hat \phi_{\alpha A}(\bd x) = \1(\hat s_A(\bd x_A) > \widehat C_{\alpha A}) =  \1(\hat p_{0A}(\bd x_A)/\hat p_{1A}(\bd x_A) > \widehat C_{\alpha A})$, where $\widehat C_{\alpha A}$ is determined by the NP umbrella classification algorithm. We always assume that the cumulative distribution function of $\hat s_{A} (\bd X_A), \text{ where } \bd X\sim P_0$, is continuous.   

\begin{lemma} \label{lem:kprime} 
Let $\alpha, \delta_1,\delta_2 \in (0,1)\,.$ 
If $m_2 = \left| \mathcal{S}_{\rm lo}^0 \right| \geq  \frac{4}{\alpha\delta_1}\,$, then the classifier $\hat{\phi}_{\alpha A}$ satisfies with probability at least $1-\delta_1-\delta_2 \,,$ 
\begin{align} \label{eq: R0_concentration} 
	\left|R_0(\hat{\phi}_{\alpha A})  - R_0(\varphi^*_{\alpha A}) \right|\leq \xi\,,
\end{align}
where
\[
	\xi = \sqrt{\frac{\ceil*{ d_{\alpha,\delta_1,m_2} \left(m_2+1\right)}\left(m_2+1-\ceil*{ d_{\alpha,\delta_1,m_2} \left(m_2+1\right)}\right)}{(m_2+2)(m_2+1)^2\,\delta_2}} + d_{\alpha,\delta_1,m_2}  + \frac{1}{m_2+1} - (1-\alpha)\,,
\]
\[
	 d_{\alpha,\delta_1,m_2} = \frac{1+ 2\delta_1 (m_2+2) (1-\alpha) + \sqrt{1+ 4\delta_1(m_2+2)(1-\alpha)\alpha}}{2\left\{ \delta_1(m_2+2)+1\right\}}\,,
\]
and $\ceil*{z}$ denotes the smallest integer larger than or equal to $z$.  Moreover, if $m_2 \geq \max(\delta_1^{-2}, \delta_2^{-2})$, we have 
$
\xi \leq  ({5}/{2}){m_2^{-1/4}}.
$	\end{lemma}

The next proposition is a result of Lemma \ref{lem:kprime} and a minor modification to the proof of Proposition 2.4 in \cite{zhao2016neyman}. We can derive an upper bound for $\left|R_1(\hat{\phi}_{\alpha A}) - R_1({\varphi}^*_{\alpha A})\right|$ the same as that for the excess type II error $R_1(\hat{\phi}_{\alpha A}) - R_1({\varphi}^*_{\alpha A})$ in \cite{zhao2016neyman}.

\begin{proposition} \label{prop:2}
Let $\alpha, \delta_1, \delta_2 \in (0,1)$. Assume that the density ratio $s_A(\cdot) = p_{1A}(\cdot)/p_{0A}(\cdot)$ satisfies the margin assumption of order $\bar\gamma$ at level $C^*_{\alpha A}$ (with constant $\bar C$) and detection condition of order $\underaccent{\bar}\gamma$ at 
level $(C^*_{\alpha A}, \delta^*)$ (with constant $\underaccent{\bar} C$), both with respect to the distribution $P_{0A}$.  
\noindent
If $m_2 \geq \max\left\{\frac{4}{\alpha \delta_1}, \delta_1^{-2}, \delta_2^{-2}, \left(\frac{2}{5}\underaccent{\bar}C{\delta^*}^{\uderbar\gamma}\right)^{-4}\right\}$,  the excess type II error of the classifier $\hat{\phi}_{\alpha A}$  satisfies with probability at least $1-\delta_1-\delta_2$,
\begin{align*}
&\left|R_1(\hat{\phi}_{\alpha A}) - R_1({\varphi}^*_{\alpha A})\right|\\
&\leq\, 
2\bar C \left[\left\{\frac{|R_0( \hat{\phi}_{\alpha A}) - R_0( \varphi^*_{\alpha A})|}{\underaccent{\bar}C}\right\}^{1/\uderbar{\gamma}} + 2  \| \hat s_A - s_A \|_{\infty} \right]^{1 + \bar\gamma} 
+ C^*_{\alpha A} |R_0( \hat{\phi}_{\alpha A}) - R_0( \varphi^*_{\alpha A})|\\
&\leq\,
2\bar C \left[\left(\frac{2}{5}m_2^{1/4}\underaccent{\bar}C\right)^{-1/\uderbar{\gamma}} + 2  \| \hat s_A - s_A \|_{\infty} \right]^{1 + \bar\gamma} 
+ C^*_{\alpha A} \left(\frac{2}{5} m_2^{1/4}\right)^{-1}\,.
\end{align*}
\end{proposition}

Propositions \ref{lem:bound_s_shat_for_plugin} and \ref{prop:2} lead to the following result.

\begin{theorem}\label{thm:1}
Let $\alpha$, $\delta_1$, $\delta_2$, $\delta_3$, $\delta_4$ $\in (0,1)$, and $l = |A|$. In addition to the assumptions of Propositions \ref{lem:bound_s_shat_for_plugin} and \ref{prop:2}, assume $n_2 \geq \left(\log\frac{2}{\delta_4}\right)^2$,	
then we have with probability at least $1-(\delta_1+\delta_2+\delta_3 +\delta_4),$  
$$
\left| \mathrm{NPC}_{\alpha A} - R_1 \left( {\varphi}^*_{\alpha A} \right)\right|\leq \widetilde C_2 \left[\left( \frac{\log m_1}{m_1}\right)^{\frac{\beta(1+\bar\gamma)}{2\beta+l}} + \left( \frac{\log n_1}{n_1}\right)^{\frac{\beta(1+\bar\gamma)}{2\beta+l}} + m_2^{-(\frac{1}{4}\wedge \frac{1+\bar\gamma}{\underaccent{\bar}\gamma})} + n_2^{-\frac{1}{4}} \right]\,,
$$	 
for some positive constant $\widetilde C_2$ that does not depend on $A$. 
\end{theorem}

%
%


Under smoothness, regularity, and sample size conditions, Theorem \ref{thm:1} shows the concentration of $\mathrm{NPC}_{\alpha A}$ around $R_1 \left( {\varphi}^*_{\alpha A}\right)$ with probability at least $1-(\delta_1+\delta_2+\delta_3+\delta_4)$. The user-specified violation rate $\delta_1$ represents the uncertainty that the type I error of an NP classifier $\hat \phi_{\alpha A}$ exceeds $\alpha$, leading to the underestimation of $R_1 ( {\varphi}^*_{\alpha A} )$; $\delta_2$ accounts for the possibility of unnecessarily stringent control on the type I error, which results in the overestimation of $R_1 ( {\varphi}^*_{\alpha A} )$; $\delta_3$ accounts for the uncertainty in training the scoring function $\hat s_A(\cdot)$ on a finite sample; $\delta_4$ represents the uncertainty of using leave-out class $1$ observations $\mathcal{S}^1_{\rm lo}$ to estimate $R_1(\hat\phi_{\alpha A})$. Note that while $\delta_1$ is both an input parameter for the construction of s-NPC and a constraint on the sample sizes, the other parameters $\delta_2$, $\delta_3$, and $\delta_4$ only have the latter role.  Like the constant $C_0$ in Proposition \ref{lem:bound_s_shat_for_plugin}, the generic constant $\widetilde C_2$ in Theorem \ref{thm:1} can be provided more explicitly, but it would be too cumbersome to do so.

\rv{Note that the upper bound in Theorem \ref{thm:1} involves $\uderbar{\gamma}$ while that in \rvv{Theorem} \ref{prop: CC1} does not. This is expected as the detection condition (that involves $\uderbar{\gamma}$) is a condition for diminishing excess type II errors under NP paradigm. Here we make some \rvv{example} simplifications to digest the bounds in \rvv{Theorems} \ref{prop: CC1} and \ref{thm:1}. \rvv{If we assume that $m_1, m_2, n_1, n_2 \sim N$ (the total sample size), $\beta = 2$, $l=1$, and $\bar \gamma  = 1$, then} the high probability upper bound for $|\mathrm{CC}_{A} - R(\varphi^*_A)|$ is $O(N^{-1/4})$, while that of $\left| \mathrm{NPC}_{\alpha A} - R_1 \left( {\varphi}^*_{\alpha A} \right)\right|$ is $O(N^{-(\frac{1}{4} \wedge \frac{2}{\uderbar \gamma})})$.  Hence, when $\uderbar \gamma > 8$, i.e., when there are not many points around the NP oracle decision boundary and thus the boundary is difficult to detect, the convergence rate of the upper bound is slower for NPC. } 

\rv{Although \rvv{Theorems} \ref{prop: CC1} and \ref{thm:1} both assume bounded supports in their conditions, we regard this as just a way to streamline the proofs. In Appendix~\ref{sec::bounds_vs_n}, we conduct a simulation study where features are generated from distributions with unbounded supports, and there is still clear concentration of $\mathrm{CC}_{A}$ and $\mathrm{NPC}_{\alpha A}$.}




\begin{theorem}\label{thm:selection_consistency_plugin}
Let $\alpha$, $\delta_1$, $\delta_2$, $\delta_3$, $\delta_4 \in (0,1)\,,$ $A_1, \ldots, A_J \subseteq\left\{1,\ldots, d \right\}$, $|A_1| = |A_2|=\cdots = |A_J| = l$, and $R_1(\varphi^*_{\alpha A_1}) < \cdots < R_1(\varphi^*_{\alpha A_J})$, ordered by p-NPC. Both $J$ and $l$ are constants that do not diverge with the sample sizes. In addition to the assumptions in Theorem \ref{thm:1}, assume that  the p-NPC's of $A_1, \ldots, A_J$ are separated by some margin $g>0$; in other words,  
$$
	 \min \limits_{i \in \{1,\dots, J-1\}}\left\{ R_1\left( {\varphi}^*_{\alpha A_{i+1}}\right) - R_1\left( {\varphi}^*_{\alpha A_i}\right) \right\} > g\,. 
$$ 
In addition, assume that $m_1, m_2, n_1, n_2$ satisfy 
\begin{equation}\label{eqn:sample size requirement NPC}
\widetilde C_2 \left[\left( \frac{\log m_1}{m_1}\right)^{\frac{\beta(1+\bar\gamma)}{2\beta+l}} + \left( \frac{\log n_1}{n_1}\right)^{\frac{\beta(1+\bar\gamma)}{2\beta+l}} + m_2^{-(\frac{1}{4}\wedge \frac{1+\bar\gamma}{\underaccent{\bar}\gamma})} + n_2^{-\frac{1}{4}} \right]  < \frac{g}{2}\,, 
\end{equation}
where $\widetilde C_2$ is the generic constant in Theorem \ref{thm:1}.  
Then with probability at least $1 - J(\delta_1+\delta_2+\delta_3+\delta_4)$, $\mathrm{NPC}_{\alpha A_i} < \mathrm{NPC}_{\alpha A_{i+1}}$ for all $i = 1, \ldots, J-1$. That is, s-NPC ranks $A_1, \ldots, A_J$ the same as p-NPC does.  
\end{theorem}

\begin{remark}
The conclusion in Theorem \ref{thm:selection_consistency_plugin} also holds under sampling bias, i.e., when the sample sizes $n$ (of class $1$) and $m$ (of class $0$) do not reflect the population proportions $\pi_0$ and $\pi_1$. 	
\end{remark}

Here we offer some intuition about the the robustness of s-NPC against sampling bias. Note that the objective and constraint of the NP paradigm only involve the class-conditional feature distributions, not the class proportions. Hence, p-NPC does not rely on the class proportions.  Furthermore, in s-NPC, each class-conditional density is estimated separately within each class, so s-NPC does not depend on the class proportions either. It is also worth noting that the proof of Theorem \ref{thm:selection_consistency_plugin} (in Appendix) does not use the relation between the ratio of class sizes in the sample and that in the population. 

\rv{Moreover, we derive partial consistency results for s-CC and s-NPC in Appendix~\ref{sec: partial consistency}, where we show that if the top $J$ feature subsets and the other feature subsets have p-CC or p-NPC differ by a margin, then s-CC or s-NPC can distinguish the top $J$ feature subsets.}


\section{Simulation studies} \label{sec:simulation}

\rvv{This section contains six simulation studies to verify the practical performance of s-CC and s-NPC in ranking features and to compare s-CC and s-NPC against multiple commonly used criteria for marginal and joint feature ranking. Table~\ref{tab:sim_design} summarizes the designs and purposes of the six studies.}

\begin{table}[htbp]
\begin{center}
\caption{Designs and purposes of six simulation studies\label{tab:sim_design}}
\rvv{
\begin{tabularx}{\textwidth}{ccrrccX}
\toprule
\multirow{2}{*}{Study} & \multirow{2}{*}{Distribution} & \multirow{2}{*}{$N$} & \multirow{2}{*}{$d$} & Sampling & Correlated & \multirow{2}{*}{Purpose}\\
 &  &  &  & bias & features & \\
\midrule
\multirow{2}{*}{S1} & \multirow{2}{*}{Gaussian} & $400$ & \multirow{2}{*}{$30$} & \multirow{2}{*}{No} & \multirow{2}{*}{No} & \multirow{2}{*}{Verify s-CC and s-NPC}\\
 &  & $1000$ &  &  &  & \\
 \midrule
\multirow{2}{*}{S2} & \multirow{2}{*}{Chi-squared} & $400$ & \multirow{2}{*}{$30$} & \multirow{2}{*}{No} & \multirow{2}{*}{No} & \multirow{2}{*}{Verify s-CC and s-NPC}\\
 &  & $1000$ &  &  &  & \\
  \midrule
\multirow{5}{*}{S3} & \multirow{5}{*}{Gaussian} & \multirow{5}{*}{$1000$} & \multirow{5}{*}{$30$} & \multirow{5}{*}{Yes} & \multirow{2}{*}{No} & Compare s-CC and s-NPC with SVM variants\\
\cmidrule{6-7}
 &  &  &  &  & \multirow{3}{*}{Yes} & Compare s-CC and s-NPC with multivariate feature ranking criteria \\
 \midrule
 S4 & Gaussian & $400$ & $500$ & No & No & Verify s-CC and s-NPC\\
 \midrule
 \multirow{4}{*}{S5} & \multirow{4}{*}{Gaussian} & \multirow{4}{*}{$200$} & \multirow{4}{*}{$10{,}000$} & \multirow{4}{*}{No} & \multirow{4}{*}{No} & Verify s-CC and s-NPC; compare them with SVM variants and multivariate feature ranking criteria\\
 \midrule
 \multirow{3}{*}{S6} & \multirow{3}{*}{\shortstack{(Mixture)\\Gaussian}} & \multirow{3}{*}{$400$} & \multirow{3}{*}{$2$} & \multirow{3}{*}{No} & \multirow{3}{*}{No} & Compare s-CC and s-NPC with marginal feature ranking criteria\\
\bottomrule
\end{tabularx}
}
\end{center}
\vspace{-.2in}
$N$: sample size (number of observations)\\
$d$: number of features
\end{table}

\rvv{In detail, we verify the performance of s-CC and s-NPC in ranking features under low-dimensional settings with the class-conditional distributions as Gaussian (studies S1 \& S3) or chi-squared (study S2), as well as under high-dimensional settings (studies S4--S5). Furthermore, in studies S3 and S5, we compare s-CC and s-NPC with three commonly used measures of feature importance in a multivariate classifier trained by the RF algorithm---the SHAP value \citep{lundberg2017unified} and two feature importance measures (the mean decrease in accuracy and the mean decrease in Gini index). Moreover, we design study S6 to demonstrate the advantages of s-CC and s-NPC over four commonly used measures for marginal feature ranking---the Pearson correlation, the distance correlation \citep{szekely2009brownian}, the two-sample $t$ test, and the two-sample Wilcoxon rank-sum test. Besides, motivated by \citep{lin2002support, guyon2002gene}, we implement variants of s-CC and s-NPC based on classifiers trained by the support vector machine (SVM) algorithm (Appendix~\ref{sec:s-CC_s-NPC_SVM_variants}), and we show in study S3 that these variants are not robust to sampling biases, unlike s-NPC.} 

\jjl{In all the simulation studies, we set the number of random splits $B=11$ (which we show in Figure~\ref{fig:1feature_varB_varn} in Appendix as a reasonable choice) for s-CC and s-NPC, \rv{as well as their SVM variants}, so that we can obtain reasonably stable criteria and meanwhile finish thousands of simulation runs within reasonable time. Regarding the RF algorithm, we use the \texttt{randomForest()} function in R package \texttt{randomForest}. The number of trees is set to \texttt{ntree=500} by default.}

\subsection{Ranking low-dimensional features at the sample level}\label{sec:sim_low_dim}
We first demonstrate the performance of s-CC and s-NPC in ranking features when $d$, the number of features, is much smaller than $N$ (the total sample size). We design \rvv{simulation studies S1 and S2} to support our theoretical results in Theorems \ref{thm:selection_consistency_cc} and \ref{thm:selection_consistency_plugin} in the absence of sampling bias. \rvv{Using simulation study S3, we demonstrate that s-NPC is robust to sampling bias, while s-CC and the SVM variants of s-CC and s-NPC are not; furthermore, we show that the RF algorithm's three feature ranking criteria (mean decrease in accuracy, mean decrease in Gini index, and SHAP value) cannot capture features' marginal ranking in the presence of feature correlations.}

\rvv{There is no sampling bias in simulation studies S1 and S2}. In \rvv{study S1}, we generate data from the following two-class Gaussian model with $d=30$ features, among which we set the first $s=10$ features to be informative (a feature is informative if and only if it has different marginal distributions in the two classes).  
\begin{align}\label{eq:best_subset}
	\bd X \given (Y=0) &\sim \mathcal{N}(\bd\mu^0, \bd\Sigma)\,, & \bd X \given (Y=1) &\sim \mathcal{N}(\bd\mu^1, \bd\Sigma)\,, & \pi_1 = \p(Y=1) = .5\,,
\end{align}
where $\bd\mu^0 = (\underbrace{-1.5,\ldots,-1.5}_{10}, \mu_{11}, \ldots, \mu_{30})\tran$, $\bd\mu^1 = (\underbrace{1,.9,\ldots,.2,.1}_{10}, \mu_{11}, \ldots, \mu_{30})\tran$, with $\mu_{11}, \ldots, \mu_{30}$ independently and identically drawn from $\mathcal N(0,1)$ and then held fixed, and $\bd\Sigma = 4 \, \mathbf{I}_{30}$. In terms of population-level criteria p-CC and p-NPC, a clear gap exists between the first $10$ informative features and the rest features, yet the $10$ features themselves have increasing criterion values but no obvious gaps. That is, the first $10$ features have true ranks going down from $1$ to $10$, and the rest of features have a tied true rank of $20.5$, i.e., the average of $11, \ldots, 30$ \footnote{Why the uninformative $20$ features should receive a true rank of $20.5$ instead of $11$ is because a reasonable ranking criterion, with data randomness, would assign these $20$ features with ranks uniformly distributed between $11$ and $30$.}.  

\begin{figure}[htbp]
\includegraphics[width=\textwidth]{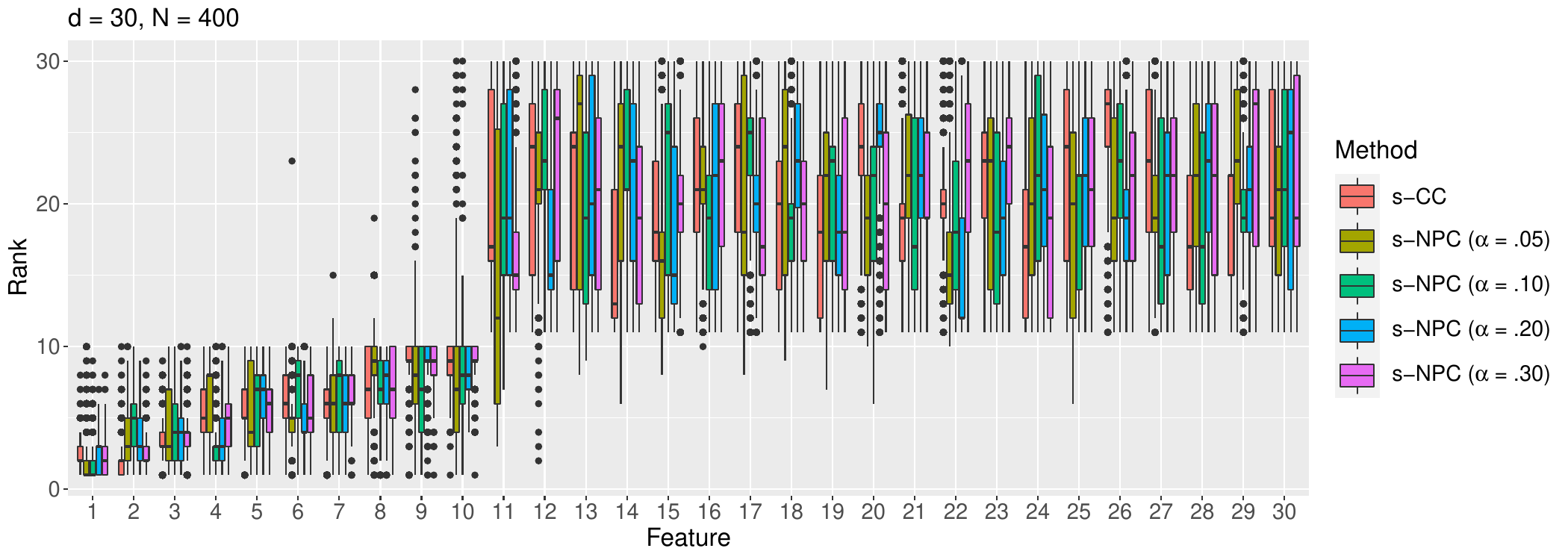}
\caption{Rank distributions of the $d=30$ features by s-CC or s-NPC (with varying $\alpha$) with sample size $N=400$ under the Gaussian setting \eqref{eq:best_subset}---simulation study S1.\label{fig:avg_rank_d30_n400}}	
\end{figure}

\begin{figure}[htbp]
\includegraphics[width=\textwidth]{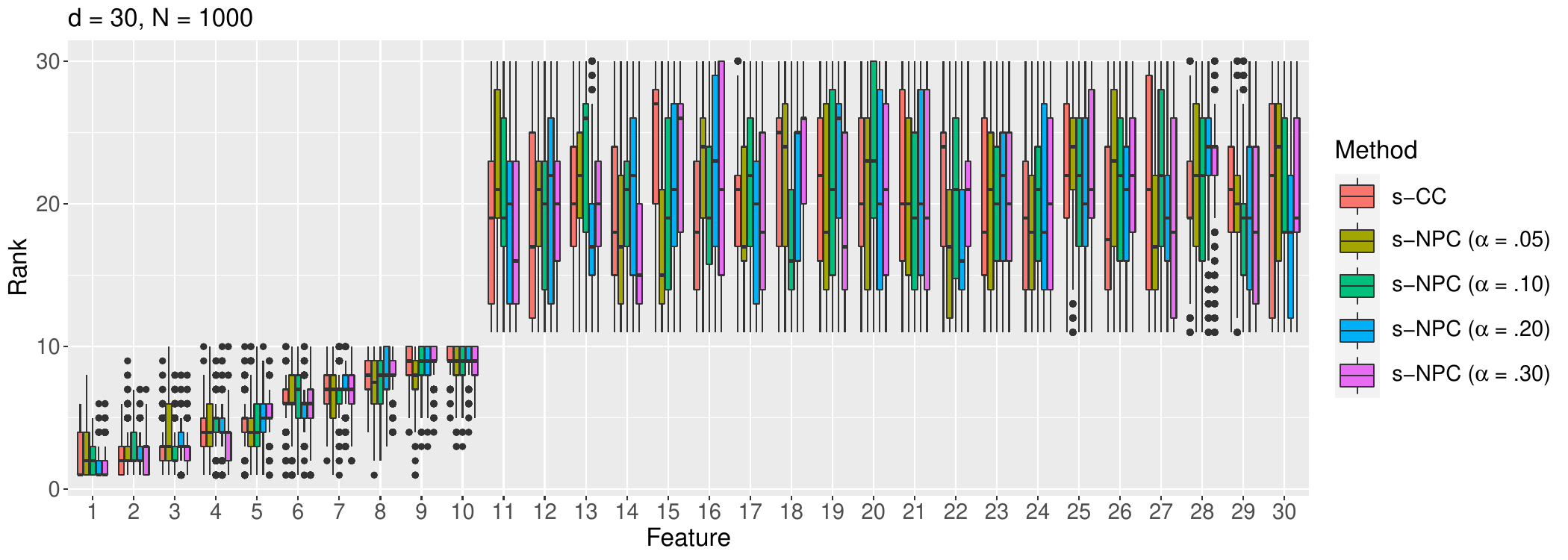}
\caption{Rank distributions of the $d=30$ features by s-CC or s-NPC (with varying $\alpha$) with sample size $N=1000$ under the Gaussian setting \eqref{eq:best_subset}---simulation study S1.\label{fig:avg_rank_d30_n1000}}	
\end{figure}

We simulate $1000$ samples of size $N=400$ \footnote{In the NP umbrella algorithm, $m_2$, i.e., the class $0$ sample size reserved for estimating the threshold, must be at least $59$ when $\alpha = \delta_1 = .05$. We set the overall sample size to $N=400$ so that  the expected $m_2$ is $100$; then the realized $m_2$ is larger than $59$ with high probability.} or $1000$ from the above model. We apply s-CC (\ref{CC}) and s-NPC with $\delta_1 = .05$ and four $\alpha$ levels $.05$, $.10$, $.20$, and $.30$ (\ref{Npscore}), five criteria in total, to each sample to rank the $30$ features. That is, for each feature, we obtain $1000$ ranks by each criterion. We summarize the average rank of each feature by each criterion in Tables \ref{tab:avg_rank_d30_n400} and \ref{tab:avg_rank_d30_n1000} \rvv{in Appendix}, and we plot the distribution of ranks of each feature in Figures \ref{fig:avg_rank_d30_n400} and \ref{fig:avg_rank_d30_n1000}. The results show that all criteria clearly distinguish the first 10 informative features from the rest. For s-NPC, we observe that its ranking is more variable for a smaller $\alpha$ (e.g., $0.05$). This is expected because, when $\alpha$ becomes smaller, the threshold in the NP classifiers would have an inevitably larger variance and lead to a more variable type II error estimate, i.e., s-NPC. As the sample size $N$ increases from $400$ (Table~\ref{tab:avg_rank_d30_n400} and Figure~\ref{fig:avg_rank_d30_n400}) to $1000$ (Table~\ref{tab:avg_rank_d30_n1000} and Figure~\ref{fig:avg_rank_d30_n1000}), all criteria achieve greater agreement with the true ranks. 

In \rvv{study S2}, we generate data from the following two-class Chi-squared distributions of $d=30$ features, among which we still set the first $s=10$ features to be informative.
\begin{align}\label{eq:chisq}
	\bd X_{\{j\}} \given (Y=0) &\sim \chi^2_1\,, \; j=1,\ldots,30, \quad \p(Y=1) = .5\,,  \\
	\bd X_{\{1\}} \given (Y=1) &\sim \chi^2_{11}\,, \; \bd X_{\{2\}} \given (Y=1) \sim \chi^2_{10}\,, \cdots \,, \bd X_{\{10\}} \given (Y=1) \sim \chi^2_{2}\,, \notag \\
	\bd X_{\{j\}} \given (Y=1) &\sim \chi^2_1\,, \; j=11,\ldots,30\,. \notag
\end{align}
Similar to the previous Gaussian setting, the first $10$ features have true ranks going down from $1$ to $10$, and the rest of features are tied with a true rank of $20.5$. We simulate $1000$ samples of size $N=400$ or $1000$ from this model, and we apply s-CC (\ref{CC}) and s-NPC with $\delta_1 = .05$ and four $\alpha$ levels $.05$, $.10$, $.20$, and $.30$ (\ref{Npscore}), five criteria in total, to each sample to rank the $30$ features. We summarize the average rank of each feature by each criterion in Tables \ref{tab:avg_rank_d30_n400_chisq} and \ref{tab:avg_rank_d30_n1000_chisq} (in Appendix), and we plot the distribution of ranks of each feature in Figures \ref{fig:avg_rank_d30_n400_chisq} and \ref{fig:avg_rank_d30_n1000_chisq} (in Appendix). The results and conclusions are consistent with those under the Gaussian setting. 

\begin{figure}[htbp]
     \centering
     \begin{subfigure}[b]{\textwidth}
         \centering
         \includegraphics[width=\textwidth]{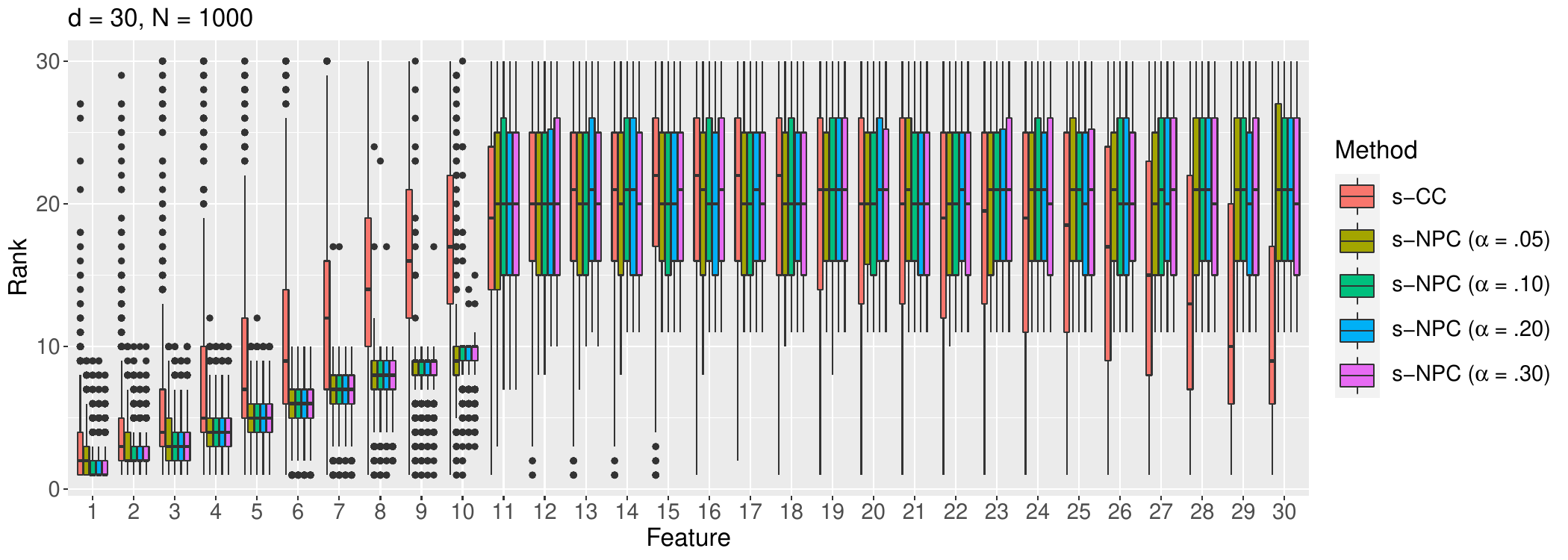}
         \caption{s-CC and s-NPC}
     \end{subfigure}
     \hfill
     \begin{subfigure}[b]{\textwidth}
         \centering
         \includegraphics[width=\textwidth]{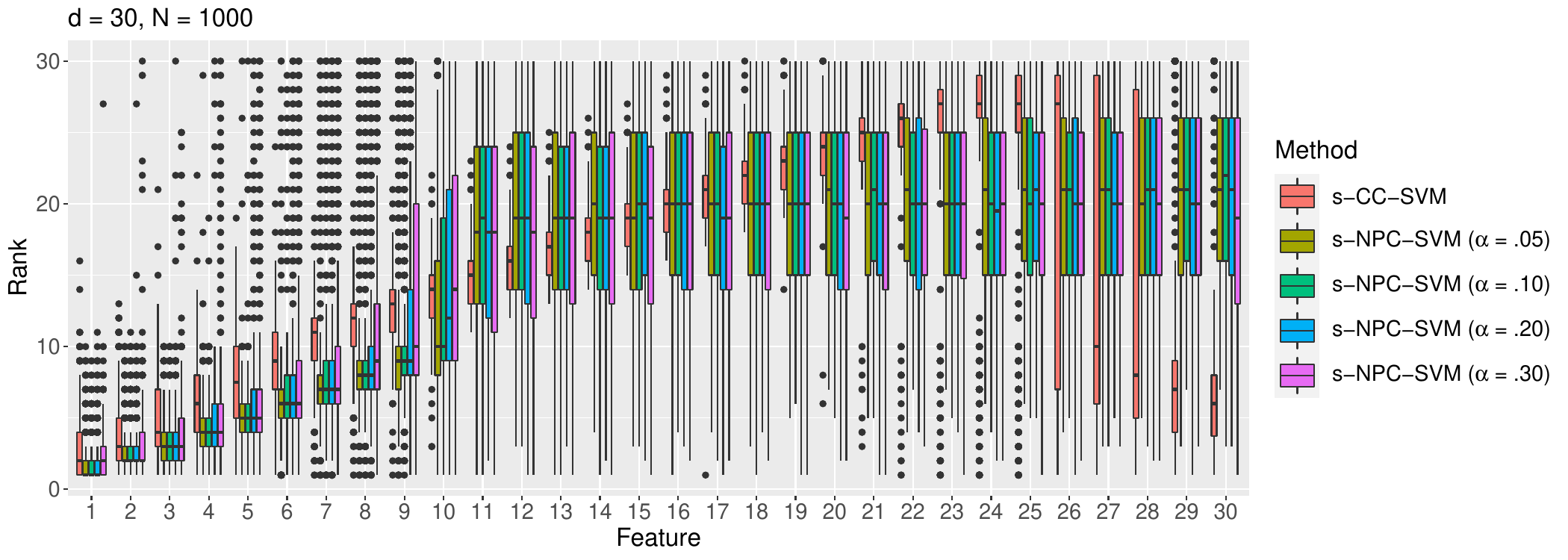}
         \caption{SVM variants of s-CC and s-NPC}
     \end{subfigure}
     \hfill
     \begin{subfigure}[b]{\textwidth}
         \centering
         \includegraphics[width=\textwidth]{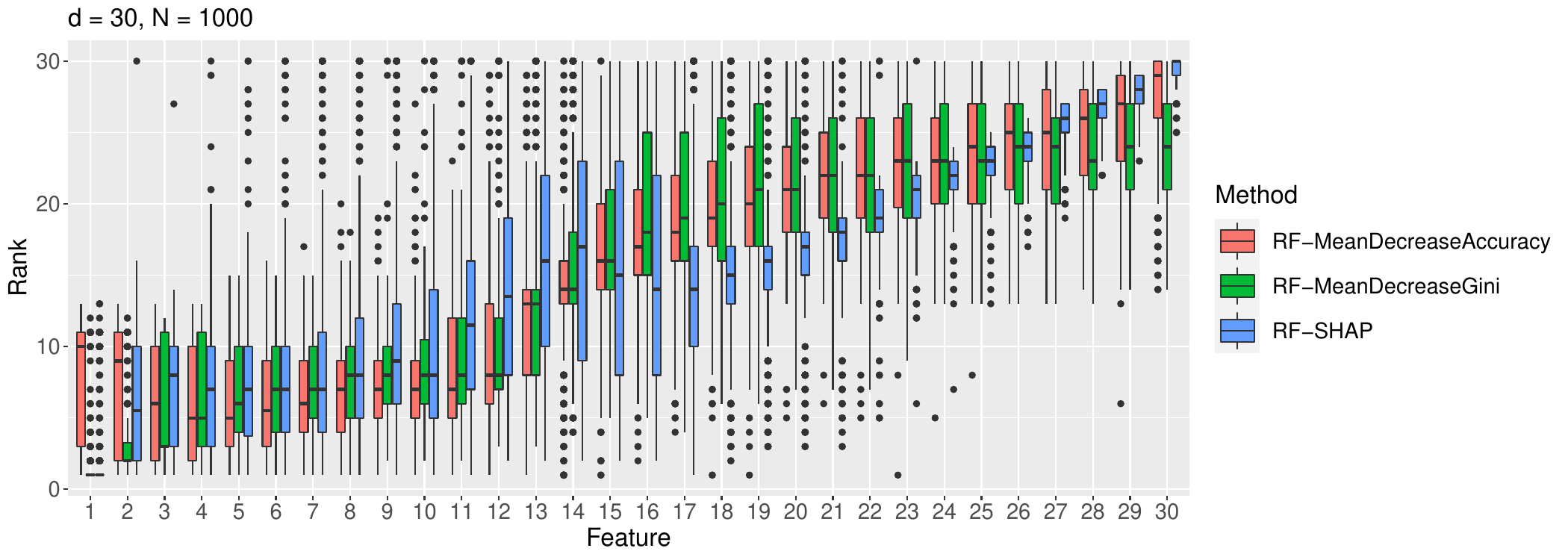}
         \caption{RF's two feature importance measures and SHAP value}
     \end{subfigure}
        \caption{Rank distributions of the features under the Gaussian setting with $d=30$, $N=1000$, sampling bias ($\pi_1^{\mathrm{population}}=.5$ and $\pi_1^{\mathrm{sample}}=.1$) \eqref{eq:gauss_bias}, and a Toeplitz-type feature covariance matrix: features $i$ and $j$ have a correlation $\rho_{ij}=.9^{|i-j|}$, $i,j=1,\ldots,30$---simulation study S3.}
        \label{fig:avg_rank_d30_n1000_bias_cor}
\end{figure}

\rv{Next, we design \rvv{study S3} with sampling bias, i.e., the two classes have proportions in the sample different from those in the population. We use the following Gaussian setting:
\begin{align}\label{eq:gauss_bias}
	\bd X \given (Y=0) \sim \mathcal{N}(\bd\mu^0, \bd\Sigma)\,, & \quad \bd X \given (Y=1) \sim \mathcal{N}(\bd\mu^1, \bd\Sigma)\,, \notag \\ \pi_1^{\textrm{population}} = \p^{\textrm{population}}(Y=1) = .5\,, & \quad \pi_1^{\textrm{sample}} = \p^{\textrm{sample}}(Y=1) = .1\,,
\end{align}
that is, class $1$ has a proportion $.5$ in the population but is undersampled with probability $.1$ in the sample. Same as in our first simulation study, we set $\bd\mu^0 = (\underbrace{-1.5,\ldots,-1.5}_{10}, \mu_{11}, \ldots, \mu_{30})\tran$, $\bd\mu^1 = (\underbrace{1,.9,\ldots,.2,.1}_{10}, \mu_{11}, \ldots, \mu_{30})\tran$, with $\mu_{11}, \ldots, \mu_{30}$ independently and identically drawn from $\mathcal N(0,1)$ and then held fixed, and we still set the diagonal entries of $\bd\Sigma$ to $4$. What is different here is that we add a scenario with feature correlations: conditional on each class, features $i$ and $j$ have a correlation $\rho_{ij}=.9^{|i-j|}$, $i,j=1,\ldots,30$; that is, $\bd\Sigma$ is a Toeplitz-type matrix with the $(i,j)$-th entry equal to $.9^{|i-j|} \times 4$. Here features $1$ to $10$ still have their true ranks going down from $1$ to $10$, while the other features still have a tied true rank of $20.5$. We simulate $1000$ samples of size $N=1000$ from this model, and we apply s-CC (\ref{CC}); s-NPC with $\delta_1 = .05$ and four $\alpha$ levels $.05$, $.10$, $.20$, and $.30$ (\ref{Npscore}); their corresponding SVM variants (Appendix C); and the RF algorithm's feature importance measures (mean decrease in accuracy and mean decrease in Gini index) and SHAP value---$13$ criteria in total---to each sample to rank the $30$ features. In Appendix, we summarize the average rank of each feature by each criterion in Table \ref{tab:avg_rank_d30_n1000_bias} (for the uncorrelated-feature scenario) and Table \ref{tab:avg_rank_d30_n1000_bias_cor} (for the correlated-feature scenario), and we plot the distribution of ranks of each feature in Figure \ref{fig:avg_rank_d30_n1000_bias} (for the uncorrelated-feature scenario) and Figure \ref{fig:avg_rank_d30_n1000_bias_cor} (for the correlated-feature scenario). The results show that, in the presence of sampling bias, only s-NPC can distinguish the first 10 informative features from the rest, while s-CC and the SVM variants of s-CC and s-NPC cannot. These results highlight the unique robustness of s-NPC to sampling bias, an advantage that even its SVM variant does not embrace. The reason is that sampling bias affects the training of the SVM scoring function. Moreover, s-CC, s-NPC, and their SVM variants---all marginal feature ranking criteria---are unaffected by feature correlations, as expected. In contrast, the RF algorithm's three feature ranking criteria, which are based on multivariate classifiers and thus not marginal, cannot accurately capture the features' marginal ranking in the presence of feature correlations (Figure \ref{fig:avg_rank_d30_n1000_bias_cor}(c)). 
}

\subsection{Ranking high-dimensional features at the sample level}
\begin{figure}[htbp]
\includegraphics[width=\textwidth]{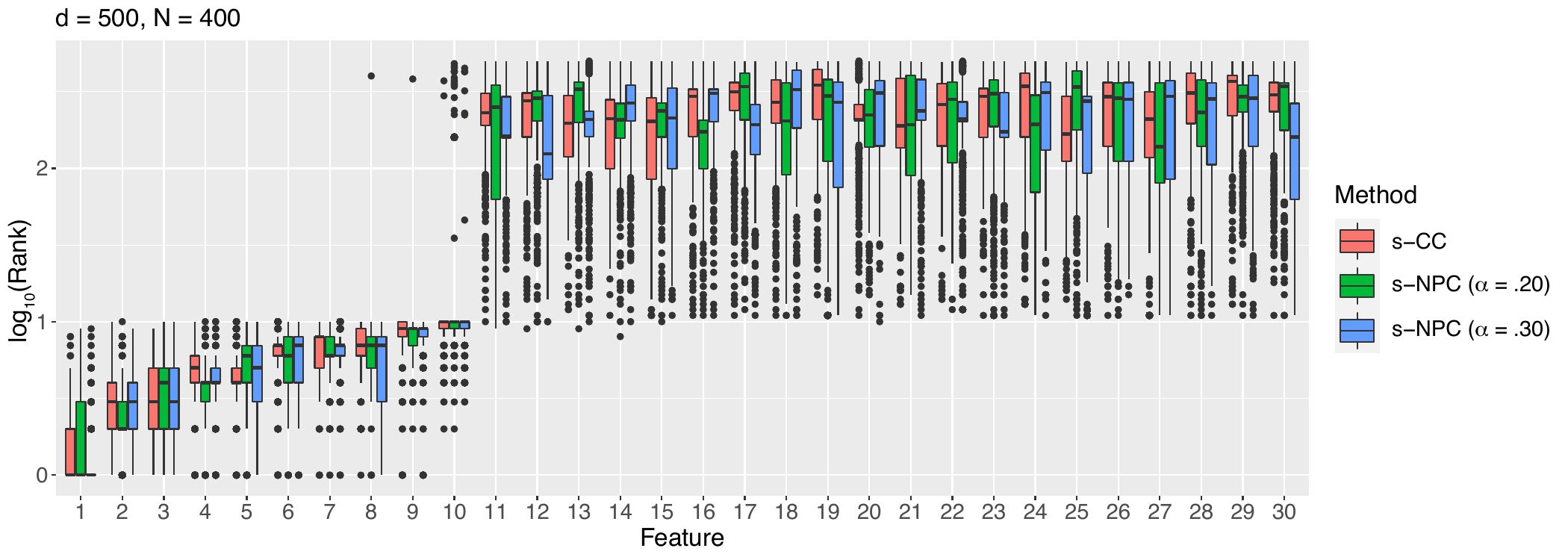}
\caption{Rank distributions of the first $30$ features by s-CC or s-NPC with $d=500$ and $N=400$ under the Gaussian setting \eqref{eq:best_subset}---simulation study S4. The vertical axis is on the $\log_{10}$ scale.\label{fig:avg_rank_d500_n400}}	
\end{figure}

We next examine the performance of s-CC and s-NPC when $d > N$, \rvv{using simulation studies S4 and S5}. In \rvv{study S4}, we set $d=500$ and $N=400$. The generative model is the same as \eqref{eq:best_subset}, and $\bd\mu^0 = (\underbrace{-1.5,\ldots,-1.5}_{10}, \mu_{11}, \ldots, \mu_{500})\tran$, $\bd\mu^1 = (\underbrace{1,.9,\ldots,.2,.1}_{10}, \mu_{11}, \ldots, \mu_{500})\tran$, with $\mu_{11}, \ldots, \mu_{500}$ independently and identically drawn from $\mathcal N(0,1)$ and then held fixed, and $\bd\Sigma = 4 \, \mathbf{I}_{500}$. Same as in the low-dimensional settings (Section \ref{sec:sim_low_dim}), p-CC and p-NPC have a clear gap between the first $10$ informative features and the rest of features. In terms of both p-CC and p-NPC, the first $10$ features have true ranks going down from $1$ to $10$, and the rest of features are tied with a true rank of $255.5$. We simulate $1000$ samples of size $N=400$ and apply s-CC (\ref{CC}) and s-NPC with $\delta_1 = .05$ and four $\alpha$ levels $.05$, $.10$, $.20$, and $.30$ (\ref{Npscore}) to each sample to rank the $500$ features. We summarize the average rank of each feature by each criterion in Table \ref{tab:avg_rank_d500_n400} (in Appendix), and we plot the distribution of ranks of each feature in Figure \ref{fig:avg_rank_d500_n400}. The results show that ranking under this high-dimensional setting is more difficult than under the low-dimensional setting. However, s-CC and s-NPC with $\alpha = 0.2$ or $0.3$ still clearly distinguish the first 10 informative features from the rest, while s-NPC with $\alpha = 0.05$ or $0.1$ have worse performance on features 8--10, demonstrating again that ranking becomes more difficult for s-NPC when $\alpha$ is smaller and requires a larger sample size $N$. 

\rv{In the more high-dimensional \rvv{study S5}, we further decrease the $N/d$ ratio by setting $d=10{,}000$ and $N=200$, a scenario that resembles many biomedical datasets. The generative model is the same as \eqref{eq:best_subset}, and $\bd\mu^0 = (\underbrace{-2.5,\ldots,-2.5}_{10}, \mu_{11}, \ldots, \mu_{10{,}000})\tran$, $\bd\mu^1 = (\underbrace{1,.9,\ldots,.2,.1}_{10}, \mu_{11}, \ldots, \mu_{10{,}000})\tran$, with $\mu_{11}, \ldots, \mu_{10{,}000}$ independently and identically drawn from $\mathcal N(0,1)$ and then held fixed, and $\bd\Sigma = 4 \, \mathbf{I}_{10{,}000}$. Same as in \rvv{study S4}, the first $10$ features have true ranks going down from $1$ to $10$, and the rest of features are tied with a true rank of $5005.5$. We simulate $1000$ samples of size $N=200$, and we apply s-CC (\ref{CC}); s-NPC with $\delta_1 = .05$ and three $\alpha$ levels $.10$, $.20$, and $.30$ (\ref{Npscore}); their SVM variants (Appendix C); and the RF's two feature importance measures (the mean decrease in accuracy and the mean decrease in Gini index)---ten criteria in total---to each sample to rank the $10{,}000$ features. Note that the SHAP value is inapplicable because it requires substantial computational time in this scenario. We summarize the average rank of each feature by each criterion in Table \ref{tab:avg_rank_d10000_n200} (in Appendix). The results show that s-NPC outperforms its SVM variant at each $\alpha$ in terms of the ranking accuracy for the top $10$ features. Moreover, five criteria, including s-CC, s-NPC ($\alpha = .30$), their SVM variants, and RF's mean decrease in accuracy, can distinguish the first $10$ informative features by assigning them average ranks no greater than $10$. The fact that s-NPC and its SVM variant with $\alpha = .10$ or $.20$ have worse performance is because of the small $N/d$ ratio. This result echos the importance of using a not-too-small $\alpha$ for s-NPC when $N$ is not too large and $N/d$ is small. Interestingly, RF's mean decrease in Gini index performs worse than its mean decrease in accuracy in ranking the $10$-th feature.}

\subsection{Comparison with other marginal feature ranking criteria} 

 In \rvv{simulation study S6}, we compare s-CC and s-NPC with four criteria that have been widely used to rank features marginally: the Pearson correlation, the distance correlation \citep{szekely2009brownian}, the two-sample $t$ test, and the two-sample Wilcoxon rank-sum test. None of these existing approaches rank features based on a prediction objective; as a result, the feature ranking they give may not reflect the prediction performance of features under a particular objective. Here we use an example to demonstrate this phenomenon. We generate data with $d=2$ features from the following model:
 \begin{align}\label{eq:gauss_mixture}
	X_1 \given (Y=0) &\sim \mathcal{N}(0, 1)\,, & X_1 \given (Y=1) &\sim \mathcal{N}(1, 1)\,, & \p(Y=1) = .5\,,	\notag\\
	X_2 \given (Y=0) &\sim \mathcal{N}(0, 1)\,, & X_2 \given (Y=1) &\sim .5\,\mathcal{N}(-2, 1) + .5\,\mathcal{N}(2, 1)\,. & 	
 \end{align}
 To calculate p-CC and p-NPC with $\delta_1=.05$ at four $\alpha$ levels $.05$, $.10$, $.20$, and $.30$ on these two features, we use a large sample with size $10^6$ for approximation, and the results in Table~\ref{tab:gauss_mixture_pop} show that all the five population-level criteria rank feature 2 as the top feature.
 
\begin{table}[htbp]
\caption{\label{tab:gauss_mixture_pop}Values of p-CC and p-NPC of the two features in \eqref{eq:gauss_mixture}.}
\centering
\small
\begin{tabular}{rrrrrr}
  \hline
Feature & p-CC & p-NPC ($\alpha$ = .05) & p-NPC ($\alpha$ = .10) & p-NPC ($\alpha$ = .20) & p-NPC ($\alpha$ = .30) \\ 
  \hline
1 & .31 & .74 & .61 & .44 & .32 \\ 
  2 & .22 & .49 & .36 & .24 & .17 \\ 
   \hline
\end{tabular}
\end{table}

Then we simulate $1000$ samples of size $N=400$ from the above model and apply nine ranking approaches: s-CC, s-NPC with $\delta_1=.05$ at four $\alpha$ levels ($.05$, $.10$, $.20$, and $.30$), the Pearson correlation, the distance correlation, the two-sample $t$ test, and the two-sample Wilcoxon rank-sum test, to each sample to rank the two features. From this we obtain $1000$ rank lists for each ranking approach, and we calculate the frequency that each approach correctly finds the true rank order. The frequencies are summarized in Table~\ref{tab:gauss_mixture_freq}, which shows that none of the four common criteria identifies feature 2 as the better feature for prediction. In other words, if users wish to rank features based on a prediction objective under the classical or NP paradigm, these criteria are not suitable. 

\begin{table}[htbp]
\caption{\label{tab:gauss_mixture_freq}The frequency that each ranking approach identifies the true rank order.}
\centering
\small
\begin{tabular}{rrrrr}
  \hline
s-CC & s-NPC ($\alpha$ = .05) & s-NPC ($\alpha$ = .10) & s-NPC ($\alpha$ = .20) & s-NPC ($\alpha$ = .30) \\ 
100\% & 99.9\% & 99.3\% & 99.7\% & 100\% \\ 
   \hline
Pearson cor & distance cor & two-sample $t$ & two-sample Wilcoxon &\\
0\% & 0.5\% & 0\% & 0\% &\\
	\hline
\end{tabular}
\end{table}

\section{Real data applications}\label{simu:realdata}
We apply s-CC and s-NPC to two real datasets to demonstrate their wide application potential in biomedical research. \jjl{Here we set the number of random splits $B=1000$ for s-CC and s-NPC, as allowed by our computational resource.} 

\subsection{Application 1: classification of breast cancer and normal tissues based on genes' DNA methylation levels}

The first dataset contains genome-wide DNA methylation profiles of $285$ breast tissues measured by the Illumina HumanMethylation450 microarray technology. This dataset includes $46$ normal tissues and $239$ breast cancer tissues. Methylation levels are measured at $468{,}424$ CpG probes in every tissue \citep{fleischer2014genome}. We download the preprocessed and normalized dataset from the Gene Expression Omnibus (GEO) \citep{edgar2002gene} with the accession number GSE60185. The preprocessing and normalization steps are described in detail in \cite{fleischer2014genome}. To facilitate the interpretation of our analysis results, we further process the data as follows. First, we discard a CpG probe if it is mapped to no gene or more than one genes. Second, if a gene contains multiple CpG probes, we calculate its methylation level as the average methylation level of these probes. This procedure leaves us with $19{,}363$ genes with distinct methylation levels in every tissue. We consider the tissues as data points and the genes as features, so we have a sample with size $N=285$ and number of features $d=19{,}363$. Since misclassifying a patient with cancer to be healthy leads to more severe consequences than the other way around, we code the $239$ breast cancer tissues as the class $0$ and the $46$ normal tissues as the class $1$ to be aligned with the NP paradigm. After applying s-CC (\ref{CC}) and s-NPC with $\delta_1 = .05$ and four $\alpha$ levels ($.05$, $.10$, $.20$, and $.30$) (\ref{Npscore}) to this sample, we summarize the top $10$ genes found by each criterion in Table \ref{tab:bc_rank}.  Most of these top ranked genes have been reported associated with breast cancer, suggesting that our proposed criteria can indeed help researchers find meaningful features. Meanwhile, although other top ranked genes do not yet have experimental validation, they have weak literature indication and may serve as potentially interesting targets for cancer researchers. For a detailed list of literature evidence, please see \textit{the Supplementary Excel File}. The fact that these five criteria find distinct sets of top genes is in line with our rationale that feature importance depends on prediction objective. By exploring top features found by each criterion, researchers will obtain a comprehensive collection of features that might be scientifically interesting. 

\begin{table}[htbp]
\begin{threeparttable}
\caption{\label{tab:bc_rank}Top 10 genes\tnote{\dag} found by each criterion in breast cancer methylation data \citep{fleischer2014genome}. Genes with strong literature evidence to be breast-cancer-associated are marked in bold; see the Supplementary Excel File.}
\centering
\small
\begin{tabular}{rccccc}
  \hline
Rank & s-CC & s-NPC ($\alpha$ = .05) & s-NPC ($\alpha$ = .10) & s-NPC ($\alpha$ = .20) & s-NPC ($\alpha$ = .30) \\ 
  \hline
1 & \textbf{HMGB2} & \textbf{HMGB2} & \textbf{HMGB2} & \textbf{ABHD14A} & \textbf{ABHD14A} \\ 
  2 & \textbf{MIR195} & MICALCL & \textbf{ABHD14A} & \textbf{ABL1} & \textbf{ABL1} \\ 
  3 & MICALCL & NR1H2 & ZFPL1 & \textbf{BAT2} & \textbf{ACTN1} \\ 
  4 & \textbf{AIM2} & \textbf{AGER} & \textbf{AGER} & \textbf{BATF} & AKAP8 \\ 
  5 & AGER & \textbf{BATF} & RILPL1 & \textbf{CCL8} & AP4M1 \\ 
  6 & KCNJ14 & ZFP106 & SKIV2L & \textbf{COG8} & \textbf{ARHGAP1} \\ 
  7 & \textbf{HYAL1} & CTNNAL1 & \textbf{TP53} & FAM180B & \textbf{ATG4B} \\ 
  8 & SKIV2L & \textbf{MIR195} & \textbf{RELA} & \textbf{HMGB2} & \textbf{BAT2} \\ 
  9 & \textbf{RUSC2} & \textbf{AIM2} & \textbf{MIR195} & \textbf{HSF1} & BAT5 \\ 
  10 & DYNC1H1 & ZFPL1 & \textbf{CCL8} & KIAA0913 & \textbf{BATF} \\ 
   \hline
\end{tabular}
\footnotesize
\begin{tablenotes}
\item[\dag] Note that $20$ genes have zero s-NPC ($\alpha$ = .20) values, and $119$ genes have zero s-NPC ($\alpha$ = .30) values. Hence, the listed top $10$ genes by either of these two criteria are the first $10$ in the alphabetical order.
\end{tablenotes}
\end{threeparttable}
\end{table}

\begin{figure}
\includegraphics[width=\textwidth]{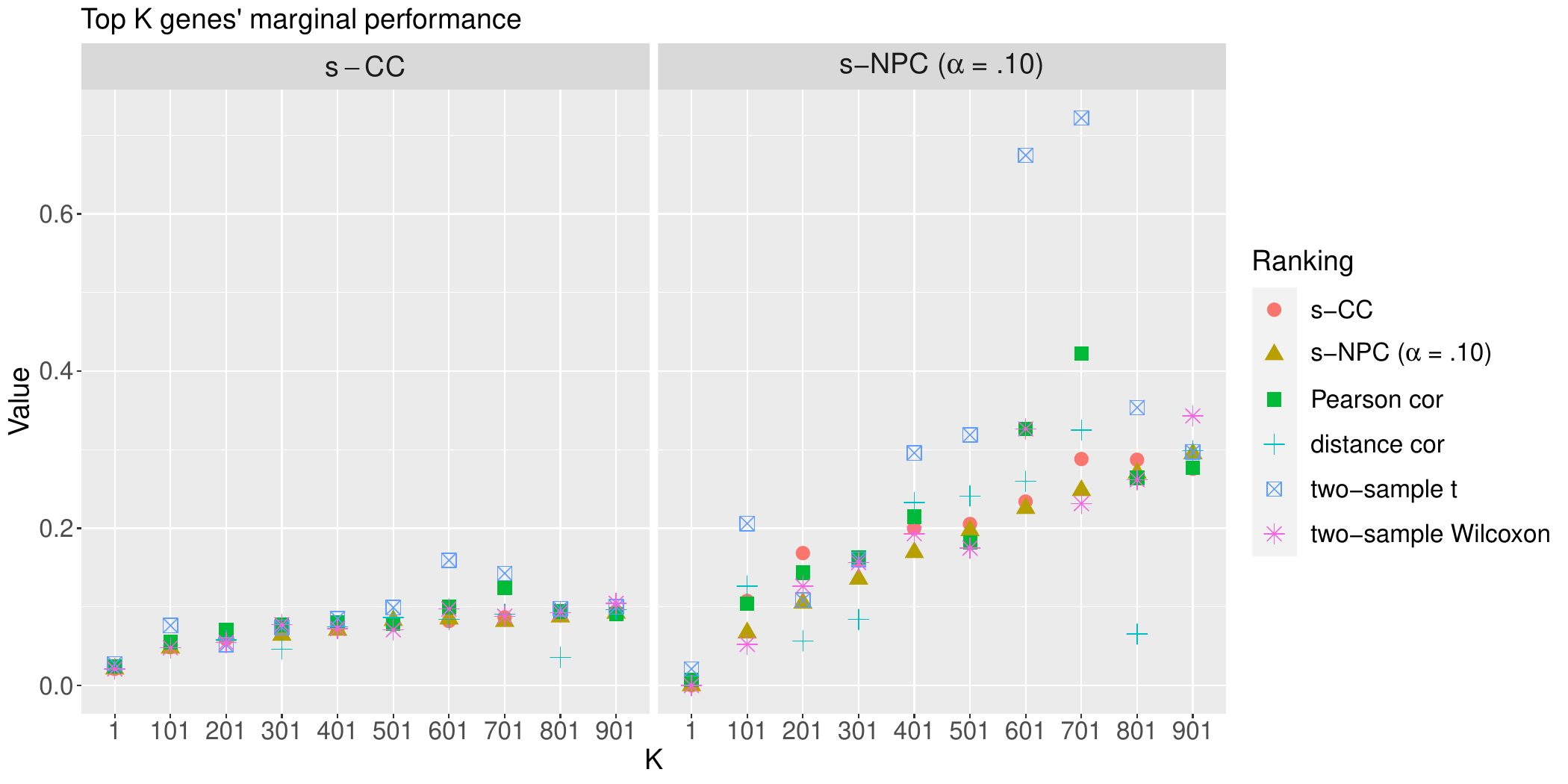}
\caption{Values of s-CC and s-NPC ($\alpha = .10$) of the top $K=1$, $101$, $201$, $401$, $501$, $601$, $701$, $801$, or $901$ features ranked by each criterion in Application 1. As expected, the ranks by s-CC are monotone in s-CC values (left), and the ranks by s-NPC ($\alpha = .10$) are monotone in s-NPC ($\alpha = .10$) values (right). We focus on the $301$st and $801$st genes ranked by the distance correlation, whose ranks by s-CC and s-NPC ($\alpha = .10$) are much better, and the $601$st and $701$st genes ranked by the two-sample $t$ test, whose ranks by s-CC and s-NPC ($\alpha = .10$) are much worse. \label{fig:real_bc_top_features_marginally}}
\end{figure}

\rv{Moreover, we apply the four widely-used but non-prediction-based marginal ranking criteria---the Pearson correlation, the distance correlation, the two-sample $t$ test, and the two-sample Wilcoxon rank-sum test, to rank the $d=19{,}363$ genes. For demonstration purpose, we check the s-CC and s-NPC ($\alpha = .10$) values of the genes ranked as the $1$st, $101$st, $201$st, $301$st, $401$st, $501$st, $601$st, $701$st, $801$st, and $901$st by each criterion. Figure~\ref{fig:real_bc_top_features_marginally} shows the cases where the ranks assigned by other criteria differ tremendously from those assigned by s-CC or s-NPC ($\alpha = .10$), including the $301$st and $801$st genes ranked by the distance correlation, whose ranks by s-CC and s-NPC ($\alpha = .10$) are much better, and the $601$st and $701$st genes ranked by the two-sample $t$ test, whose ranks by s-CC and s-NPC ($\alpha = .10$) are much worse.} 

\begin{figure}
\includegraphics[width=\textwidth]{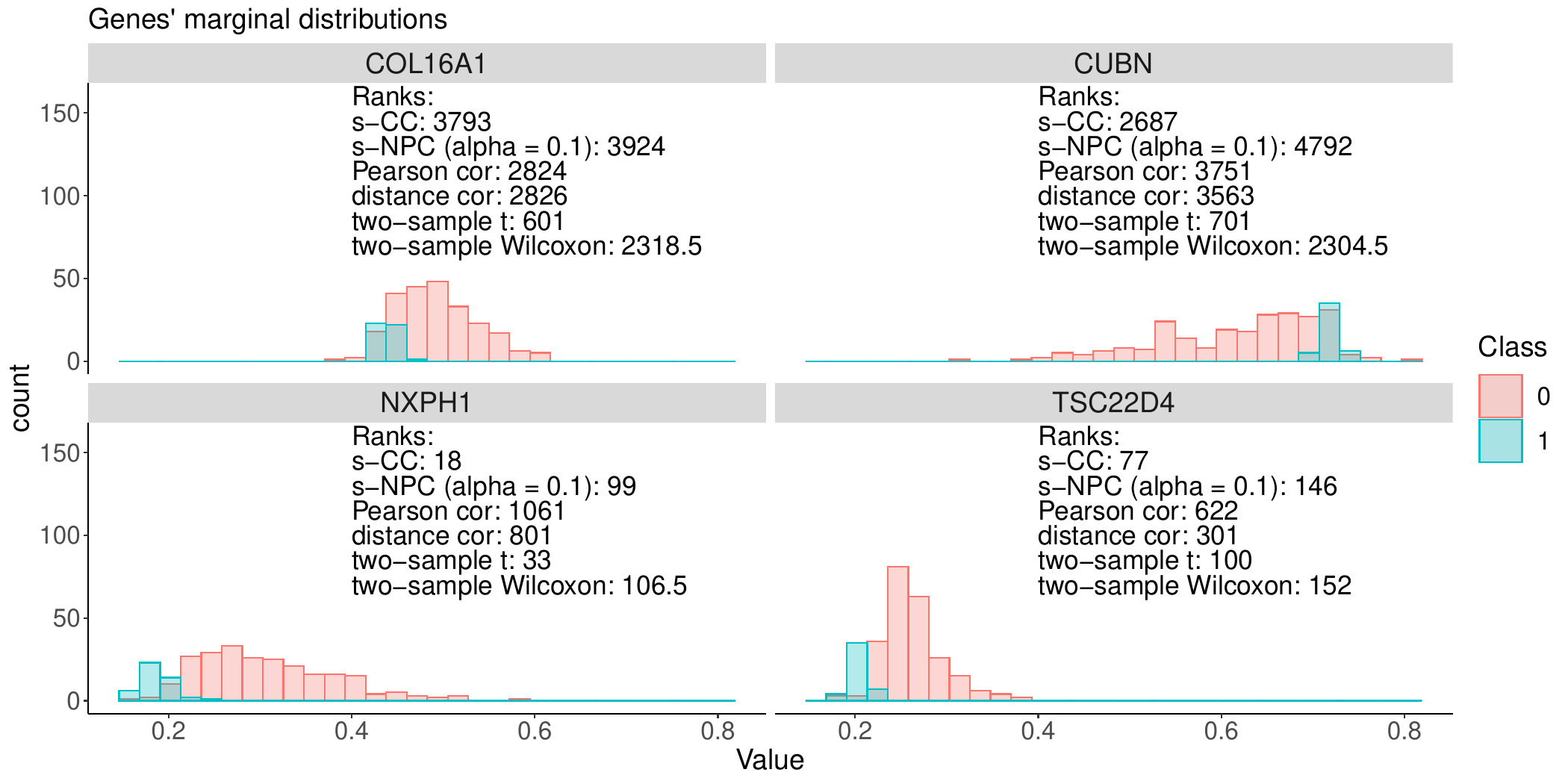}
\caption{Class-conditional distributions of four example genes in Application 1. Out of $d=19{,}363$ genes, the top genes are ranked much better by the two-sample $t$ test than by s-CC or s-NPC ($\alpha = .10$); the bottom genes are ranked top by s-CC and s-NPC ($\alpha = .10$) but much worse by Pearson correlation and distance correlation. \label{fig:real_bc_top_features_marginally_supp}}
\end{figure}

\rv{Figure~\ref{fig:real_bc_top_features_marginally_supp} illustrates the class-conditional distributions of these four genes. Interestingly, the two genes \textit{NXPH1} and \textit{TSC22D4} are ranked top by s-CC (with ranks $18$ and $77$) and s-NPC ($\alpha = .10$) (with ranks $99$ and $146$), while they are ranked worse by both the Pearson correlation (with ranks $1061$ and $622$) and the distance correlation (with ranks $801$ and $301$). Their class-conditional distributions show distinct, non-overlapping density peaks between classes $0$ and $1$, confirming their top ranks assigned by s-CC and s-NPC. Literature also suggests the two genes' potential roles in breast cancer: a methylation study found \textit{NXPH1} as a candidate biomarker gene for HER2+ breast cancer \citep{lindqvist2014whole}; an RNA-seq study found \textit{TSC22D4} up-regulated in a BRCA1 mutated cell line (SL) compared to a BRCA1 wild-type cell line (SB) \citep{privat2018antioxydation}; another study identified \textit{TSC22D2}, an isoform of \textit{TSC22D4}, as a novel cancer-associated gene in a rare multi-cancer family \citep{liang2016tsc22d2}.} 

\rv{In Figure~\ref{fig:real_bc_top_features_marginally_supp}, another two genes, \textit{COL16A1} and \textit{CUBN}, are ranked much lower by s-CC (with ranks $3793$ and $2687$) and s-NPC ($\alpha = .10$) (with ranks $3924$ and $4792$) than by the two-sample $t$ test (with ranks $601$ and $701$). Inspecting the two genes' class-conditional distributions, we find that their class-$1$ conditional distributions have high-density domains largely contained in the high-density domains of their respective class-$0$ conditional distributions, an observation consistent with the low rankings assigned by s-CC and s-NPC ($\alpha = .10$). Both genes do not seem to have direct associations with breast cancer: \textit{COL16A1} encodes the alpha chain of type XVI collagen; \textit{CUBN} encodes a protein called cubilin, which is involved in the uptake of vitamin B12 from food into the body.}

\subsection{Application 2: classification of high-risk and low-risk prostate cancer patients based on microRNA expression levels in urine samples}

The second dataset contains microRNA (miRNA) expression levels in urine samples of prostate cancer patients, downloaded from the GEO with accession number GSE86474 \citep{jeon2019temporal}. This dataset is composed of $78$ high-risk and $61$ low-risk patients. To align with the NP paradigm, we code the high-risk and low-risk patients as class $0$ and $1$, respectively, so $m/n=78/61$. In our data pre-processing, we retain miRNAs that have at least $60\%$ non-zero expression levels across the $N=139$ patients, resulting in $d=112$ features. We use this dataset to demonstrate that s-NPC is robust to sampling bias that results in disproportional training data; that is, training data have different class proportions from those of the population. \rvv{Since in many biomedical datasets, the proportions of diseased patients do not reflect the true proportions in the population, a desirable feature ranking criterion should be robust to such sampling bias so that the selected features would maintain good out-of-sample predictive power.} 

\begin{figure}
\includegraphics[width=\textwidth]{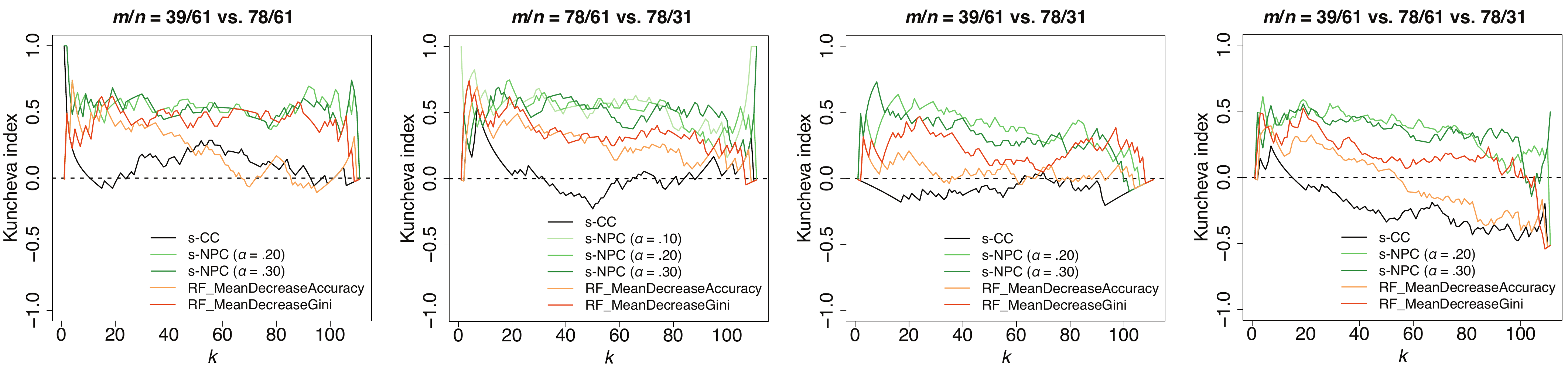}
\caption{\rv{Kuncheva indices of s-CC, s-NPC, and the two RF feature importance measures in ranking the top $k$ features in Application 2.}\label{fig:consistency}}	
\end{figure}

We create two sub-datasets by randomly removing one half of the data points in class $0$ or $1$, so that one sub-dataset has $m/n=39/61$ and the other has $m/n=78/31$. We apply s-CC, s-NPC with $\delta_1 = .05$, and \rv{the RF algorithm's feature importance measures (the mean decrease in accuracy and the mean decrease in Gini index) to the full dataset and each sub-dataset to rank features. To evaluate each criterion's robustness to disproportional data, we compare its rank lists from these three datasets with different $m/n$ ratios. For this comparison, we use the Kuncheva index \citep{kuncheva2007stability}, which quantifies the overlap of top $k$ feature sets and accounts for the overlap by chance.
\[ \text{Kuncheva index}(A_k, B_k) = \frac{|A_k \cap B_k| - k^2/d}{k - k^2/d}\,, \;k=1,\ldots,d\,,
\]
where $A_k$ and $B_k$ are the top $k$ features from two rank lists. The Kuncheva index has range $[-1,1]$, and its value is monotone increasing in $|A_k \cap B_k|$. If $A_k$ and $B_k$ overlap by chance, the Kuncheva index would be close to $0$. For more than two rank lists, their Kuncheva index for the top $k$ features is defined as the average of their pairwise Kuncheva indices for the top $k$ features. In our case, the larger the Kuncheva indices are for varying $k$, the more robust a criterion is to disproportional data. We illustrate the Kuncheva indices of s-CC, s-NPC, and the two RF feature importance measures in Figure~\ref{fig:consistency}, which shows that s-NPC is the most robust criterion.}  

\section{Conclusion and perspectives}\label{sec:conclusions}

This work introduces model-free objective-based marginal feature ranking criteria---s-CC and s-NPC---for the purpose of binary decision-making. The explicit use of a prediction objective to rank features is demonstrated to \rvv{select more predictive features than existing practices for marginal or multivariate feature ranking do. The reason is that commonly used marginal feature ranking criteria are association measures not reflecting the prediction objective or capturing certain data distributional characteristics, and that popular multivariate feature ranking criteria rely on multivariate classification algorithms such as SVM and RF, which are sensitive to sampling bias and feature correlations.} 

\rv{It is worth nothing that s-CC and s-NPC are extendable to multi-class classification. For s-NPC, as it is based on the Neyman-Pearson paradigm for binary classification, its extension to multi-class classification will rely on the one-vs-rest approach.  Concretely, we single out the most important class as the class $0$ and combine the rest of classes into the class $1$, converting the multi-class problem into a binary classification problem; then s-NPC can be applied.  For s-CC, with $K$ classes, the classical oracle classifier is known to be $\varphi^*(\cdot) = \arg\max_{k\in\{1, \ldots, K\}} \pi_k p_k(\cdot)$, where $\pi_k = \p(Y=k)$ and $p_k(\cdot)$ denotes the the class-$k$ conditional density of feature(s). Then we can define s-CC as the average classification error of plug-in classifiers of this oracle, similar to equations (7) and (8) in our manuscript. We have implemented the multiple-class extensions of s-CC and s-NPC in our R package \texttt{frc}.}   

In addition to the illustrated CC and NP paradigms, the same marginal ranking idea extends to other prediction objectives such as the cost-sensitive learning and global paradigms. Another extension direction is to rank feature pairs in the same model-free fashion.  In addition to the biomedical examples we show in this paper, model-free objective-based marginal feature ranking is also useful for finance applications, among others. For example, a loan company has successful business in region A and would like to establish new business in region B.  To build a loan-eligibility model for region B, which has a much smaller fraction of eligible applicants than region A, the company may use the top ranked features by s-NPC in region A, thanks to the robustness of s-NPC to sampling bias.   

Both s-CC and s-NPC involve sample splitting. The default option is a half-half split for both class $0$ and class $1$ observations. It remains an open question whether a refined splitting strategy may lead to a better ranking agreement between the sample-level and population-level criteria. Intuitively, there is a trade-off between classifier training and objective evaluation: using more data for training can result in a classifier closer to the oracle, while saving more data to evaluate the objective can lead to a less variable criterion.  

\section*{Acknowledgements}

This work was supported by the following grants: National Science Foundation (NSF) DMS-1613338 (to X.T. and J.J.L.), DMS-2113500 (to X.T.), and DMS-2113754 (to J.J.L). In addition, J.J.L. and Y.E.C. were supported by NSF DBI-1846216, NIH/NIGMS R01GM120507, Johnson \& Johnson WiSTEM2D Award, Sloan Research Fellowship, and UCLA David Geffen School of Medicine W.M. Keck Foundation Junior Faculty Award.

\appendix
\renewcommand\thefigure{\thesection.\arabic{figure}}    
\setcounter{figure}{0}    

\renewcommand\thetable{\thesection.\arabic{table}}    
\setcounter{table}{0}    

\renewcommand\thetheorem{\thesection.\arabic{theorem}}    
\setcounter{theorem}{0}    

\renewcommand\theequation{\thesection.\arabic{equation}}    
\setcounter{equation}{0}

\section{A condition where CC and NPC agree. }

\begin{lemma}\label{lem: toy example1}
\rv{Suppose that two features $\bd X_{\{1\}}$ and $\bd X_{\{2\}}$  have class-conditional densities
\begin{align*}
	\bd X_{\{1\}} | (Y=0) &\sim \mathcal{N}\left(\mu_1^0, (\sigma_1)^2\right)\,, & \bd X_{\{1\}} | (Y=1) &\sim \mathcal{N}\left(\mu_1^1, (\sigma_1)^2\right)\,,\\
	\bd X_{\{2\}} | (Y=0) &\sim \mathcal{N}\left(\mu_2^0, (\sigma_2)^2\right)\,, & \bd X_{\{2\}}| (Y=1) &\sim \mathcal{N}\left(\mu_2^1, (\sigma_2)^2\right)\,.
\end{align*}
That is, each feature has the same class-conditional variance under the two classes. For $\alpha\in(0,1)$\,, let $\varphi^*_{\alpha\{1\}}$ or $\varphi^*_{\alpha\{2\}}$ be the level-$\alpha$ NP oracle classifier using only the feature $\bd X_{\{1\}}$ or $\bd X_{\{2\}}$ respectively, and let $\varphi^*_{\{1\}}$ or $\varphi^*_{\{2\}}$ be the corresponding classical oracle classifier. Then
we have simultaneously for all $\alpha$, 
\begin{align*}
	\text{\rm{sign}}\left\{R_1\left(\varphi^*_{\alpha \{2\}}\right)  - R_1\left({\varphi}^*_{\alpha \{1\}}\right)\right\} = &\text{\rm{sign}}\left\{ R\left(\varphi^*_{\{2\}} \right) -R\left(\varphi^*_{\{1\}} \right) \right\} =   \text{\rm{sign}}\left\{\frac{|\mu_1^1 -\mu_1^0| }{\sigma_1} - \frac{|\mu_2^1 -\mu_2^0 | }{\sigma_2}\right\}\,,
\end{align*}
where $\rm{sign}(\cdot)$ is the sign function.}

\end{lemma}\par

\section{Partial consistency results of s-CC and s-NPC}\label{sec: partial consistency}

In this section, we present partial consistency results for s-CC (Theorem \ref{thm:partial__consistency_cc}) and s-NPC (Theorem \ref{thm:partial_consistency_plugin}).  The proofs are skipped due to the similarities to those of Theorem \ref{thm:selection_consistency_cc}  and Theorem \ref{thm:selection_consistency_plugin}.  

\begin{theorem}\label{thm:partial__consistency_cc}
\rv{Let $\delta_3$, $\delta_4$, $\delta_5\in (0,1)\,,$ $A_1, \ldots, A_J, A_{J+1}, \ldots, A_{J+H} \subseteq\left\{1,\ldots, d \right\}$ and $|A_1| = |A_2|=\ldots = |A_{J+H}| = l$. We consider $J$, $H$ and $l$ to be constants that do not diverge with the sample sizes. In addition to the assumptions in \rvv{Theorem} \ref{prop: CC1}, assume that  the \jjl{p-CC's} separate the first $J$ feature sets and the last $H$ by some margin $g>0$; in other words,  
$$
	 \min \limits_{i \in \{J+1,\dots, J+H\}} R\left( {\varphi}^*_{A_i}\right) - \max \limits_{i \in \{1,\dots, J\}} R\left( {\varphi}^*_{A_i}\right)    > g\,. 
$$ 
In addition,  assume $m_1, m_2, n_1, n_2$ satisfy that 
\begin{equation}\label{eqn:sample size requirement CC partial}
\widetilde C_1 \left[\left( \frac{\log m_1}{m_1}\right)^{\frac{\beta\bar\gamma}{2\beta+l}} + \left( \frac{\log n_1}{n_1}\right)^{\frac{\beta\bar\gamma}{2\beta+l}} + m_2^{-\frac{1}{4}} + n_2^{-\frac{1}{4}} \right]  < \frac{g}{2}\,, 
\end{equation}
where $\widetilde C_1$ is the generic constant in \rvv{Theorem} \ref{prop: CC1}.  
Then with probability at least $1 - (J+H)(\delta_3+\delta_4+\delta_5)$, $\max_{i\in\{1, \ldots, J\}}\mathrm{CC}_{A_i} <\min_{i\in\{J+1, \ldots, J+H\}} \mathrm{CC}_{A_{i}}$.} 
\end{theorem}

\begin{theorem}\label{thm:partial_consistency_plugin}
\rv{Let $\alpha$, $\delta_1$, $\delta_2$, $\delta_3$, $\delta_4 \in (0,1)\,,$ $A_1, \ldots, A_J, A_{J+1}, \ldots, A_{J+H} \subseteq\left\{1,\ldots, d \right\}$ and $|A_1| = |A_2|=\ldots = |A_{J+H}| = l$. We consider  $J$, $H$, $l$ to be constants that do not diverge with the sample sizes. In addition to the assumptions in Theorem \ref{thm:1}, assume that  the p-NPC's separate the first $J$ feature sets and the last $H$ by some margin $g>0$; in other words,  
$$
	 \min \limits_{i \in \{J+1,\dots, J+H\}} R_1\left( {\varphi}^*_{\alpha A_i}\right) - \max \limits_{i \in \{1,\dots, J\}} R_1\left( {\varphi}^*_{\alpha A_i}\right)    > g\,. 
$$ 
In addition,  assume $m_1, m_2, n_1, n_2$ satisfy that 
\begin{equation}\label{eqn:sample size requirement NPC partial}
\widetilde C_2 \left[\left( \frac{\log m_1}{m_1}\right)^{\frac{\beta(1+\bar\gamma)}{2\beta+l}} + \left( \frac{\log n_1}{n_1}\right)^{\frac{\beta(1+\bar\gamma)}{2\beta+l}} + m_2^{-(\frac{1}{4}\wedge \frac{1+\bar\gamma}{\underaccent{\bar}\gamma})} + n_2^{-\frac{1}{4}} \right]  < \frac{g}{2}\,, 
\end{equation}
where $\widetilde C_2$ is the generic constant in Theorem \ref{thm:1}.  
Then with probability at least $1 - (J+H)(\delta_3+\delta_4+\delta_5)$, $\max_{i\in\{1, \ldots, J\}}\mathrm{NPC}_{\alpha A_i} <\min_{i\in\{J+1, \ldots, J+H\}} \mathrm{NPC}_{\alpha A_{i}}$.} 
\end{theorem}

\section{Variants of s-CC and s-NPC based on SVM classifiers}\label{sec:s-CC_s-NPC_SVM_variants}

\rv{In parallel to s-CC and s-NPC, which are defined based on plug-in classifiers whose scoring function is the ratio of kernel density estimates, we implement variants of s-CC and s-NPC based on SVM classifiers. Specifically, we change the scoring function $\hat{s}^{(b)}_A (\cdot)$ from $\hat{p}_{1A}^{(b)}(\cdot)/ \hat{p}_{0A}^{(b)}(\cdot)$ to the predicted probability function $\p(Y=1 | \bd X_A = \cdot)$ of SVM, which we implement using the \texttt{R} function \texttt{svm()} in the \texttt{e1071} package. With the new scoring function, we construct the classical classifier $\hat\phi_A^{(b)}(\bd x) = \1\left( \hat{s}^{(b)}_A(\bd x_A) > 1/2\right)$ and define the s-CC variant, called s-CC-SVM, by following \eqref{CC}. Similarly, we construct the NP classifier $\hat{\phi}_{\alpha A}^{(b)}(\cdot) = \1 \left(\hat{s}^{(b)}_A (\cdot) >  \widehat{C}_{\alpha A}^{(b)} \right)$ in \eqref{eq:NP_classifier} using the NP umbrella algorithm; then we define the s-NPC variant, called s-NPC-SVM, by following \eqref{Npscore}.
}

\section{Proofs}

\subsection{Proof of Lemma \ref{lem: toy example1}}

First we realize that the following three statements are equivalent:
	\begin{enumerate}
		\item[(1)] Feature importance ranking under the NP paradigm is invariant to $\alpha\,;$
		\item[(2)] Feature importance ranking under the classical paradigm is invariant to $\pi_0\,;$ 
		\item[(3)] Feature importance ranking under the NP paradigm for $\forall \alpha \in (0,1)$ is the same as feature importance ranking under the classical paradigm $\forall \pi_0 \in (0,1)$.
	\end{enumerate}
We explore conditions for statement (1) to hold. We will divide our analysis into four scenarios (i)-(iv) regarding distribution means.  

\noindent\textbf{Scenario (i)}: suppose $\mu_1^0 \leq \mu_1^1$ and $\mu_2^0 \leq  \mu_2^1$. Let $c_1$, $c_2 \in \R$ be such that	\begin{align*}
		1- \alpha = \Phi\left( \frac{c_1 - \mu_1^0}{\sigma_1}\right)\,,\quad
		1- \alpha = \Phi\left( \frac{c_2 - \mu_2^0}{\sigma_2}\right)\,,
	\end{align*} 
	where $\Phi(\cdot)$ denotes the cumulative distribution function of $\mathcal N (0,1)$\,.
Then the NP oracle classifier using the feature $\bd X_{\{1\}}$ or $\bd X_{\{2\}}$ can be written as 
\begin{align*}
	{\varphi}^*_{\alpha \{1\}} (\bd X) = \1 \left(\bd X_{\{1\}} > c_1\right)\quad \text{or} \quad
	{\varphi}^*_{\alpha \{2\}} (\bd X) = \1\left(\bd X_{\{2\}} > c_2\right)\,.
\end{align*}
These oracle classifiers have type II errors 
 \begin{align*}
 	R_1\left({\varphi}^*_{\alpha \{1\}}\right)=\Phi \left(\frac{c_1- \mu_1^1}{\sigma_1} \right)\,,\quad 
 	R_1\left({\varphi}^*_{\alpha \{2\}}\right)=\Phi \left(\frac{c_2- \mu_2^1}{\sigma_2} \right)\,.
 \end{align*}
The chain of equivalence holds
	 \begin{align}
	  	 &R_1\left(\varphi^*_{\alpha \{2\}}\right)  \geq R_1\left({\varphi}^*_{\alpha \{1\}}\right)\notag \\ 
   \Leftrightarrow \quad & \Phi \left(\frac{c_2- \mu_2^1}{\sigma_2} \right)  \geq \Phi \left(\frac{c_1- \mu_1^1}{\sigma_1} \right)\notag \\
 \Leftrightarrow \quad & \frac{c_2- \mu_2^1}{\sigma_2}   \geq \frac{c_1- \mu_1^1}{\sigma_1}\notag  \\
  \Leftrightarrow \quad &  \frac{c_2- \mu_2^0}{\sigma_2}  + \frac{\mu_2^0 - \mu_2^1}{\sigma_2}   \geq \frac{c_1- \mu_1^0}{\sigma_1}  + \frac{\mu_1^0 - \mu_1^1}{\sigma_1} \notag\\
  \Leftrightarrow \quad & \Phi^{-1}(1-\alpha) + \frac{\mu_2^0 - \mu_2^1}{\sigma_2}   \geq  \Phi^{-1}(1-\alpha)  + \frac{\mu_1^0 - \mu_1^1}{\sigma_1} \notag\\
 \Leftrightarrow \quad &    \frac{\mu_1^1 -\mu_1^0 }{\sigma_1} - \frac{\mu_2^1 -\mu_2^0  }{\sigma_2} \geq 0\,, \notag
  \end{align}

Therefore, 

$$
\text{\rm{sign}}\left\{R_1\left(\varphi^*_{\alpha \{2\}}\right)  - R_1\left({\varphi}^*_{\alpha \{1\}}\right)\right\} = \text{\rm{sign}}\left\{\frac{\mu_1^1 -\mu_1^0 }{\sigma_1} - \frac{\mu_2^1 -\mu_2^0  }{\sigma_2}\right\}\,.
$$


\noindent\textbf{Scenario (ii)}: suppose $\mu_1^0 > \mu_1^1$ and $\mu_2^0 > \mu_2^1$. Let $c_1$, $c_2 \in \R$ be such that	
\begin{align*}
		\alpha = \Phi\left( \frac{c_1 - \mu_1^0}{\sigma_1}\right)\,,\quad
		\alpha = \Phi\left( \frac{c_2 - \mu_2^0}{\sigma_2}\right)\,,
	\end{align*} 
then the NP oracle classifier using the feature $\bd X_{\{1\}}$ or $\bd X_{\{2\}}$ can be written as 
\begin{align*}
	{\varphi}^*_{\alpha \{1\}} (\bd X) = \1 \left(\bd X_{\{1\}} < c_1\right) \quad \text{or} \quad	{\varphi}^*_{\alpha \{2\}} (\bd X) = \1\left(\bd X_{\{2\}} < c_2\right)\,.
\end{align*}
These oracle classifiers have type II errors 
 \begin{align*}
 	R_1\left({\varphi}^*_{\alpha \{1\}}\right)=1-\Phi \left(\frac{c_1- \mu_1^1}{\sigma_1} \right)\,,\quad 
 	R_1\left({\varphi}^*_{\alpha \{2\}}\right)=1-\Phi \left(\frac{c_2- \mu_2^1}{\sigma_2} \right)\,.
 \end{align*}
The chain of equivalence holds
	 \begin{align}
	  	 &R_1\left(\varphi^*_{\alpha \{2\}}\right)  \geq R_1\left({\varphi}^*_{\alpha \{1\}}\right)\notag \\ 
   \Leftrightarrow \quad & \Phi \left(\frac{c_2- \mu_2^1}{\sigma_2} \right)  \leq \Phi \left(\frac{c_1- \mu_1^1}{\sigma_1} \right)\notag \\
 \Leftrightarrow \quad & \frac{c_2- \mu_2^1}{\sigma_2}   \leq \frac{c_1- \mu_1^1}{\sigma_1}\notag  \\
  \Leftrightarrow \quad &  \frac{c_2- \mu_2^0}{\sigma_2}  + \frac{\mu_2^0 - \mu_2^1}{\sigma_2}   \leq \frac{c_1- \mu_1^0}{\sigma_1}  + \frac{\mu_1^0 - \mu_1^1}{\sigma_1} \notag\\
  \Leftrightarrow \quad & \Phi^{-1}(\alpha) + \frac{\mu_2^0 - \mu_2^1}{\sigma_2}   \leq  \Phi^{-1}(\alpha)  + \frac{\mu_1^0 - \mu_1^1}{\sigma_1} \notag\\
 \Leftrightarrow \quad &   \frac{\mu_1^1 -\mu_1^0 }{\sigma_1} - \frac{\mu_2^1 -\mu_2^0  }{\sigma_2} \leq 0\,. \notag
  \end{align}
Hence, 
$$
\text{\rm{sign}}\left\{R_1\left(\varphi^*_{\alpha \{2\}}\right)  - R_1\left({\varphi}^*_{\alpha \{1\}}\right)\right\} = \text{\rm{sign}}\left\{-\frac{\mu_1^1 -\mu_1^0 }{\sigma_1} + \frac{\mu_2^1 -\mu_2^0  }{\sigma_2}\right\}\,.
$$

\noindent\textbf{Scenario (iii)}: suppose $\mu_1^0 \leq \mu_1^1$ and $\mu_2^0 > \mu_2^1$. Let $c_1$, $c_2 \in \R$ be such that	
\begin{align*}
		1-\alpha = \Phi\left( \frac{c_1 - \mu_1^0}{\sigma_1}\right)\,,\quad
		\alpha = \Phi\left( \frac{c_2 - \mu_2^0}{\sigma_2}\right)\,,
	\end{align*} 
then the NP oracle classifier using the feature $\bd X_{\{1\}}$ or $\bd X_{\{2\}}$ can be written as 
\begin{align*}
	{\varphi}^*_{\alpha \{1\}} (\bd X) = \1 \left(\bd X_{\{1\}} > c_1\right)\quad \text{or} \quad
	{\varphi}^*_{\alpha \{2\}} (\bd X) = \1\left(\bd X_{\{2\}} < c_2\right)\,.
\end{align*}
These oracle classifiers have type II errors 
 \begin{align*}
 	R_1\left({\varphi}^*_{\alpha \{1\}}\right)=\Phi \left(\frac{c_1- \mu_1^1}{\sigma_1} \right)\,,\quad 
 	R_1\left({\varphi}^*_{\alpha \{2\}}\right)=1-\Phi \left(\frac{c_2- \mu_2^1}{\sigma_2} \right)\,.
 \end{align*}
Because $\Phi(a)+\Phi(-a) =1$ for any $a\in \R$ and $\Phi^{-1}(1-\alpha)= -\Phi^{-1}(\alpha)$ for all $\alpha \in (0,1)$, we have the chain of equivalence
	 \begin{align}
	  	 &R_1\left(\varphi^*_{\alpha \{2\}}\right)  \geq R_1\left({\varphi}^*_{\alpha \{1\}}\right)\notag \\ 
   \Leftrightarrow \quad & 1-\Phi \left(\frac{c_2- \mu_2^1}{\sigma_2} \right)  \geq \Phi \left(\frac{c_1- \mu_1^1}{\sigma_1} \right)\notag \\
 \Leftrightarrow \quad &  - \frac{c_2- \mu_2^1}{\sigma_2}   \geq \frac{c_1- \mu_1^1}{\sigma_1}\notag  \\
  \Leftrightarrow \quad &  -\frac{c_2- \mu_2^0}{\sigma_2}  -\frac{\mu_2^0 - \mu_2^1}{\sigma_2}   \geq \frac{c_1- \mu_1^0}{\sigma_1}  + \frac{\mu_1^0 - \mu_1^1}{\sigma_1} \notag\\
  \Leftrightarrow \quad & -\Phi^{-1}(\alpha) - \frac{\mu_2^0 - \mu_2^1}{\sigma_2}   \geq  \Phi^{-1}(1-\alpha)  + \frac{\mu_1^0 - \mu_1^1}{\sigma_1} \notag\\
 \Leftrightarrow \quad &    \frac{\mu_1^1 -\mu_1^0 }{\sigma_1} + \frac{\mu_2^1 -\mu_2^0  }{\sigma_2} \geq 0\,. \notag
  \end{align}
Hence, 
$$
\text{\rm{sign}}\left\{R_1\left(\varphi^*_{\alpha \{2\}}\right)  - R_1\left({\varphi}^*_{\alpha \{1\}}\right)\right\} = \text{\rm{sign}}\left\{\frac{\mu_1^1 -\mu_1^0 }{\sigma_1} + \frac{\mu_2^1 -\mu_2^0  }{\sigma_2}\right\}\,.
$$

\noindent\textbf{Scenario (iv)}: suppose $\mu_1^0 > \mu_1^1$ and $\mu_2^0 \leq \mu_2^1$. Let $c_1$, $c_2 \in \R$ be such that	
\begin{align*}
		\alpha = \Phi\left( \frac{c_1 - \mu_1^0}{\sigma_1}\right)\,,\quad
		1-\alpha = \Phi\left( \frac{c_2 - \mu_2^0}{\sigma_2}\right)\,,
	\end{align*} 
then the NP oracle classifier using the feature $\bd X_{\{1\}}$ or $\bd X_{\{2\}}$ can be written as 
\begin{align*}
	{\varphi}^*_{\alpha \{1\}} (\bd X) = \1 \left(\bd X_{\{1\}} < c_1\right)\quad \text{or} \quad
	{\varphi}^*_{\alpha \{2\}} (\bd X) = \1\left(\bd X_{\{2\}} > c_2\right)\,.
\end{align*}
These oracle classifiers have type II errors 
 \begin{align*}
 	R_1\left({\varphi}^*_{\alpha \{1\}}\right)=1-\Phi \left(\frac{c_1- \mu_1^1}{\sigma_1} \right)\,,\quad 
 	R_1\left({\varphi}^*_{\alpha \{2\}}\right)=\Phi \left(\frac{c_2- \mu_2^1}{\sigma_2} \right)\,.
 \end{align*}
Because $\Phi(a)+\Phi(-a) =1$ for any $a\in \R$ and $\Phi^{-1}(\alpha)= -\Phi^{-1}(1-\alpha)$ for all $\alpha \in (0,1)$, we have the chain of equivalence
	 \begin{align}
	  	 &R_1\left(\varphi^*_{\alpha \{2\}}\right)  \geq R_1\left({\varphi}^*_{\alpha \{1\}}\right)\notag \\ 
   \Leftrightarrow \quad & \Phi \left(\frac{c_2- \mu_2^1}{\sigma_2} \right)  \geq 1-\Phi \left(\frac{c_1- \mu_1^1}{\sigma_1} \right)\notag \\
 \Leftrightarrow \quad &   \frac{c_2- \mu_2^1}{\sigma_2}   \geq -\frac{c_1- \mu_1^1}{\sigma_1}\notag  \\
  \Leftrightarrow \quad &  \frac{c_2- \mu_2^0}{\sigma_2}   +\frac{\mu_2^0 - \mu_2^1}{\sigma_2}   \geq  - \frac{c_1- \mu_1^0}{\sigma_1}  - \frac{\mu_1^0 - \mu_1^1}{\sigma_1} \notag\\
  \Leftrightarrow \quad & \Phi^{-1}(1-\alpha) + \frac{\mu_2^0 - \mu_2^1}{\sigma_2}   \geq  -\Phi^{-1}(\alpha)  - \frac{\mu_1^0 - \mu_1^1}{\sigma_1} \notag\\
 \Leftrightarrow \quad &   - \frac{\mu_1^1 -\mu_1^0 }{\sigma_1} - \frac{\mu_2^1 -\mu_2^0  }{\sigma_2} \geq 0\,. \notag
  \end{align}
Hence, 
$$
\text{\rm{sign}}\left\{R_1\left(\varphi^*_{\alpha \{2\}}\right)  - R_1\left({\varphi}^*_{\alpha \{1\}}\right)\right\} = \text{\rm{sign}}\left\{-\frac{\mu_1^1 -\mu_1^0 }{\sigma_1} - \frac{\mu_2^1 -\mu_2^0  }{\sigma_2}\right\}\,.
$$

Finally to sum up scenarios (i)-(iv), we conclude that 
$$
\text{\rm{sign}}\left\{R_1\left(\varphi^*_{\alpha \{2\}}\right)  - R_1\left({\varphi}^*_{\alpha \{1\}}\right)\right\} = \text{\rm{sign}}\left\{\frac{|\mu_1^1 -\mu_1^0| }{\sigma_1} - \frac{|\mu_2^1 -\mu_2^0 | }{\sigma_2}\right\}\,.
$$

\bigskip
\subsection{Proof of Proposition \ref{lem:bound_s_shat_for_plugin}}


 	let $h_{m_1}=\left(\frac{\log m_1}{m_1}\right)^{\frac{1}{2\beta+l}}$. By Lemma 1 in \cite{tong2013plug}, there exists some constant $C_{0}$ that does not depend on $m_1$ and $\delta_3$,  such that with probability at least $1-\delta_3/2$, 
 	\[ \left\|\hat p_{0A}(\bd x_A) - p_{0A}(\bd x_A) \right\|_{\infty}\le \varepsilon_{0} \,,
 	\] where $\varepsilon_{0} = C_{0}\sqrt{\frac{\log(2m_1/\delta_3)}{m_1h_{m_1}^l}}$\,,
where $C_{0}=\sqrt{48c_{1}} + 32c_{2}+2Lc_{3}+L'+L+C\sum_{1\leq|\bd q|\leq\lfloor\beta\rfloor}\frac{1}{\bd q!}$, in which $c_{1}=\mu_{\max}\|K\|^2$, $c_{2}=\|K\|_{\infty}+\mu_{\max}+\int|K||\bd t|^{\beta}d\bd t$, $c_{3}=\int |K||\bd t|^{\beta}d \bd t$ and $C$ is such that $C \geq \sup_{1\leq|\bd q|\leq\lfloor \beta\rfloor}\sup_{\bd x_A\in[-1, 1]^l}|p^{(\bd q)}_{0A}(\bd x_A)|$. 

Similarly let $h_{n_1}=\left(\frac{\log n_1}{n_1}\right)^{\frac{1}{2\beta+l}}$, there exists some constant $C_{1}$ that does not depend on $n_1$ and $\delta_3$,  such that with probability at least $1-\delta_3/2$, 
 	\[ \left\|\hat p_{1A}(\bd x_A) - p_{1A}(\bd x_A) \right\|_{\infty}\le \varepsilon_1 \,,
 	\] where $\varepsilon_1= C_{1}\sqrt{\frac{\log(2n_1/\delta_3)}{n_1h_{n_1}^l}}$\,. Then, we have with probability at least $1-\delta_3$,
 	\begin{align*}
 		&{}\left\|\frac{ \hat p_{1A}(\bd x_A)}{\hat p_{0A}(\bd x_A) }- \frac{p_{1A}(\bd x_A)}{p_{0A}(\bd x_A)}\right\|_{\infty}\\
 		\le{}& \left\|\frac{ \hat p_{1A}(\bd x_A)}{\hat p_{0A}(\bd x_A) }- \frac{p_{1A}(\bd x_A)}{\hat p_{0A}(\bd x_A)}\right\|_{\infty}+\left\|\frac{  p_{1A}(\bd x_A)}{\hat p_{0A}(\bd x_A) }- \frac{p_{1A}(\bd x_A)}{ p_{0A}(\bd x_A)}\right\|_{\infty}\\
 		\le{}& \left\|\frac{1}{\hat p_{0A}(\bd x_A)}\right\|_{\infty} \left\|\hat p_{1A}(\bd x_A) - p_{1A}(\bd x_A) \right\|_{\infty} + \left\|\frac{p_{1A}}{p_{0A}}\right\|_{\infty}\left\|\frac{ p_{0A}}{\hat p_{0A}}-1\right\|_{\infty}\\
 		\le{}& \left\|\frac{1}{\hat p_{0A}(\bd x_A)}\right\|_{\infty} \left\|\hat p_{1A}(\bd x_A) - p_{1A}(\bd x_A) \right\|_{\infty} + \left\|\frac{p_{1A}}{p_{0A}}\right\|_{\infty}\left\|\frac{ p_{0A} - \hat p_{0A}}{\hat p_{0A}}\right\|_{\infty}\\
 		\le{}& \frac{\varepsilon_1 + \varepsilon_0 \mu_{\max}/\mu_{\rm min}}{\mu_{\rm min} - \varepsilon_0 }=: b_{m_1, n_1}\,.
 	\end{align*}
When $n_1 \wedge m_1 \geq 2/\delta_3\,,$ 
\[ 
\varepsilon_0 \leq \sqrt{2}{C}_0 \left( \frac{\log m_1}{m_1}\right)^{\beta/(2\beta+l)}\,, \quad\varepsilon_1 \leq \sqrt{2}{C}_1 \left( \frac{\log n_1}{n_1}\right)^{\beta/(2\beta+l)}\,.
\]
These combined with $\sqrt{\frac{\log\left(2m_1/\delta_3\right)}{m_1h_{m_1}^{l}}} < \frac{\mu_{\min}}{2 C_0} $ imply that

\begin{equation}\label{equ:bound on bm1n1}
  b_{m_1,n_1} \leq \widetilde{C}\left[\left( \frac{\log m_1}{m_1}\right)^{\beta/(2\beta+l)} + \left( \frac{\log n_1}{n_1}\right)^{\beta/(2\beta+l)} \right]\,,
\end{equation} for some positive constant $\widetilde{C}$ that does not depend on the subset $A$.

%
%
%
%
%

\subsection{Proof of \rvv{Theorem} \ref{prop: CC1}}

We can bound $\left|\text{CC}_{A} - R \left( {\varphi}^*_{A} \right)\right|$ as:
 \begin{align}\label{eqn:cc decomposition}
 	{} & \left|\mathrm{CC}_{A} - R \left( {\varphi}^*_{A} \right)\right|
 	\le {}  \left| \rm{CC}_{A} - R \left( \hat{\phi}_{A} \right)\right| + \left|R \left( \hat{\phi}_{A} \right) - R \left( {\varphi}^*_{A} \right)\right|\,.
 \end{align}
The second part of the right hand side of \eqref{eqn:cc decomposition} can be bounded as 
$$
\left|R \left( \hat{\phi}_{A} \right) - R \left( {\varphi}^*_{A} \right)\right|\leq \pi_0  \left|R_0 \left( \hat{\phi}_{A} \right) - R_0 \left( {\varphi}^*_{A} \right)\right| + \pi_1 \left|R_1 \left( \hat{\phi}_{A} \right) - R_1 \left( {\varphi}^*_{A} \right)\right|\,.
$$
Recall that $\hat \phi_A(\bd x) = \1(\hat s_A(\bd x_A) > m_1 /n_2) =  \1(\hat p_{1A}(\bd x_A)/\hat p_{0A}(\bd x_A)> m_1/n_1)$ and $\varphi^*_A(\bd x) = \1(s_A(\bd x_A) > \pi_0 / \pi_1) =  \1(p_{1A}(\bd x_A)/p_{0A}(\bd x_A)> \pi_0/\pi_1)$. Define an event $\mathcal{E}$ by 
$$\mathcal{E} = \left\{\left\| \hat p_{1A}(\bd x_A)/\hat p_{0A}(\bd x_A) - p_{1A}(\bd x_A)/p_{0A}(\bd x_A)\right\|_{\infty} \leq \widetilde{C}\left[\left( \log m_1 / m_1\right)^{\beta/(2\beta+l)} + \left( \log n_1/ n_1\right)^{\beta/(2\beta+l)} \right]\right\}\,,$$
where $\widetilde C$ is a universal constant as in Proposition \ref{lem:bound_s_shat_for_plugin}.  Proposition \ref{lem:bound_s_shat_for_plugin} guarantees that $\p(\mathcal{E})\geq 1- \delta_3$.  Denote $f(m_1, n_1) = \widetilde{C}\left[\left( \log m_1 / m_1\right)^{\beta/(2\beta+l)} + \left( \log n_1/ n_1\right)^{\beta/(2\beta+l)} \right]$. When $m_1/n_1 = \pi_0 / \pi_1$, on the event  $\mathcal{E}$, we have

\begin{align*}
R_0 \left( \hat{\phi}_{A} \right) - R_0 \left( {\varphi}^*_{A} \right) &= P_{0A}\left(\frac{\hat p_{1A}(\bd X_A)}{\hat p_{0A}(\bd X_A)} > \frac{\pi_0}{\pi_1}\right) - P_{0A}\left(\frac{p_{1A}(\bd X_A)}{p_{0A}(\bd X_A)} > \frac{\pi_0}{\pi_1}\right)\\
&\leq  P_{0A} \left( \frac{\pi_0}{\pi_1}  \geq \frac{p_{1A}(\bd X_A)}{p_{0A}(\bd X_A)} >  \frac{\pi_0}{\pi_1} -  f(m_1, n_1) \right)\\
&\leq \bar C \left(f(m_1, n_1)\right)^{\bar{\gamma}}\,,
\end{align*}
where the last inequality follows from the margin assumption around $\pi_0 / \pi_1$.  Similarly, 
\begin{align*}
R_0 \left( {\varphi}^*_{A} \right) - R_0 \left( \hat{\phi}_{A} \right)  &= P_{0A}\left(\frac{ p_{1A}(\bd X_A)}{ p_{0A}(\bd X_A)} > \frac{\pi_0}{\pi_1}\right) - P_{0A}\left(\frac{\hat p_{1A}(\bd X_A)}{\hat p_{0A}(\bd X_A)} > \frac{\pi_0}{\pi_1}\right)\\
&\leq  P_{0A} \left( \frac{\pi_0}{\pi_1}  < \frac{p_{1A}(\bd X_A)}{p_{0A}(\bd X_A)} \leq   \frac{\pi_0}{\pi_1} +  f(m_1, n_1) \right)\\
&\leq \bar C \left(f(m_1, n_1)\right)^{\bar{\gamma}}\,.
\end{align*}

Therefore, it can be concluded that  
\begin{align*}
\left|R_0 \left( \hat{\phi}_{A} \right) - R_0 \left( {\varphi}^*_{A} \right)\right| &\leq \bar C \left(f(m_1, n_1)\right)^{\bar{\gamma}}\,,
\end{align*}
Similarly, also on the event $\mathcal{E}$, we have $\left|R_1 \left( \hat{\phi}_{A} \right) - R_1 \left( {\varphi}^*_{A} \right)\right|\leq \bar C \left(f(m_1, n_1)\right)^{\bar{\gamma}}$. Therefore, 
$$
\left|R \left( \hat{\phi}_{A} \right) - R \left( {\varphi}^*_{A} \right)\right|\leq (\pi_0+\pi_1) \bar C \left(f(m_1, n_1)\right)^{\bar{\gamma}} = \bar C \left(f(m_1, n_1)\right)^{\bar{\gamma}}\,.
$$

The first part of the right hand side of \eqref{eqn:cc decomposition} can be bounded as 
\begin{align*}
	\left| \mathrm{CC}_{A} - R \left( \hat{\phi}_{A} \right)\right| &= \left|\frac{1}{m_2+n_2}\left\{ \sum_{i=n_1+1}^{n_1+n_2} \left[ 1-\hat{\phi}_{A}\left(\bd X_i^{1}\right) \right]  + \sum_{i'=m_1+1}^{m_1+m_2} \hat{\phi}_A\left(\bd X_{i'}^{0}\right) \right\} - R(\hat \phi_A)\right|\\
	&\leq \pi_1 \left|\frac{1}{n_2}\sum_{i=n_1+1}^{n_1+n_2} \left[ 1-\hat{\phi}_{A}\left(\bd X_i^{1}\right) \right] - R_1(\hat \phi_A)\right|   + \pi_0 \left| \frac{1}{m_2}\sum_{i'=m_1+1}^{m_1+m_2} \hat{\phi}_A\left(\bd X_{i'}^{0}\right) - R_0(\hat \phi_A)\right|\,.
\end{align*}

Let $D > 0$, then conditioning on $\hat{s}_A(\cdot)$, by Hoeffding's inequality, we have
\begin{align*}
& \p\left( \left|\frac{1}{n_2} \sum_{i=n_1 + 1}^{n_1 + n_2} \1\left(\hat{s}_A(\bd {X}_{iA}^1)< \frac{\pi_0}{\pi_1} \right)  - \E\left[\1\left(\hat{s}_A(\bd {X}_A^1) < \frac{\pi_0}{\pi_1}\right) \right]\right| > D \given[\Big] \hat{s}_A(\cdot)\, \right) 
  \\ \leq  & 2e^{ -2n_2D^2} \,. 
\end{align*}
This implies the following unconditional result, $$\p\left( \left|\frac{1}{n_2} \sum_{i=n_1+ 1}^{n_1 + n_2} \1\left(\hat{s}_A(\bd {X}_{iA}^1)< \frac{\pi_0}{\pi_1}\right)  - \E\left[\1\left(\hat{s}_A(\bd {X}_A^1) < \frac{\pi_0}{\pi_1}\right) \right]\right| \leq D \right) \geq 1-  2e^{ -2n_2D^2}\,.$$
Let $ 2e^{ -2n_2D^2} = \delta_4$ and then $D = \sqrt{\frac{1}{2n_2} \log \frac{2}{\delta_4}}$. So we have with probability at least $1-\delta_4\,,$
$$
\left|\frac{1}{n_2}\sum_{i=n_1+1}^{n_1+n_2} \left[ 1-\hat{\phi}_{A}\left(\bd X_i^{1}\right) \right] - R_1(\hat \phi_A)\right|\leq \sqrt{\frac{1}{2n_2} \log \frac{2}{\delta_4}}\,.
$$

When $n_2 \geq (\log\frac{2}{\delta_4})^2$, $\left|\frac{1}{n_2}\sum_{i=n_1+1}^{n_1+n_2} \left[ 1-\hat{\phi}_{A}\left(\bd X_i^{1}\right) \right] - R_1(\hat \phi_A)\right|\leq \frac{1}{\sqrt{2}} n_2^{-\frac{1}{4}}$.

Similarly, we have with probability $1-\delta_5$, 
$$
\left| \frac{1}{m_2}\sum_{i'=m_1+1}^{m_1+m_2} \hat{\phi}_A\left(\bd X_{i'}^{0}\right) - R_0(\hat \phi_A)\right|\leq \sqrt{\frac{1}{2m_2} \log \frac{2}{\delta_5}}\,.
$$
When $m_2 \geq (\log\frac{2}{\delta_5})^2$, $\left| \frac{1}{m_2}\sum_{i'=m_1+1}^{m_1+m_2} \hat{\phi}_A\left(\bd X_{i'}^{0}\right) - R_0(\hat \phi_A)\right|\leq\frac{1}{\sqrt{2}} m_2^{-\frac{1}{4}}$.

Therefore, with probability at least $1- \delta_3 - \delta_4 - \delta_5$, 
$$
\left|\mathrm{CC}_{A} - R \left( {\varphi}^*_{A} \right)\right| \leq \bar C \left(f(m_1, n_1)\right)^{\bar\gamma} + \pi_0 \cdot \frac{1}{\sqrt{2}} m_2^{-\frac{1}{4}} + \pi_1\cdot  \frac{1}{\sqrt{2}} n_2^{-\frac{1}{4}}\,.
$$
Then we have 
$$
\left| \mathrm{CC}_{A} - R \left( {\varphi}^*_{A} \right)\right|\leq \widetilde C_1 \left[\left( \frac{\log m_1}{m_1}\right)^{\frac{\beta\bar\gamma}{2\beta+l}} + \left( \frac{\log n_1}{n_1}\right)^{\frac{\beta\bar\gamma}{2\beta+l}} + m_2^{-\frac{1}{4}} + n_2^{-\frac{1}{4}} \right]\,,
$$	 
for some positive constant $\widetilde C_1$ that does not depend on $A$. 

\subsection{Proof of Theorem \ref{thm:selection_consistency_cc}}

\rvv{Theorem} \ref{prop: CC1} and the sample size condition imply that for each $i\in\{1, \ldots, J\}$, we have with probability at least $1-(\delta_3+\delta_4+\delta_5)$, 
\begin{align*}
 \left|\mathrm{CC}_{A_i} - R(\varphi^*_{A_i})\right|\leq  \frac{g}{2}\,. 
\end{align*}
Also considering the separation conditions that 

$$
	 \min \limits_{j \in \{1,\dots, J-1\}}\left\{ R\left( {\varphi}^*_{A_{j+1}}\right) - R\left( {\varphi}^*_{A_j}\right) \right\} > g\,, 
$$ 
we have the conclusion with probability at least $1-J(\delta_3+\delta_4 + \delta_5)$.

\subsection{Proof of Lemma \ref{lem:kprime}}
	
Given any feature set $A$\,, let $\{{T}_{i A} := \hat s_{A} (\bd X^0_{iA}), \bd X^0_{i} \in \mathcal{S}_{\rm lo}^0\}$ be the scores by applying the scoring function $\hat s_A(\cdot) = \hat p_{0A}(\cdot)/\hat p_{1A}(\cdot)$ to  $\mathcal{S}^0_{\rm lo }$. Sort $\{{T}_{i A}\}$ in an increasing order such that ${T}_{(1)A}\leq, \dots, {T}_{(m_2)A}\,.$ Let $\widehat C^\prime_{\alpha A} = {T}_{(k^\prime)A}$ be a score threshold using $k^\prime$-th order statistic, where $k^\prime = \ceil*{(m_2+1)d_{\alpha,\delta_1, m_2}}$, in which
\[
d_{\alpha,\delta_1,m_2} = \frac{1+ 2\delta_1 (m_2+2) (1-\alpha) + \sqrt{1+ 4\delta_1(m_2+2)(1-\alpha)\alpha}}{2\left\{ \delta_1(m_2+2)+1\right\}}\,,
\]
and $\ceil*{z}$ denotes the smallest integer larger than or equal to $z$. Denote the corresponding NP classifier as 
\[
\hat{\phi}^\prime_{\alpha A}(\bd X)=\1\left(\hat s_{A}(\bd X_A) > \widehat C^\prime_{\alpha A}\right).
	\]
Because we use kernel density estimates and the kernels are $\beta$-valid, the scoring function $\hat s_{A}(\cdot)$ is continuous. Therefore, by Proposition $1$ in \cite{tong2016neyman}, we have 
\begin{align*}
	\p\left(R_0 \left( \hat \phi^\prime_{\alpha A}\right) > \alpha \right) &=  \sum_{j=k'}^{m_2}\binom{m_2}{j} (1-\alpha)^j {\alpha}^{m_2-j}\,,\\
	\p\left(R_0 \left( \hat \phi_{\alpha A}\right) > \alpha \right) &= \sum_{j=k^*}^{m_2}\binom{m_2}{j} (1-\alpha)^j {\alpha}^{m_2-j}\,.
\end{align*}

Note that by the definition of $k^*$, 
\[k^*=\min\left\{k: \sum_{j=k}^{m_2}\binom{m_2}{j} (1-\alpha)^j {\alpha}^{m_2-j}\leq \delta_1 \right\}\,.
\] 
Proposition 2.2 in \cite{zhao2016neyman} implies  
$\p\left(R_0 \left( \hat \phi^\prime_{\alpha A}\right) > \alpha \right)\leq \delta_1$. 
So we also have $\sum_{j=k'}^{m_2}\binom{m_2}{j} (1-\alpha)^j {\alpha}^{m_2-j}\leq \delta_1$.  This together with the definition of $k^*$ implies that $k'\geq k^*$,  and therefore $ R_0(\hat{\phi}_{\alpha A}) \geq R_0(\hat{\phi}^\prime_{\alpha A})$.

By Lemma 2.1 in \cite{zhao2016neyman}, for any $\delta_2 \in (0,1)\,,$ if $m_2 \geq \frac{4}{\alpha\delta_1}\, $,
\[\p \left( \left| R_0 \left( \hat{\phi}^\prime_{\alpha A}\right) - R_0(\varphi^*_{\alpha A})\right|> \xi \right) \leq \delta_2\,, 
	\]
where $\xi$ is defined by \[
	\xi = \sqrt{\frac{\ceil*{ d_{\alpha,\delta_1,m_2} \left(m_2+1\right)}\left(m_2+1-\ceil*{ d_{\alpha,\delta_1,m_2} \left(m_2+1\right)}\right)}{(m_2+2)(m_2+1)^2\,\delta_2}} + d_{\alpha,\delta_1,m_2}  + \frac{1}{m_2+1} - (1-\alpha)\,.
\]

Let $\mathcal{E}_1 := \left\{ R_0 \left( \hat{\phi}_{\alpha A}\right) \leq \alpha\right\}$ and $\mathcal{E}_2 := \left\{ \left| R_0 \left( \hat{\phi}^\prime_{\alpha A}\right) - R_0(\varphi^*_{\alpha A})\right| \leq \xi\right\}$. On the event $\mathcal{E}_1 \cap \mathcal{E}_2$, which has probability at least $1-\delta_1 - \delta_2$, we have 
\[ \alpha = R_0(\varphi^*_{\alpha A})  \geq R_0(\hat{\phi}_{\alpha A}) \geq R_0(\hat{\phi}^\prime_{\alpha A}) \geq R_0(\varphi^*_{\alpha A}) - \xi\,, 
	\]
	which implies$$ \left|R_0(\hat{\phi}_{\alpha A})  - R_0(\varphi^*_{\alpha A}) \right|\leq \xi\,.$$ 
If $m_2 \geq \max(\delta_1^{-2}, \delta_2^{-2})$, we have 
$
\xi \leq  ({5}/{2}){m_2^{-1/4}},
$	
also by Lemma 2.1 of \cite{zhao2016neyman}.

\subsection{Proof of Theorem \ref{thm:1}}
We can bound $\left|\text{NPC}_{\alpha A} - R_1 \left( {\varphi}^*_{\alpha A} \right)\right|$ as follows:
 \begin{align*}
 	{} & \left|\mathrm{NPC}_{\alpha A} - R_1 \left( {\varphi}^*_{\alpha A} \right)\right|
 	\le {}  \left| \mathrm{NPC}_{\alpha A} - R_1 \left( \hat{\phi}_{\alpha A} \right)\right| + \left|R_1 \left( \hat{\phi}_{\alpha A} \right) - R_1 \left( {\varphi}^*_{\alpha A} \right)\right|\,.
 \end{align*}
First we derive a bound for $|\text{NPC}_{\alpha A} - R_1(\hat \phi_{\alpha A})|$.
 Let $D > 0$, then conditioning on $\hat{s}_A(\cdot)$ and $\widehat{C}_{\alpha A}$, by Hoeffding's inequality, we have
\begin{align*}
& \p\left( \left|\frac{1}{n_2} \sum_{i=n_1 + 1}^{n_1 + n_2} \1\left(\hat{s}_A(\bd {X}_{iA}^1)< \widehat{C}_{\alpha A}\right)  - \E\left[\1\left(\hat{s}_A(\bd {X}_A^1) < \widehat{C}_{\alpha A}\right) \right]\right| > D \given[\Big] \hat{s}_A(\cdot)\,,\widehat{C}_{\alpha A}\, \right) 
  \\ \leq  & 2e^{ -2n_2D^2} \,. \notag\label{equ:E3}
\end{align*}
This implies the following unconditional result, $$\p\left( \left|\frac{1}{n_2} \sum_{i=n_1+ 1}^{n_1 + n_2} \1\left(\hat{s}_A(\bd {X}_{iA}^1)< \widehat{C}_{\alpha A}\right)  - \E\left[\1\left(\hat{s}_A(\bd {X}_A^1) < \widehat{C}_{\alpha A}\right) \right]\right| \leq D \right) \geq 1-  2e^{ -2n_2D^2}\,.$$
Let $ 2e^{ -2n_2D^2} = \delta_4$ and then $D = \sqrt{\frac{1}{2n_2} \log \frac{2}{\delta_4}}$. So we have with probability at least $1-\delta_4\,,$
\begin{equation*}
	 \left| \mathrm{NPC}_{\alpha A} - R_1 \left( \hat{\phi}_{\alpha A} \right)\right| \leq  \sqrt{\frac{1}{2n_2} \log \frac{2}{\delta_4}}\,.
\end{equation*}	 
When $n_2 \geq (\log\frac{2}{\delta_4})^2$, $\left| \mathrm{NPC}_{\alpha A} - R_1 \left( \hat{\phi}_{\alpha A} \right)\right|\leq \frac{1}{\sqrt{2}} n_2^{-\frac{1}{4}}$. 

Propositions \ref{lem:bound_s_shat_for_plugin} and \ref{prop:2} imply that, it holds with probability at least $1-\delta_1 - \delta_2 - \delta_3$, 
\begin{align*}
&\left|R_1 \left( \hat{\phi}_{\alpha A} \right) - R_1 \left( {\varphi}^*_{\alpha A} \right)\right|\\
\leq& 2\bar C \left[\left(\frac{2}{5}m_2^{1/4}\underaccent{\bar}C\right)^{-1/\uderbar{\gamma}} + \widetilde{C}\left[\left( \frac{\log m_1}{m_1}\right)^{\beta/(2\beta+l)} + \left( \frac{\log n_1}{n_1}\right)^{\beta/(2\beta+l)} \right] \right]^{1 + \bar\gamma} 
+ C^*_{\alpha A} \left(\frac{2}{5} m_2^{1/4}\right)^{-1}\\
\leq& \widetilde C \left[\left( \frac{\log m_1}{m_1}\right)^{\frac{\beta(1+\bar\gamma)}{2\beta+l}} + \left( \frac{\log n_1}{n_1}\right)^{\frac{\beta(1+\bar\gamma)}{2\beta+l}} + m_2^{-(\frac{1}{4}\wedge \frac{1+\bar\gamma}{\underaccent{\bar}\gamma})}  \right]\,.
\end{align*}
for some generic constant $\widetilde C$.  The thresholds $C^*_{\alpha A}$'s are bounded from above by a single constant that does not depend on $A$; indeed, we can just take the upper bound to be $\mu_{\max} / \mu_{\min}$. Therefore, we have with probability at least $1-\delta_1 - \delta_2 - \delta_3 - \delta_4$, 
$$
\left| \mathrm{NPC}_{\alpha A} - R_1 \left( \hat{\phi}_{\alpha A} \right)\right|\leq \widetilde C_2 \left[\left( \frac{\log m_1}{m_1}\right)^{\frac{\beta(1+\bar\gamma)}{2\beta+l}} + \left( \frac{\log n_1}{n_1}\right)^{\frac{\beta(1+\bar\gamma)}{2\beta+l}} + m_2^{-(\frac{1}{4}\wedge \frac{1+\bar\gamma}{\underaccent{\bar}\gamma})} + n_2^{-\frac{1}{4}} \right]\,,
$$	 
for some generic constant $\widetilde C_2$ that does not depend on $A$.

%
%
%
%

\subsection{Proof of Theorem \ref{thm:selection_consistency_plugin}}

Theorem \ref{thm:1} and the sample size condition imply that for each $i\in\{1, \ldots, J\}$, we have with probability at least $1-(\delta_1+\delta_2+\delta_3+\delta_4)$, 
\begin{align*}
 \left|\mathrm{NPC}_{\alpha A_i} - R_1(\varphi^*_{\alpha A_i})\right|\leq  \frac{g}{2}\,. 
\end{align*}
Also considering the separation conditions that 

$$
	 \min \limits_{j \in \{1,\dots, J-1\}}\left\{ R_1\left( {\varphi}^*_{\alpha A_{j+1}}\right) - R_1\left( {\varphi}^*_{\alpha A_j}\right) \right\} > g\,, 
$$ 
we have the conclusion with probability at least $1-J(\delta_1+\delta_2+\delta_3+\delta_4)$.  


\section{Time complexity of s-CC and s-NPC}\label{sec::time_complexity}

\rv{The time complexity discussion is for one feature and $B=1$.} 

\rv{For s-CC, its time complexity depends on the univariate kernel density estimation (KDE) and evaluation. In s-CC, we apply KDE to the $m_1$ observations in class $0$ and the $n_1$ observations in class $1$ to obtain the density estimates of the two class-conditional distributions; then we evaluate the two density estimates on the $m_2$ observations in class $0$ and the $n_2$ observations in class $1$ to obtain $m_2+n_2$ density ratio estimates, which, together with the threshold $m_1/n_1$, gives the s-CC (Equation (8) in manuscript). Without approximation, the time complexity of s-CC is $O\left((m_1+n_1)\cdot (m_2+n_2)\right)$, which is what we have for the numerical examples in our manuscript. For datasets with large sample sizes, we may use approximate methods for KDE, such as  \cite{Raykar.Duraiswami.Zhao.2010}, which can reduce the time complexity to $O(m+n)$. (Note that $m=m_1+m_2$ and $n=n_1+n_2$.)}

\rv{For s-NPC, in addition to the time complexity of KDE and evaluation, additional complexity arises from the threshold searching step in the NP umbrella algorithm \citep{tong2016neyman}, which involves sorting the $m_2$ density ratio estimates of the left-out class $0$ observations and has time complexity $O(m_2 \log m_2)$.}

\rv{In conclusion, when the training sample size is large, we can use approximate KDE methods to reduce the time complexity of s-CC to $O(N)$, where $N = m+n$ is the total sample size. For s-NPC, we can set an upper bound on $m_2$ so that the time complexity is also $O(N)$.
}

\section{Relationships between numerical high-probability error bounds and sample sizes}\label{sec::bounds_vs_n}

\begin{figure}
\includegraphics[width=\textwidth]{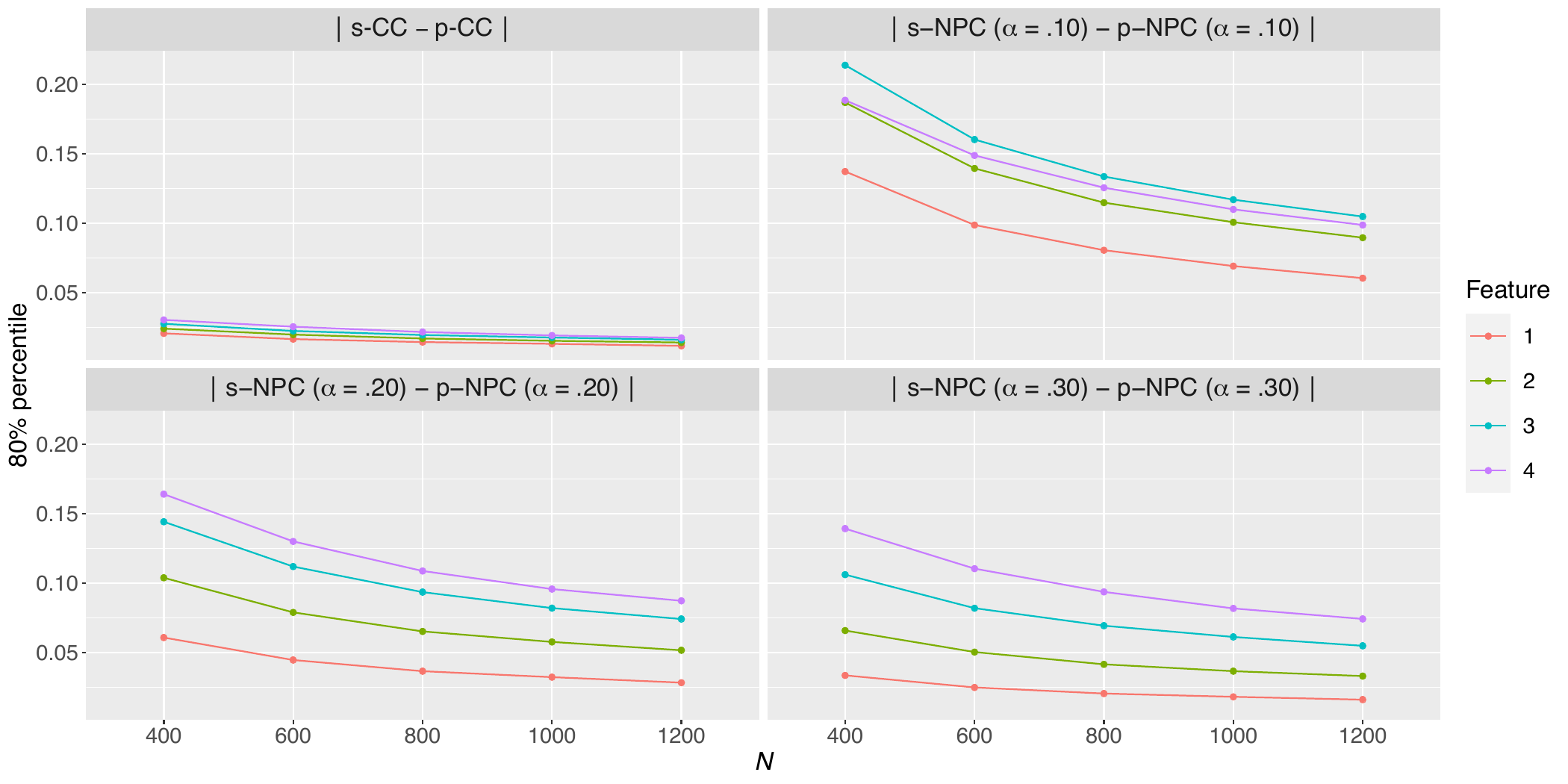}
\caption{80\% percentile (from $5000$ independent simulations) of absolute differences between sample-level criteria and their population counterparts for four features \eqref{eq:bounds_vs_n} with varying sample size $N$. \rvv{The decay rate is close to $O(N^{-1/2})$.}\label{fig:bounds_vs_n}}	
\end{figure}

\rv{In practice, we cannot realize the theoretical high-probability bounds on the differences between s-CC (or s-NPC) and p-CC (or p-NPC) due to unspecified constants (\rvv{Theorem}~\ref{prop: CC1} and Theorem~\ref{thm:1}). To provide guidance for practitioners, here we investigate the relationships between the numerical, realized high-probability error bounds and sample sizes. We consider the following four features $\bd X_{\{1\}}, \bd X_{\{2\}}, \bd X_{\{3\}}, \bd X_{\{4\}} \in \R$,\footnote{Usually, we denote the two scalar-valued features by $X_1$ and $X_2$, but here we use $\bd X_{\{1\}}$ and $\bd X_{\{2\}}$ to be consistent with the notation $\bd X_{A}$.} whose class-conditional distributions are \jjl{the following Gaussians}: 
\begin{align}\label{eq:bounds_vs_n}
	\bd X_{\{1\}} \given (Y=0) &\sim \mathcal{N}(0, 1)\,, & \bd X_{\{1\}}\given (Y=1) &\sim \mathcal{N}(2.5, 1)\,,\\
	\bd X_{\{2\}} \given (Y=0) &\sim \mathcal{N}(0, 1)\,, & \bd X_{\{2\}} \given (Y=1) &\sim \mathcal{N}(2, 1)\,, \notag \\ 
	\bd X_{\{3\}} \given (Y=0) &\sim \mathcal{N}(0, 1)\,, & \bd X_{\{3\}} \given (Y=1) &\sim \mathcal{N}(1.5, 1)\,, \notag \\ 
	\bd X_{\{4\}} \given (Y=0) &\sim \mathcal{N}(0, 1)\,, & \bd X_{\{4\}} \given (Y=1) &\sim \mathcal{N}(1, 1)\,, \notag
\end{align}
and the class priors are equal, i.e., $\pi_0 = \pi_1 =  .5$. 
It can be calculated that the four features have the following p-CC and p-NPC values.
\begin{center}
\begin{tabular}{rrrrr}
  \hline
Feature & 1 & 2 & 3 & 4 \\ 
  \hline
p-CC & 0.106 & 0.159 & 0.227 & 0.309 \\ 
  p-NPC ($\alpha = .10$) & 0.112 & 0.236 & 0.414 & 0.611 \\ 
  p-NPC ($\alpha = .20$) & 0.049 & 0.123 & 0.255 & 0.437 \\ 
  p-NPC ($\alpha = .30$) & 0.024 & 0.070 & 0.165 & 0.317 \\ 
   \hline
\end{tabular}
\end{center}
We design a simulation study with five sample sizes $N=400$, $600$, $800$, $1000$, and $1200$, i.e., the total number of observations from both classes. For each sample size, we simulate $5000$ independent training datasets, and we calculate s-CC and s-NPC with $\alpha = .10$, $.20$, and $.30$ for each feature on each dataset. Figure~\ref{fig:bounds_vs_n} shows the trends of $80\%$ percentiles (i.e., 80\%-probability upper bounds) of absolute differences between sample-level criteria and their population counterparts for each feature and each $N$. As expected, all the upper bounds decrease as $N$ increases. Moreover, to reach the same high-probability error bound, s-CC requires a smaller sample size than s-NPC does, and s-NPC with a larger $\alpha$ requires a smaller sample size than s-NPC with a smaller $\alpha$ does. This observation is consistent with our numerical results in Section~\ref{sec:simulation}. In addition, we observe that the bounds are overall better for the stronger features, a desirable phenomenon as the stronger features are usually of more interests.}

\section{Neyman-Pearson Lemma}\label{sec::np lamma}

\begin{lemma}[Neyman-Pearson Lemma \citep{neyman1933problem}]\label{lem: neyman_pearson}
	Let $P_0$ and $P_1$ be probability distributions possessing densities $p_0$ and $p_1$ respectively. Let $P$ be the probability distribution of a random feature vector $\bd X \in \mathcal X \subseteq \R^d$. The null and alternative hypotheses are $H_0: P = P_0$ and $H_1: P = P_1$. Let $s^*(\cdot) = p_1(\cdot)/p_0(\cdot)$\,. For a given level $\alpha \in (0, 1)$, let $C_\alpha^* \in \R$ be such that
	\begin{align*}\label{eq:np_lemma_condition}
		P_0\left(s^*(\bd X)> C_{\alpha}^*\right) \leq \alpha \text{\quad and \quad}  P_0\left(s^*(\bd X)\ge C_{\alpha}^*\right) \geq \alpha\,.
	\end{align*}
When $P_0\left(s^*(\bd X) =  C_{\alpha}^*\right) = 0$, the most powerful test of level $\alpha$ is 
 	\begin{equation*}
 	\varphi_{\alpha}^*(\bd x) = \1\left(s^*(\bd x) > C_\alpha^*\right)\,. \end{equation*}
 \end{lemma}

\section{Additional tables and figures}\label{sec:add_tabs_figs}

\clearpage
\begin{figure}
    \centering
    \makebox{\includegraphics[width = \textwidth]{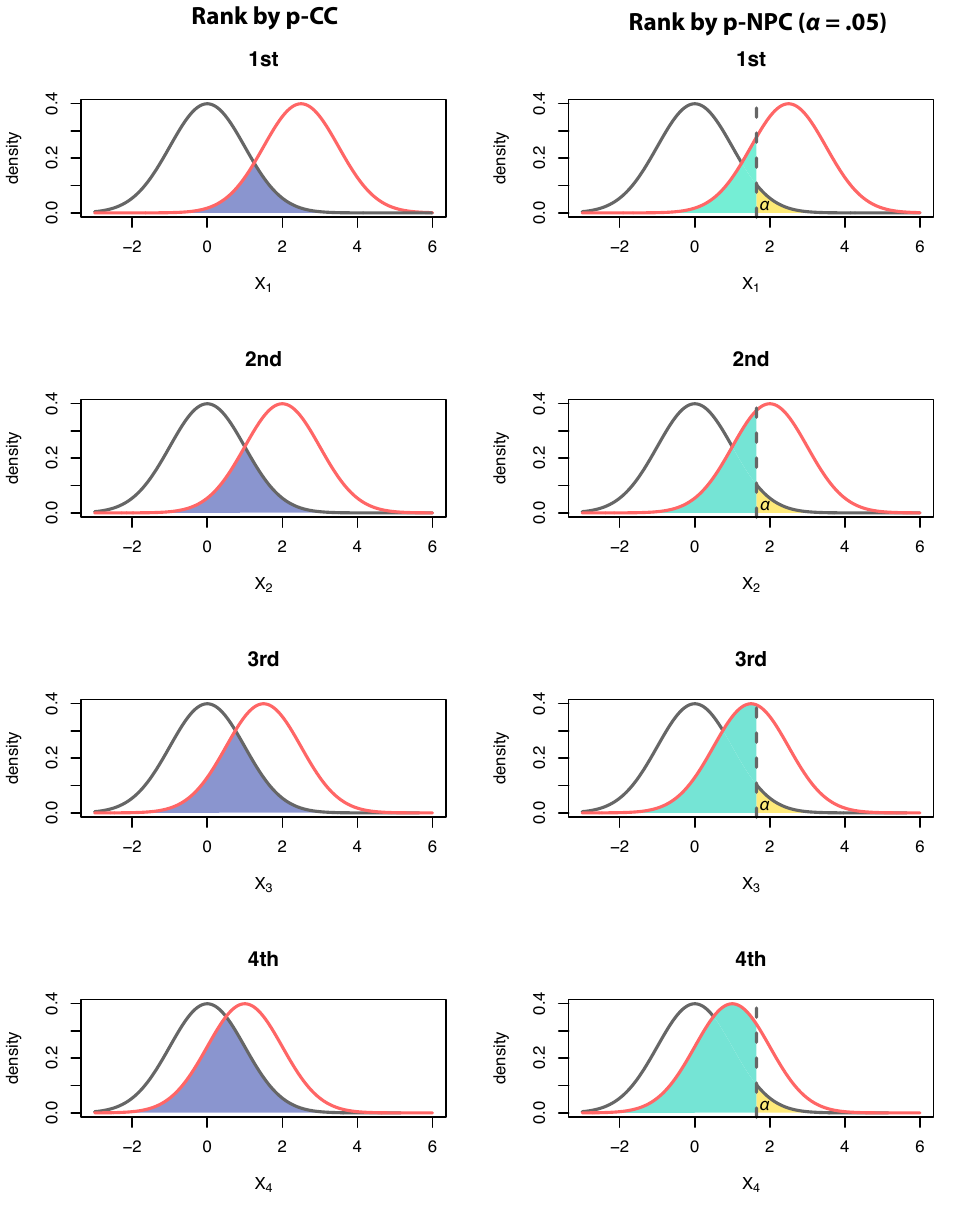}}
    \caption{\small{An illustration of marginal feature ranking by p-CC (left) and p-NPC (right). In this example, $\pi_0 = \pi_1 = .5$. The purple areas indicate p-CC values, the green areas indicate p-NPC values, and the yellow areas indicate $\alpha$ (the type I error upper bound in the NP paradigm). For both p-CC and p-NPC, a smaller value gives a better ranking.}}\label{fig:marginal_ranking_illustration}
\end{figure}

\begin{figure}
    \centering
    \makebox{\includegraphics[width = 0.75\textwidth]{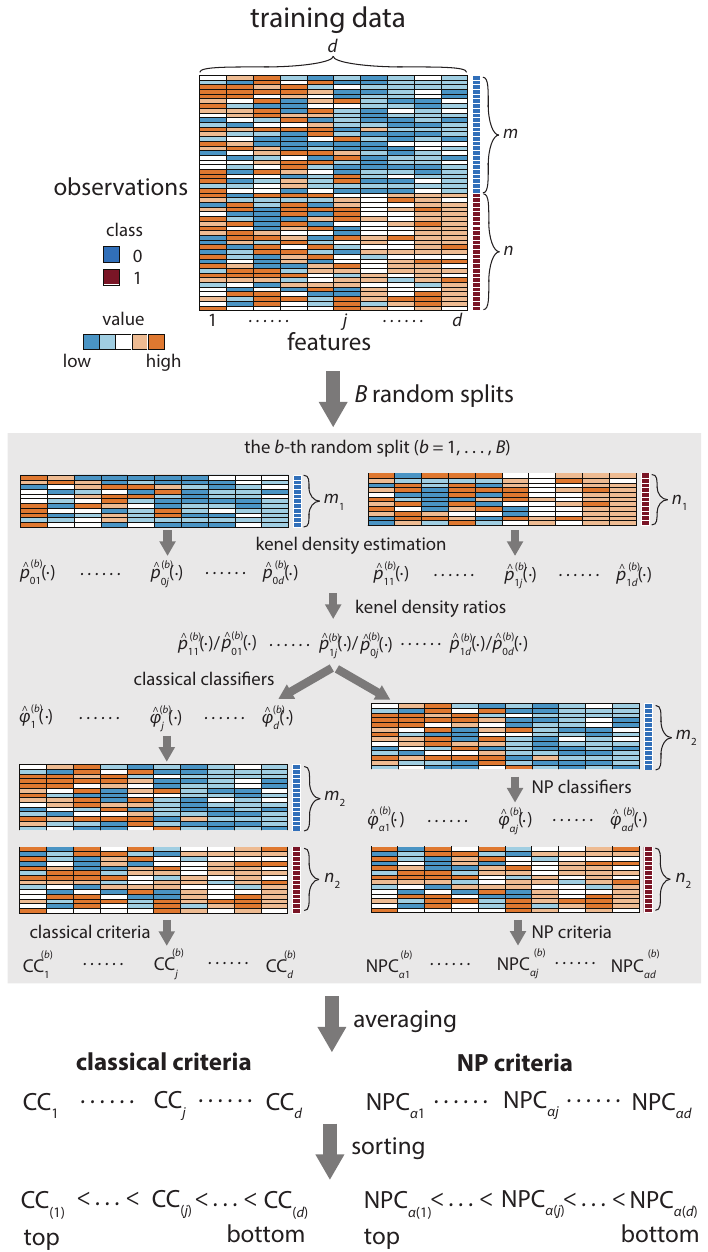}}
    \caption{\small{An illustration of the calculation of s-CC and s-NPC.}}\label{fig:illustration}
\end{figure}

\clearpage
\begin{table}[htbp]
\caption{\label{tab:avg_rank_d30_n400}Average ranks of the $d=30$ features by s-CC or s-NPC (with varying $\alpha$) with sample size $N=400$ under the Gaussian setting \eqref{eq:best_subset}---simulation study S1.}
\centering
\small
\begin{tabular}{lrrrrrrrrrr}
  \hline
 & 1 & 2 & 3 & 4 & 5 & 6 & 7 & 8 & 9 & 10 \\ 
  \hline
s-CC & 2.19 & 2.03 & 3.45 & 4.94 & 5.60 & 6.28 & 5.80 & 7.05 & 8.84 & 8.82 \\ 
  s-NPC ($\alpha = .05$) & 2.17 & 3.73 & 4.04 & 6.43 & 5.37 & 5.11 & 6.21 & 9.35 & 8.97 & 8.54 \\ 
  s-NPC ($\alpha = .10$) & 1.91 & 4.43 & 4.34 & 3.26 & 5.99 & 6.93 & 6.39 & 7.17 & 6.89 & 7.85 \\ 
  s-NPC ($\alpha = .20$) & 2.39 & 3.67 & 3.50 & 3.51 & 6.35 & 4.70 & 5.91 & 7.82 & 8.84 & 8.32 \\ 
  s-NPC ($\alpha = .30$) & 1.96 & 2.54 & 3.86 & 4.40 & 5.65 & 5.21 & 6.53 & 7.14 & 8.67 & 9.04 \\ 
   \hline
 & 11 & 12 & 13 & 14 & 15 & 16 & 17 & 18 & 19 & 20 \\ 
  \hline
s-CC & 19.80 & 21.75 & 21.36 & 16.34 & 18.79 & 21.53 & 22.60 & 18.89 & 17.26 & 23.31 \\ 
  s-NPC ($\alpha = .05$) & 15.38 & 21.58 & 22.65 & 21.47 & 17.09 & 21.30 & 20.79 & 21.65 & 20.96 & 18.15 \\ 
  s-NPC ($\alpha = .10$) & 20.66 & 23.62 & 18.73 & 23.01 & 21.69 & 19.03 & 23.05 & 18.83 & 20.77 & 20.33 \\ 
  s-NPC ($\alpha = .20$) & 20.81 & 17.65 & 21.73 & 21.67 & 17.50 & 21.30 & 20.30 & 22.75 & 18.18 & 23.84 \\ 
  s-NPC ($\alpha = .30$) & 16.72 & 22.23 & 19.93 & 19.27 & 19.80 & 21.97 & 19.29 & 19.92 & 18.95 & 19.75 \\ 
   \hline  
 & 21 & 22 & 23 & 24 & 25 & 26 & 27 & 28 & 29 & 30 \\ 
  \hline
s-CC & 19.34 & 19.63 & 22.10 & 17.26 & 22.18 & 25.03 & 22.69 & 18.33 & 20.51 & 21.31 \\ 
  s-NPC ($\alpha = .05$) & 21.47 & 16.29 & 21.36 & 20.56 & 19.39 & 20.32 & 19.84 & 21.70 & 23.34 & 19.79 \\ 
  s-NPC ($\alpha = .10$) & 19.62 & 18.80 & 19.10 & 21.55 & 19.98 & 22.81 & 18.93 & 18.44 & 19.91 & 20.95 \\ 
  s-NPC ($\alpha = .20$) & 21.59 & 15.89 & 20.22 & 20.88 & 21.44 & 18.87 & 21.03 & 21.77 & 20.49 & 22.06 \\ 
  s-NPC ($\alpha = .30$) & 20.50 & 21.82 & 22.03 & 18.52 & 20.88 & 21.26 & 21.26 & 21.13 & 23.13 & 21.65 \\ 
   \hline
\end{tabular}
\end{table}

\begin{table}[htbp]
\caption{\label{tab:avg_rank_d30_n1000}Average ranks of the $d=30$ features by s-CC or s-NPC (with varying $\alpha$) with sample size $N=1000$ under the Gaussian setting \eqref{eq:best_subset}---simulation study S1.}
\centering
\small
\begin{tabular}{lrrrrrrrrrr}
  \hline
 & 1 & 2 & 3 & 4 & 5 & 6 & 7 & 8 & 9 & 10 \\ 
  \hline
s-CC & 2.21 & 2.28 & 2.73 & 4.09 & 4.64 & 6.14 & 6.93 & 7.93 & 8.71 & 9.34 \\ 
  s-NPC ($\alpha$ = .05) & 2.55 & 2.60 & 4.21 & 4.44 & 4.28 & 6.43 & 6.48 & 6.99 & 8.22 & 8.80 \\ 
  s-NPC ($\alpha$ = .10) & 1.97 & 2.76 & 2.72 & 4.49 & 4.26 & 6.63 & 6.74 & 7.67 & 8.72 & 9.04 \\ 
  s-NPC ($\alpha$ = .20) & 1.36 & 2.35 & 3.23 & 4.19 & 4.67 & 5.93 & 7.02 & 8.24 & 8.75 & 9.24 \\ 
  s-NPC ($\alpha$ = .30) & 1.85 & 2.73 & 2.71 & 3.58 & 5.18 & 6.11 & 6.80 & 8.04 & 9.01 & 8.99 \\ 
   \hline
 & 11 & 12 & 13 & 14 & 15 & 16 & 17 & 18 & 19 & 20 \\ 
  \hline
s-CC & 18.65 & 18.19 & 20.78 & 19.92 & 23.99 & 18.60 & 19.87 & 22.16 & 21.70 & 21.61 \\ 
  s-NPC ($\alpha$ = .05) & 22.07 & 20.25 & 21.63 & 18.63 & 17.00 & 22.16 & 19.80 & 23.05 & 19.68 & 20.84 \\ 
  s-NPC ($\alpha$ = .10) & 20.37 & 19.67 & 22.67 & 20.15 & 19.31 & 19.58 & 21.61 & 18.53 & 20.51 & 22.49 \\ 
  s-NPC ($\alpha$ = .20) & 19.10 & 20.26 & 18.08 & 20.69 & 22.15 & 22.65 & 18.19 & 21.55 & 23.79 & 20.48 \\ 
  s-NPC ($\alpha$ = .30) & 18.19 & 19.32 & 20.80 & 16.88 & 22.97 & 21.70 & 19.81 & 23.49 & 19.24 & 20.95 \\ 
   \hline
 & 21 & 22 & 23 & 24 & 25 & 26 & 27 & 28 & 29 & 30 \\ 
  \hline
s-CC & 20.73 & 21.62 & 20.33 & 19.17 & 22.00 & 19.08 & 20.81 & 20.00 & 21.02 & 19.76 \\ 
  s-NPC ($\alpha = .05$) & 20.55 & 17.53 & 20.00 & 18.30 & 22.76 & 22.20 & 18.84 & 22.09 & 20.14 & 22.47 \\ 
  s-NPC ($\alpha = .10$) & 19.15 & 20.34 & 20.38 & 20.82 & 21.23 & 21.27 & 21.80 & 21.03 & 18.29 & 20.81 \\ 
  s-NPC ($\alpha = .20$) & 20.59 & 18.04 & 21.27 & 19.80 & 21.80 & 20.29 & 19.80 & 23.33 & 19.71 & 18.45 \\ 
  s-NPC ($\alpha = .30$) & 20.39 & 20.00 & 21.47 & 20.23 & 21.91 & 21.71 & 18.88 & 22.39 & 19.32 & 20.35 \\ 
   \hline
\end{tabular}
\end{table}

\begin{table}
\caption{\label{tab:avg_rank_d30_n400_chisq}Average ranks of the $d=30$ features by s-CC or s-NPC (with varying $\alpha$) with sample size $N=400$ under the Chi-squared setting \eqref{eq:chisq}---simulation study S2.}
\centering
\small
\begin{tabular}{lrrrrrrrrrr}
  \hline
 & 1 & 2 & 3 & 4 & 5 & 6 & 7 & 8 & 9 & 10 \\ 
  \hline
s-CC & 1.10 & 2.14 & 2.85 & 3.97 & 4.98 & 5.99 & 6.99 & 7.98 & 9.00 & 10.01 \\ 
  s-NPC ($\alpha = .05$) & 1.40 & 2.75 & 3.79 & 3.73 & 3.80 & 5.83 & 7.38 & 7.43 & 9.32 & 10.78 \\ 
  s-NPC ($\alpha = .10$) & 1.15 & 2.27 & 2.76 & 4.01 & 4.96 & 5.91 & 6.98 & 7.98 & 9.00 & 10.39 \\ 
  s-NPC ($\alpha = .20$) & 1.15 & 2.01 & 2.93 & 3.97 & 4.99 & 5.98 & 7.00 & 7.99 & 8.99 & 10.10 \\ 
  s-NPC ($\alpha = .30$) & 1.08 & 2.08 & 2.99 & 4.00 & 4.90 & 5.98 & 6.99 & 7.98 & 9.00 & 10.01 \\ 
   \hline
 & 11 & 12 & 13 & 14 & 15 & 16 & 17 & 18 & 19 & 20 \\ 
  \hline
s-CC & 20.56 & 19.80 & 22.54 & 21.97 & 20.63 & 20.80 & 18.33 & 18.26 & 20.41 & 21.65 \\ 
  s-NPC ($\alpha = .05$) & 21.00 & 17.83 & 19.21 & 21.03 & 19.86 & 19.81 & 22.79 & 21.50 & 22.56 & 23.24 \\ 
  s-NPC ($\alpha = .10$) & 17.67 & 24.46 & 18.36 & 16.72 & 22.22 & 20.05 & 23.21 & 19.76 & 21.00 & 20.20 \\ 
  s-NPC ($\alpha = .20$) & 19.10 & 25.31 & 21.22 & 19.55 & 20.19 & 20.17 & 20.45 & 19.95 & 21.28 & 19.41 \\ 
  s-NPC ($\alpha = .30$) & 19.91 & 22.46 & 20.62 & 21.77 & 20.49 & 18.79 & 19.72 & 19.77 & 22.06 & 18.42 \\ 
   \hline
 & 21 & 22 & 23 & 24 & 25 & 26 & 27 & 28 & 29 & 30 \\ 
  \hline
s-CC & 21.27 & 20.02 & 19.08 & 21.68 & 21.05 & 19.50 & 19.14 & 22.41 & 21.63 & 19.27 \\ 
  s-NPC ($\alpha = .05$) & 18.84 & 20.55 & 20.00 & 21.93 & 19.52 & 19.11 & 18.81 & 21.81 & 20.97 & 18.44 \\ 
  s-NPC ($\alpha = .10$) & 21.05 & 20.11 & 23.74 & 20.93 & 16.57 & 20.58 & 22.53 & 19.00 & 23.53 & 17.88 \\ 
  s-NPC ($\alpha = .20$) & 20.98 & 19.77 & 18.05 & 18.58 & 20.08 & 23.26 & 19.18 & 20.78 & 23.61 & 18.98 \\ 
  s-NPC ($\alpha = .30$) & 18.93 & 22.03 & 23.31 & 19.47 & 20.75 & 22.04 & 20.49 & 19.23 & 20.35 & 19.37 \\ 
   \hline
\end{tabular}
\end{table}

\begin{figure}
\includegraphics[width=\textwidth]{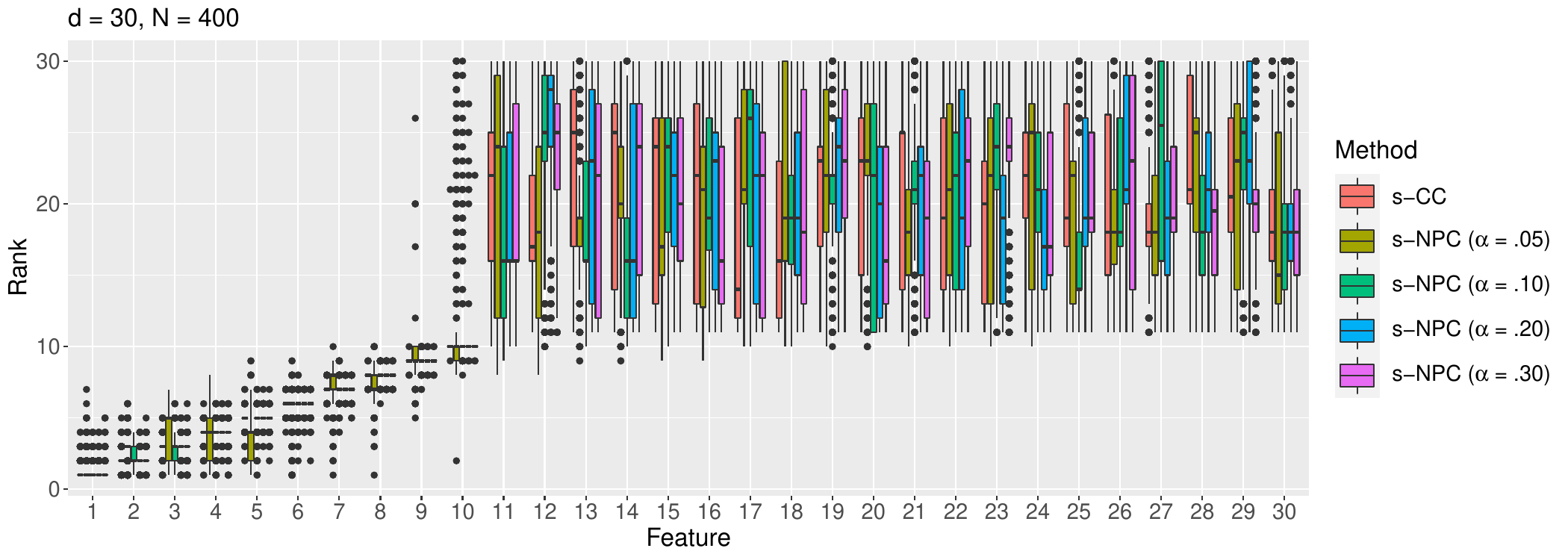}
\caption{Rank distributions of the $d=30$ features by s-CC or s-NPC (with varying $\alpha$) with sample size $N=400$ under the Chi-squared setting \eqref{eq:chisq}---simulation study S2.\label{fig:avg_rank_d30_n400_chisq}}	
\end{figure}

\begin{table}
\caption{\label{tab:avg_rank_d30_n1000_chisq}Average ranks of the $d=30$ features by s-CC or s-NPC (with varying $\alpha$) with sample size $N=1000$ under the Chi-squared setting \eqref{eq:chisq}---simulation study S2.}
\centering
\small
\begin{tabular}{lrrrrrrrrrr}
  \hline
 & 1 & 2 & 3 & 4 & 5 & 6 & 7 & 8 & 9 & 10 \\ 
  \hline
s-CC & 1.05 & 2.68 & 2.30 & 3.99 & 4.99 & 5.99 & 7.00 & 8.00 & 9.00 & 10.00 \\ 
  s-NPC ($\alpha = .05$) & 1.44 & 2.38 & 2.28 & 3.99 & 4.95 & 5.99 & 7.22 & 7.77 & 8.99 & 10.12 \\ 
  s-NPC ($\alpha = .10$) & 1.05 & 2.70 & 2.28 & 3.99 & 4.99 & 5.99 & 7.01 & 7.99 & 9.00 & 10.01 \\ 
  s-NPC ($\alpha = .20$) & 1.07 & 2.47 & 3.06 & 3.41 & 5.00 & 6.00 & 7.00 & 8.00 & 9.00 & 10.00 \\ 
  s-NPC ($\alpha = .30$) & 1.05 & 2.20 & 3.17 & 3.63 & 4.97 & 5.98 & 7.00 & 8.00 & 9.00 & 10.00 \\ 
   \hline
 & 11 & 12 & 13 & 14 & 15 & 16 & 17 & 18 & 19 & 20 \\ 
  \hline
s-CC & 19.87 & 20.05 & 24.51 & 22.30 & 23.43 & 23.26 & 18.01 & 18.22 & 23.86 & 19.85 \\ 
  s-NPC ($\alpha = .05$) & 23.85 & 21.45 & 18.73 & 18.45 & 18.49 & 17.36 & 20.65 & 21.48 & 20.10 & 18.70 \\ 
  s-NPC ($\alpha = .10$) & 24.12 & 17.48 & 20.12 & 19.22 & 21.02 & 22.58 & 21.03 & 17.97 & 17.97 & 20.29 \\ 
  s-NPC ($\alpha = .20$) & 20.19 & 24.91 & 20.15 & 22.55 & 21.36 & 18.35 & 19.82 & 20.66 & 20.76 & 19.32 \\ 
  s-NPC ($\alpha = .30$) & 22.50 & 22.12 & 23.15 & 23.31 & 20.11 & 22.51 & 19.61 & 18.88 & 19.52 & 19.91 \\ 
   \hline
 & 21 & 22 & 23 & 24 & 25 & 26 & 27 & 28 & 29 & 30 \\ 
  \hline
s-CC & 19.54 & 18.13 & 19.57 & 20.90 & 20.24 & 19.89 & 19.30 & 18.04 & 18.10 & 22.94 \\ 
  s-NPC ($\alpha = .05$) & 19.05 & 20.94 & 18.09 & 21.05 & 19.02 & 18.04 & 23.57 & 22.26 & 23.89 & 24.69 \\ 
  s-NPC ($\alpha = .10$) & 19.99 & 18.39 & 17.08 & 20.04 & 20.31 & 18.84 & 22.61 & 22.38 & 23.93 & 24.63 \\ 
  s-NPC ($\alpha = .20$) & 20.29 & 17.02 & 20.71 & 19.12 & 17.70 & 18.30 & 20.29 & 22.33 & 24.30 & 21.90 \\ 
  s-NPC ($\alpha = .30$) & 20.54 & 19.42 & 16.57 & 18.70 & 20.48 & 20.09 & 22.57 & 17.09 & 19.93 & 22.98 \\ 
   \hline
\end{tabular}
\end{table}

\begin{figure}
\includegraphics[width=\textwidth]{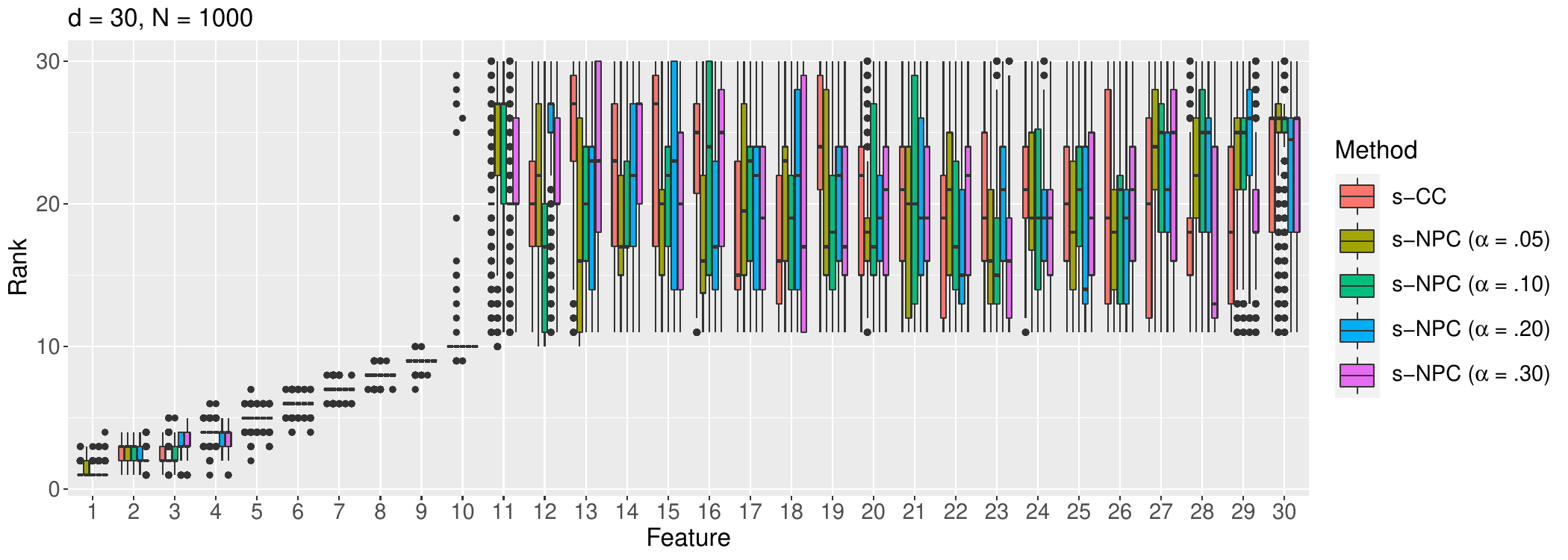}
\caption{Rank distributions of the $d=30$ features by s-CC or s-NPC (with varying $\alpha$) with sample size $N=1000$ under the Chi-squared setting \eqref{eq:chisq}---simulation study S2.\label{fig:avg_rank_d30_n1000_chisq}}	
\end{figure}

\begin{table}
\caption{\label{tab:avg_rank_d30_n1000_bias}Average ranks of the $d=30$ features by s-CC, s-NPC (with varying $\alpha$), or their SVM variants with $N=1000$ under the Gaussian setting with sampling bias ($\pi_1^{\mathrm{population}}=.5$ and $\pi_1^{\mathrm{sample}}=.1$) \eqref{eq:gauss_bias} and without feature correlations---simulation study S3. Undesirable average ranks---i.e., the average ranks of the informative (first $10$) features greater than $10$ and the average ranks of the uninformative (latter $20$) features smaller than $10$---are underlined.}
\small
\centering
\begin{tabular}{rrrrrrrrrrr}
  \hline
 & 1 & 2 & 3 & 4 & 5 & 6 & 7 & 8 & 9 & 10 \\ 
  \hline
s-CC & 2.10 & 4.05 & 6.06 & 5.08 & 6.75 & 7.89 & 8.38 & \underline{15.85} & \underline{13.04} & \underline{14.30} \\ 
  s-NPC ($\alpha = .05$) & 3.30 & 3.20 & 3.53 & 3.69 & 5.54 & 6.22 & 7.20 & 8.04 & 6.87 & 7.67 \\ 
  s-NPC ($\alpha = .10$) & 2.50 & 2.08 & 4.30 & 4.79 & 4.87 & 5.93 & 7.79 & 7.08 & 7.74 & 7.98 \\ 
  s-NPC ($\alpha = .20$) & 2.17 & 3.04 & 2.82 & 4.75 & 5.41 & 6.52 & 6.55 & 6.42 & 8.32 & 8.99 \\ 
  s-NPC ($\alpha = .30$) & 1.91 & 2.29 & 3.76 & 4.75 & 5.28 & 5.89 & 7.40 & 8.11 & 8.30 & 7.30 \\
  s-CC-SVM & 2.38 & 2.43 & 4.12 & 5.26 & 7.00 & 8.73 & \underline{10.04} & \underline{11.40} & \underline{12.44} & \underline{13.45} \\ 
  s-NPC-SVM ($\alpha = .05$) & 2.02 & 2.52 & 3.86 & 4.59 & 5.92 & 5.26 & 6.15 & 8.50 & 9.12 & \underline{12.08} \\ 
  s-NPC-SVM ($\alpha = .10$) & 1.84 & 3.58 & 2.79 & 4.79 & 5.01 & 5.86 & 7.13 & 8.98 & \underline{13.22} & \underline{13.35} \\ 
  s-NPC-SVM ($\alpha = .20$) & 2.01 & 3.41 & 2.92 & 5.06 & 5.09 & 9.40 & 5.88 & 8.84 & \underline{14.66} & \underline{15.12} \\ 
  s-NPC-SVM ($\alpha = .30$) & 2.02 & 3.26 & 3.78 & 5.36 & 8.41 & 6.66 & 7.28 & \underline{11.48} & \underline{14.17} & \underline{19.34} \\ 
  \hline
 & 11 & 12 & 13 & 14 & 15 & 16 & 17 & 18 & 19 & 20 \\ 
  \hline
s-CC & 16.13 & 18.25 & 20.37 & 22.20 & 23.07 & 20.88 & 17.31 & 18.61 & 18.83 & 20.15 \\ 
  s-NPC ($\alpha = .05$) & 21.19 & 21.63 & 17.19 & 21.28 & 19.35 & 17.98 & 18.37 & 20.21 & 19.73 & 21.66 \\ 
  s-NPC ($\alpha = .10$) & 22.31 & 21.38 & 19.78 & 22.60 & 18.04 & 17.45 & 16.55 & 19.62 & 17.76 & 21.27 \\ 
  s-NPC ($\alpha = .20$) & 23.22 & 20.31 & 18.75 & 18.42 & 18.15 & 18.71 & 21.65 & 19.06 & 16.69 & 21.55 \\ 
  s-NPC ($\alpha = .30$) & 23.23 & 19.66 & 19.66 & 18.36 & 19.32 & 20.16 & 15.93 & 15.84 & 22.74 & 21.39 \\ 
  s-CC-SVM & 14.45 & 15.46 & 16.47 & 17.48 & 18.50 & 19.50 & 20.53 & 21.54 & 22.55 & 23.57 \\ 
  s-NPC-SVM ($\alpha = .05$) & 17.78 & 20.17 & 20.39 & 18.28 & 19.14 & 20.46 & 18.75 & 20.50 & 21.01 & 19.25 \\ 
  s-NPC-SVM ($\alpha = .10$) & 18.17 & 14.39 & 19.19 & 17.82 & 19.96 & 22.91 & 21.36 & 16.34 & 20.66 & 19.80 \\ 
  s-NPC-SVM ($\alpha = .20$) & 16.62 & 17.10 & 19.87 & 20.25 & 22.43 & 17.03 & 16.44 & 21.85 & 20.60 & 20.53 \\ 
  s-NPC-SVM ($\alpha = .30$) & 21.64 & 12.28 & 19.18 & 23.97 & 18.91 & 17.76 & 17.41 & 17.58 & 20.06 & 22.68 \\ 
   \hline
 & 21 & 22 & 23 & 24 & 25 & 26 & 27 & 28 & 29 & 30 \\ 
  \hline
s-CC & 17.60 & 17.09 & 21.92 & 18.29 & 17.58 & 20.84 & 22.87 & 14.73 & 18.52 & 16.28 \\ 
  s-NPC ($\alpha = .05$) & 24.11 & 18.94 & 21.15 & 18.50 & 19.86 & 23.07 & 18.39 & 22.99 & 21.25 & 22.86 \\ 
  s-NPC ($\alpha = .10$) & 24.67 & 20.20 & 20.06 & 21.97 & 22.92 & 21.94 & 16.97 & 23.18 & 18.36 & 22.93 \\ 
  s-NPC ($\alpha = .20$) & 17.93 & 22.79 & 21.60 & 24.64 & 21.47 & 21.45 & 22.99 & 19.35 & 21.39 & 19.88 \\ 
  s-NPC ($\alpha = .30$) & 23.30 & 24.46 & 22.10 & 21.47 & 21.08 & 22.45 & 17.56 & 22.86 & 19.85 & 18.58 \\ 
  s-CC-SVM & 24.57 & 25.56 & 26.52 & 27.16 & 26.87 & 22.30 & 19.29 & 13.19 & \underline{6.00} & \underline{6.24} \\ 
  s-NPC-SVM ($\alpha = .05$) & 19.82 & 21.72 & 21.85 & 20.42 & 19.86 & 23.08 & 19.36 & 20.20 & 21.09 & 21.84 \\ 
  s-NPC-SVM ($\alpha = .10$) & 20.73 & 19.62 & 22.84 & 22.10 & 19.65 & 18.14 & 21.00 & 19.09 & 21.36 & 23.34 \\ 
  s-NPC-SVM ($\alpha = .20$) & 20.11 & 21.57 & 16.15 & 20.37 & 20.96 & 19.04 & 20.47 & 18.85 & 21.67 & 20.69 \\ 
  s-NPC-SVM ($\alpha = .30$) & 17.97 & 18.64 & 18.00 & 19.46 & 19.88 & 20.83 & 19.39 & 17.25 & 20.64 & 19.71 \\ 
   \hline
\end{tabular}
\end{table}

\begin{figure}
     \centering
     \begin{subfigure}[b]{\textwidth}
         \centering
         \includegraphics[width=\textwidth]{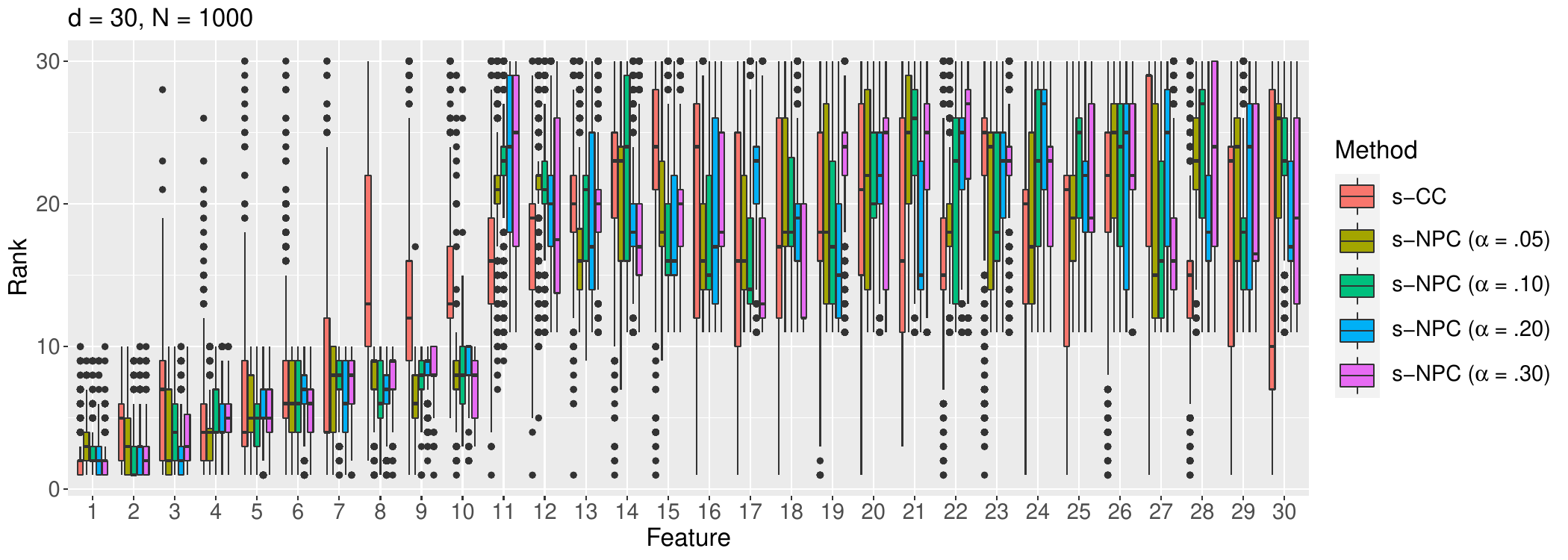}
         \caption{s-CC and s-NPC}
     \end{subfigure}
     \hfill
     \begin{subfigure}[b]{\textwidth}
         \centering
         \includegraphics[width=\textwidth]{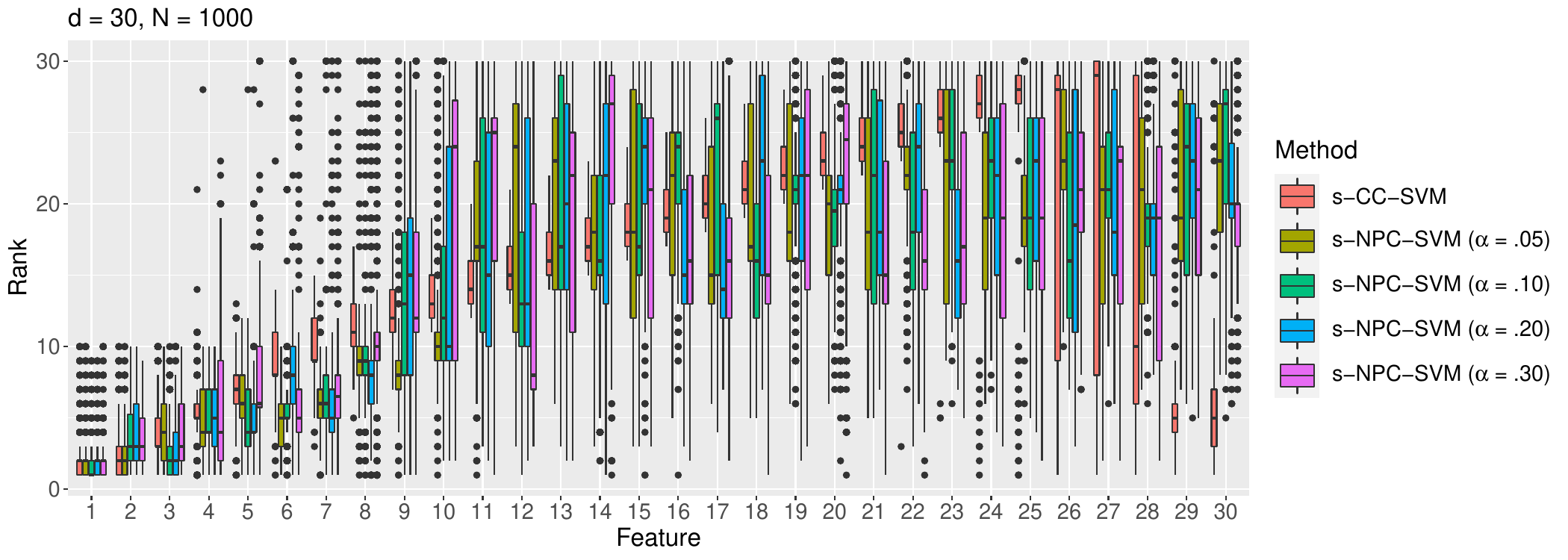}
         \caption{SVM variants of s-CC and s-NPC}
     \end{subfigure}
        \caption{Rank distributions of the features under the Gaussian setting with $d=30$, $N=1000$, and sampling bias ($\pi_1^{\mathrm{population}}=.5$ and $\pi_1^{\mathrm{sample}}=.1$) \eqref{eq:gauss_bias}, and without feature correlations---simulation study S3.}
        \label{fig:avg_rank_d30_n1000_bias}
\end{figure}

\begin{table}
\caption{\label{tab:avg_rank_d30_n1000_bias_cor}Average ranks of the $d=30$ features by s-CC, s-NPC (with varying $\alpha$), their SVM variants, or the RF algorithm's feature importance measures and SHAP value, with sample size $N=1000$ under the Gaussian setting with sampling bias ($\pi_1^{\mathrm{population}}=.5$ and $\pi_1^{\mathrm{sample}}=.1$) \eqref{eq:gauss_bias} and a Toeplitz-type feature covariance matrix: features $i$ and $j$ have a correlation $\rho_{ij}=.9^{|i-j|}$, $i,j=1,\ldots,30$---simulation study S3. Undesirable average ranks---i.e., the average ranks of the informative (first $10$) features greater than $10$ and the average ranks of the uninformative (latter $20$) features smaller than $10$---are underlined.}
\small
\centering
\begin{tabular}{rrrrrrrrrrr}
  \hline
 & 1 & 2 & 3 & 4 & 5 & 6 & 7 & 8 & 9 & 10 \\ 
  \hline
s-CC & 3.13 & 4.12 & 5.68 & 7.32 & 8.80 & \underline{10.42} & \underline{12.32} & \underline{14.34} & \underline{16.20} & \underline{17.54} \\ 
  s-NPC ($\alpha = .05$) & 2.13 & 2.77 & 3.55 & 4.27 & 5.10 & 5.94 & 6.75 & 7.51 & 8.41 & 9.24 \\ 
  s-NPC ($\alpha = .10$) & 1.85 & 2.60 & 3.31 & 4.24 & 5.16 & 6.01 & 6.77 & 7.64 & 8.38 & 9.18 \\ 
  s-NPC ($\alpha = .20$) & 1.80 & 2.48 & 3.29 & 4.13 & 5.09 & 5.96 & 6.88 & 7.65 & 8.48 & 9.25 \\ 
  s-NPC ($\alpha = .30$) & 1.80 & 2.57 & 3.27 & 4.11 & 5.05 & 6.01 & 6.79 & 7.70 & 8.52 & 9.20 \\ 
  s-CC-SVM & 2.83 & 3.75 & 4.84 & 6.24 & 7.52 & 8.85 & \underline{10.27} & \underline{11.52} & \underline{12.74} & \underline{13.80} \\ 
  s-NPC-SVM ($\alpha = .05$) & 2.03 & 2.77 & 3.50 & 4.37 & 5.26 & 6.31 & 7.52 & 8.79 & \underline{10.15} & \underline{12.61} \\ 
  s-NPC-SVM ($\alpha = .10$) & 1.85 & 2.63 & 3.50 & 4.36 & 5.35 & 6.51 & 7.81 & 9.06 & \underline{10.89} & \underline{13.59} \\ 
  s-NPC-SVM ($\alpha = .20$) & 1.96 & 2.72 & 3.60 & 4.57 & 5.81 & 7.05 & 8.68 & \underline{10.29} & \underline{11.85} & \underline{14.76} \\ 
  s-NPC-SVM ($\alpha = .30$) & 2.09 & 3.05 & 3.86 & 5.16 & 6.25 & 8.05 & 9.35 & \underline{11.24} & \underline{13.61} & \underline{15.26} \\ 
  RF-MeanDecreaseAccuracy & 7.61 & 6.73 & 6.12 & 6.00 & 5.94 & 6.05 & 6.56 & 6.82 & 7.06 & 7.43 \\ 
  RF-MeanDecreaseGini & 1.82 & 3.80 & 5.66 & 6.46 & 6.76 & 7.29 & 7.46 & 7.87 & 8.20 & 8.42 \\ 
  RF-SHAP & 2.47 & 6.12 & 6.74 & 7.04 & 7.19 & 7.92 & 8.64 & 9.36 & \underline{10.69} & \underline{10.80} \\ 
  \hline
 & 11 & 12 & 13 & 14 & 15 & 16 & 17 & 18 & 19 & 20 \\ 
  \hline
s-CC & 18.95 & 19.98 & 20.40 & 20.48 & 20.72 & 20.87 & 20.66 & 20.28 & 19.94 & 19.28 \\ 
  s-NPC ($\alpha = .05$) & 19.96 & 19.87 & 20.18 & 20.34 & 20.29 & 20.52 & 20.16 & 20.32 & 20.43 & 20.39 \\ 
  s-NPC ($\alpha = .10$) & 20.34 & 20.13 & 20.28 & 20.73 & 20.09 & 20.53 & 20.38 & 20.47 & 20.74 & 20.09 \\ 
  s-NPC ($\alpha = .20$) & 20.45 & 20.37 & 20.62 & 20.70 & 20.45 & 20.01 & 20.52 & 20.38 & 20.53 & 20.58 \\ 
  s-NPC ($\alpha = .30$) & 20.41 & 20.46 & 20.46 & 20.46 & 20.56 & 20.74 & 20.39 & 20.35 & 20.60 & 20.66 \\ 
  s-CC-SVM & 14.86 & 15.86 & 16.88 & 17.89 & 18.91 & 19.93 & 20.92 & 21.96 & 22.96 & 23.93 \\ 
  s-NPC-SVM ($\alpha = .05$) & 18.19 & 19.27 & 19.53 & 19.71 & 19.39 & 19.98 & 19.79 & 20.07 & 20.28 & 20.28 \\ 
  s-NPC-SVM ($\alpha = .10$) & 18.50 & 19.23 & 19.21 & 19.30 & 19.64 & 19.69 & 19.88 & 19.78 & 20.05 & 20.09 \\ 
  s-NPC-SVM ($\alpha = .20$) & 18.02 & 18.87 & 18.82 & 19.14 & 19.88 & 19.77 & 19.23 & 19.96 & 19.68 & 19.63 \\ 
  s-NPC-SVM ($\alpha = .30$) & 17.66 & 17.92 & 18.76 & 18.79 & 18.63 & 19.46 & 19.45 & 19.52 & 19.66 & 19.32 \\ 
  RF-MeanDecreaseAccuracy & \underline{8.06} & \underline{9.28} & 11.83 & 14.35 & 16.78 & 18.10 & 19.02 & 19.92 & 20.53 & 21.24 \\ 
  RF-MeanDecreaseGini & \underline{8.77} & \underline{9.78} & 12.88 & 15.79 & 17.88 & 19.81 & 20.30 & 20.93 & 21.81 & 21.74 \\ 
  RF-SHAP & 12.20 & 14.53 & 16.04 & 16.66 & 16.04 & 15.20 & 15.12 & 14.88 & 15.68 & 16.48 \\ 
   \hline
    & 21 & 22 & 23 & 24 & 25 & 26 & 27 & 28 & 29 & 30 \\ 
  \hline
s-CC & 19.11 & 18.56 & 18.73 & 17.88 & 17.96 & 16.66 & 15.66 & 14.71 & 12.79 & 11.49 \\ 
  s-NPC ($\alpha = .05$) & 20.61 & 20.16 & 20.36 & 20.73 & 20.96 & 20.70 & 20.38 & 20.75 & 21.09 & 21.10 \\ 
  s-NPC ($\alpha = .10$) & 20.34 & 20.21 & 20.62 & 20.78 & 20.53 & 20.38 & 20.66 & 20.77 & 21.03 & 20.78 \\ 
  s-NPC ($\alpha = .20$) & 20.39 & 20.52 & 20.62 & 20.63 & 20.35 & 20.48 & 20.40 & 20.79 & 20.32 & 20.88 \\ 
  s-NPC ($\alpha = .30$) & 20.34 & 20.19 & 20.84 & 20.44 & 20.41 & 20.48 & 20.44 & 20.62 & 20.51 & 20.63 \\ 
  s-CC-SVM & 24.73 & 25.37 & 25.34 & 24.59 & 22.94 & 20.03 & 16.53 & 12.82 & \underline{9.18} & \underline{7.00} \\ 
  s-NPC-SVM ($\alpha = .05$) & 20.04 & 20.59 & 20.14 & 20.43 & 20.70 & 20.57 & 20.63 & 20.34 & 20.63 & 21.14 \\ 
  s-NPC-SVM ($\alpha = .10$) & 20.32 & 20.18 & 20.09 & 20.18 & 20.09 & 20.23 & 20.43 & 20.67 & 20.89 & 20.99 \\ 
  s-NPC-SVM ($\alpha = .20$) & 19.77 & 19.97 & 19.82 & 19.73 & 20.37 & 20.18 & 20.02 & 20.29 & 20.12 & 20.43 \\ 
  s-NPC-SVM ($\alpha = .30$) & 19.56 & 19.87 & 19.67 & 20.00 & 19.98 & 19.72 & 19.88 & 19.95 & 20.00 & 19.27 \\ 
  RF-MeanDecreaseAccuracy & 21.71 & 22.35 & 22.54 & 23.00 & 23.33 & 23.97 & 24.37 & 25.15 & 25.81 & 27.34 \\ 
  RF-MeanDecreaseGini & 22.21 & 22.29 & 22.89 & 23.20 & 23.20 & 23.37 & 23.39 & 23.48 & 23.70 & 23.87 \\ 
  RF-SHAP & 17.66 & 19.02 & 20.30 & 21.58 & 22.87 & 24.18 & 25.48 & 26.75 & 28.02 & 29.37 \\ 
   \hline
\end{tabular}
\end{table}

\begin{table}[htbp]
\caption{\label{tab:avg_rank_d500_n400}Average ranks of the first $30$ features by s-CC or s-NPC (with varying $\alpha$) with $d=500$ and $N=400$ under the Gaussian setting \eqref{eq:best_subset}---simulation study S4.}
\centering
\scriptsize
\begin{tabular}{lrrrrrrrrrr}
  \hline
 & 1 & 2 & 3 & 4 & 5 & 6 & 7 & 8 & 9 & 10 \\ 
  \hline
s-CC & 1.51 & 3.39 & 3.25 & 4.82 & 4.43 & 6.47 & 6.59 & 6.80 & 8.53 & 9.86 \\ 
  s-NPC ($\alpha$ = .05) & 2.48 & 3.14 & 3.81 & 4.57 & 4.88 & 33.75 & 87.81 & 177.79 & 136.12 & 183.96 \\ 
  s-NPC ($\alpha$ = .10) & 2.21 & 2.34 & 3.84 & 4.08 & 5.56 & 6.70 & 6.61 & 19.97 & 116.98 & 51.27 \\ 
  s-NPC ($\alpha$ = .20) & 1.87 & 2.55 & 3.60 & 3.76 & 5.41 & 6.35 & 6.67 & 7.51 & 8.61 & 46.10 \\ 
  s-NPC ($\alpha$ = .30) & 1.43 & 3.29 & 3.44 & 4.54 & 5.52 & 6.25 & 6.86 & 5.91 & 8.34 & 11.48 \\ 
   \hline
 & 11 & 12 & 13 & 14 & 15 & 16 & 17 & 18 & 19 & 20 \\ 
  \hline
s-CC & 234.07 & 244.32 & 213.54 & 213.01 & 183.60 & 249.73 & 292.85 & 269.15 & 328.63 & 240.94 \\ 
  s-NPC ($\alpha$ = .05) & 270.19 & 252.46 & 174.22 & 211.67 & 125.66 & 241.64 & 317.62 & 340.59 & 231.31 & 205.63 \\ 
  s-NPC ($\alpha$ = .10) & 254.37 & 300.12 & 317.98 & 213.02 & 263.69 & 223.81 & 296.64 & 279.72 & 288.77 & 234.69 \\ 
  s-NPC ($\alpha$ = .20) & 223.00 & 253.27 & 287.14 & 205.65 & 249.97 & 187.17 & 312.73 & 224.19 & 265.96 & 238.16 \\ 
  s-NPC ($\alpha$ = .30) & 209.82 & 192.70 & 206.62 & 271.58 & 236.41 & 263.22 & 189.90 & 299.44 & 238.57 & 269.64 \\ 
   \hline
 & 21 & 22 & 23 & 24 & 25 & 26 & 27 & 28 & 29 & 30 \\ 
  \hline
s-CC & 237.12 & 273.42 & 276.32 & 305.19 & 207.06 & 267.22 & 219.78 & 287.79 & 315.43 & 288.44 \\ 
  s-NPC ($\alpha = .05$) & 272.94 & 259.56 & 231.43 & 253.27 & 209.29 & 238.76 & 339.55 & 301.08 & 214.29 & 286.21 \\ 
  s-NPC ($\alpha = .10$) & 292.65 & 184.08 & 305.28 & 175.42 & 199.44 & 269.48 & 238.83 & 146.56 & 298.10 & 263.97 \\ 
  s-NPC ($\alpha = .20$) & 224.42 & 243.56 & 290.61 & 203.90 & 300.27 & 243.51 & 208.30 & 241.41 & 288.29 & 277.29 \\ 
  s-NPC ($\alpha = .30$) & 276.51 & 224.59 & 223.39 & 270.17 & 208.61 & 248.46 & 236.14 & 253.09 & 295.42 & 198.65 \\ 
   \hline
\end{tabular}
\end{table}


\begin{table}
\caption{\label{tab:avg_rank_d10000_n200}Average ranks of the first $20$ features by s-CC, s-NPC (with varying $\alpha$), their SVM variants, or the RF algorithm's feature importance measures (the mean decrease in accuracy and the mean decrease in Gini index) with $d=10{,}000$ and $N=200$ under the Gaussian setting \eqref{eq:best_subset}---simulation study S5. The average ranks of the top $10$ features, if exceeding $10$ (i.e., undesirable), are underlined.}
\small
\centering
\begin{tabular}{rrrrrr}
  \hline
 & 1 & 2 & 3 & 4 & 5 \\ 
  \hline
s-CC & 1.99 & 1.75 & 4.23 & 4.93 & 6.24 \\ 
  s-NPC ($\alpha = .10$) & 1.36 & 5.38 & 3.30 & 3.74 & 4.57 \\ 
  s-NPC ($\alpha = .20$) & 1.86 & 2.03 & 5.22 & 5.02 & 5.70 \\ 
  s-NPC ($\alpha = .30$) & 2.22 & 1.42 & 5.80 & 7.22 & 6.55 \\ 
  S-CC-SVM & 1.88 & 1.74 & 4.39 & 5.31 & 6.12 \\ 
  s-NPC-SVM ($\alpha = .10$) & 2.83 & \underline{64.98} & \underline{233.79} & \underline{435.20} & \underline{1231.04} \\ 
  s-NPC-SVM ($\alpha = .20$) & 2.15 & 1.69 & 5.23 & 5.80 & 7.43 \\ 
  s-NPC-SVM ($\alpha = .30$) & 2.22 & 1.47 & 5.77 & 7.37 & 6.69 \\ 
  RF\_MeanDecreaseAccuracy & 1.97 & 2.48 & 4.73 & 5.37 & 5.92 \\ 
  RF\_MeanDecreaseGini & 1.94 & 3.02 & 4.60 & 5.50 & 5.75 \\ 
   \hline
 & 6 & 7 & 8 & 9 & 10 \\ 
  \hline
s-CC & 6.62 & 6.81 & 7.29 & 7.15 & 7.99 \\ 
  s-NPC ($\alpha = .10$) & \underline{66.50} & \underline{964.71} & \underline{2715.94} & \underline{2978.16} & \underline{3759.18} \\ 
  s-NPC ($\alpha = .20$) & 5.27 & 7.20 & 7.00 & 6.68 & \underline{601.75} \\ 
  s-NPC ($\alpha = .30$) & 5.91 & 6.34 & 6.10 & 6.86 & 6.58 \\ 
  S-CC-SVM & 7.07 & 6.70 & 6.79 & 7.05 & 7.95 \\ 
  s-NPC-SVM ($\alpha = .10$) & \underline{1416.56} & \underline{1621.85} & \underline{1660.69} & \underline{2508.79} & \underline{3242.82} \\ 
  s-NPC-SVM ($\alpha = .20$) & 7.07 & 6.51 & 6.95 & \underline{109.10} & \underline{754.96} \\ 
  s-NPC-SVM ($\alpha = .30$) & 6.79 & 6.65 & 6.19 & 6.26 & \underline{15.98} \\ 
  RF\_MeanDecreaseAccuracy & 6.08 & 6.57 & 6.66 & 7.40 & 7.82 \\ 
  RF\_MeanDecreaseGini & 5.60 & 6.75 & 6.91 & 7.22 & \underline{37.34} \\ 
   \hline
 & 11 & 12 & 13 & 14 & 15 \\ 
  \hline
s-CC & 5458.50 & 3926.18 & 4217.73 & 4814.75 & 3882.88 \\ 
  s-NPC ($\alpha = .10$) & 3852.06 & 4512.05 & 4101.91 & 4421.95 & 4479.34 \\ 
  s-NPC ($\alpha = .20$) & 4155.99 & 4768.53 & 4547.92 & 3807.99 & 4684.78 \\ 
  s-NPC ($\alpha = .30$) & 4710.67 & 4276.04 & 4762.01 & 4787.34 & 4649.83 \\ 
  S-CC-SVM & 4706.49 & 3694.43 & 4528.69 & 4490.41 & 4949.34 \\ 
  s-NPC-SVM ($\alpha = .10$) & 3136.52 & 3635.93 & 3730.52 & 4193.05 & 3933.72 \\ 
  s-NPC-SVM ($\alpha = .20$) & 4153.33 & 4368.59 & 5165.12 & 4748.92 & 5557.16 \\ 
  s-NPC-SVM ($\alpha = .30$) & 6235.73 & 5586.78 & 6055.84 & 6029.32 & 5565.73 \\ 
  RF\_MeanDecreaseAccuracy & 4764.65 & 5284.11 & 5571.87 & 5327.75 & 4956.21 \\ 
  RF\_MeanDecreaseGini & 4944.51 & 5281.69 & 5204.39 & 5292.28 & 5313.57 \\ 
   \hline
 & 16 & 17 & 18 & 19 & 20 \\ 
  \hline
s-CC & 4783.62 & 4933.65 & 4910.66 & 4917.21 & 4713.67 \\ 
  s-NPC ($\alpha = .10$) & 4265.04 & 3983.15 & 4408.34 & 4238.35 & 4422.13 \\ 
  s-NPC ($\alpha = .20$) & 4275.42 & 4867.43 & 4489.88 & 4537.64 & 4582.16 \\ 
  s-NPC ($\alpha = .30$) & 5304.28 & 4825.68 & 4861.39 & 5025.82 & 4908.58 \\ 
  S-CC-SVM & 5152.39 & 4960.78 & 5054.89 & 5181.85 & 4810.33 \\ 
  s-NPC-SVM ($\alpha = .10$) & 4481.82 & 5014.15 & 4794.49 & 4096.55 & 4571.58 \\ 
  s-NPC-SVM ($\alpha = .20$) & 5165.05 & 5517.70 & 5070.07 & 5115.56 & 4998.20 \\ 
  s-NPC-SVM ($\alpha = .30$) & 5737.40 & 5872.29 & 5109.93 & 5444.06 & 4901.91 \\ 
  RF\_MeanDecreaseAccuracy & 5382.33 & 4612.39 & 4546.59 & 4555.47 & 4549.88 \\ 
  RF\_MeanDecreaseGini & 6218.43 & 5119.48 & 5264.91 & 4459.50 & 5310.33 \\ 
   \hline
\end{tabular}
\end{table}


\begin{figure}
\includegraphics[width=\textwidth]{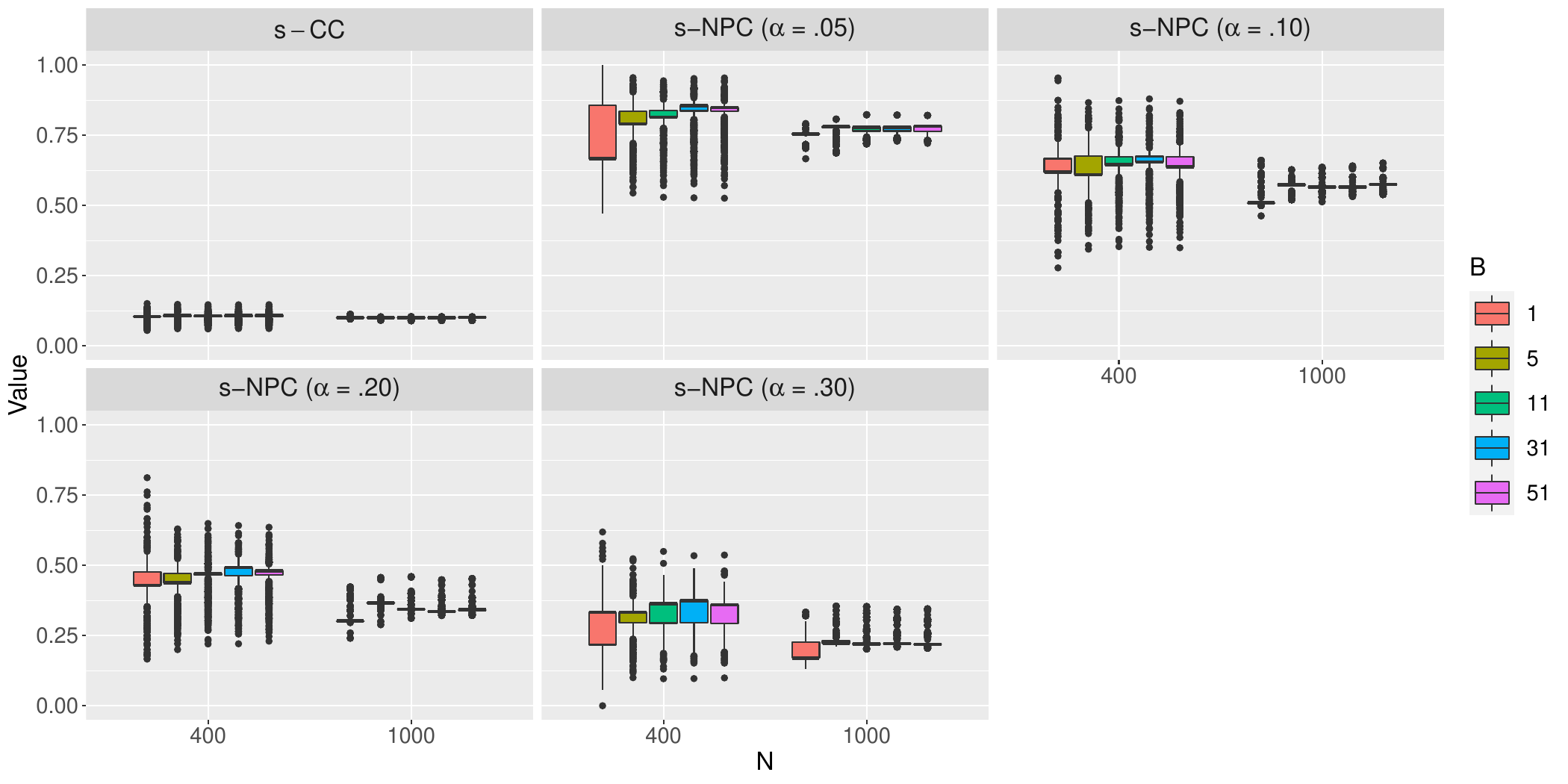}
\caption{Distributions of s-CC or s-NPC (with varying $\alpha$) values of one feature, whose class-conditional distributions are $X|(Y=0) \sim \mathcal{N}(-1.5, 2^2)$ and $X|(Y=1) \sim \mathcal{N}(1, 2^2)$ and whose class prior is $\p(Y=1)=.5$, with varying $B$ (number of random splits) and $N$ (sample size) across $1000$ independent simulations. Based on these distributions, $B=11$ is a reasonable choice.\label{fig:1feature_varB_varn}}	
\end{figure}

\end{document}